\documentclass[10pt,twocolumn,superscriptaddress,showpacs,prd,aps,amsmath,amssymb,nofootinbib]{revtex4-1}
\usepackage{graphicx}
\usepackage{longtable}
\usepackage{orcidlink}
\usepackage{color}
\usepackage{makecell}
\usepackage{appendix}
\usepackage{mathrsfs}
\usepackage{url}

\usepackage[normalem]{ulem}
\usepackage{epstopdf}
\usepackage{color}
\usepackage{soul}
\usepackage{ulem}
\usepackage{bm}
\usepackage{xspace}

\usepackage[]{amsmath}
\usepackage[]{amsfonts}
\usepackage[]{amssymb}
\usepackage[]{mathrsfs}
\usepackage[]{mathtools}
\usepackage{times}
\usepackage{hyperref}
\hypersetup{
  colorlinks=true,        
  linkcolor=blue,         
  citecolor=cyan,         
}
\newcommand{\apjl}{Astrophys. J. Lett.}

\newcommand{\ASAS}{Astron. Astrophys.}

\newcommand{\mnras}[3]{\emph{Mon. Not. Roy. Astron. Soc.} \textbf{#1}, {#2} ({#3})}

\newcommand{\PR}{Phys. Rept.}

\newcommand{\dirac}{\partial\llap{$\diagup$\kern-2pt}}

\newcommand{\fettu}[1]{\mathbf{#1}}

\newcommand{\Tr}{\mathrm{Tr}}

\def\QEQ{{%
			\setbox0\hbox{$I$}%
			\rlap{\hbox to \wd0{\hss--\hss}}\box0
		}}

\begin{document}

\title{Constraints from Gamma-ray Burst Phenomenology on the Hypothesis of Quark Star as the central engines of Gamma-ray Bursts}

\author{Xin-Ying Song\,\orcidlink{0000-0002-2176-8778}}
\email{songxy@ihep.ac.cn}
\affiliation{University of Chinese Academy of Sciences, Chinese Academy of Sciences, Beijing 100049, China}
\affiliation{Key Laboratory of Particle Astrophysics, Institute of high-energy Physics, Chinese Academy of Sciences, Beijing 100049, China}
\date{\today}

\begin{abstract}
The existence of a strange quark star (QS) predicted in the Bodmer-Witten hypothesis has been a matter of debate. The combustion from a neutron star to a strange QS in its accreted process in a low-mass X-ray binary is proposed to be a scenario that generates gamma-ray bursts (GRBs); the baryon contamination of the outflow is very low and mainly from the masses of crusts ($M_{\rm crust}$) of QSs. A special subset of GRBs detected in the past 16 years are collected and used to estimate $M_{\rm crust}$ under this assumption of QSs as central engines. Correspondingly, $M_{\rm crust}$ is calculated in the frameworks of several models for cold dense quark matter (MIT bag model and Nambu-Jona-Lasino model with or without the impacts from the formation of color superconducting condensates being considered), for comparison with the observation. In conclusion, we find that the GRB samples have so far failed to provide positive support for this hypothesis, and the Nambu-Jona-Lasino model in which the existence of hybrid stars is allowed might be more consistent with the observation.

\end{abstract}

\maketitle

\section{Introduction}
The compact star provides a good laboratory for the study of cold dense matter because it is most likely that quark matter (QM) exists in the core of compact stars, as quarks are deconfined in extremely dense baryon matter. The strange quark matter is predicted to be the ground state of QCD at a finite baryon number within the framework of the MIT bag model~\cite[e.g.,][]{1971PhRvD...4.1601B,1976PhLB...62..241B,1984PhRvD..30..272W, 1984PhRvD..30.2379F,1986ApJ...310..261A}. Based on this hypothesis, the conversion of neutron stars to strange stars is taken as a possible origin for gamma-ray bursts (GRBs)~\cite[e.g.,][]{1996PhRvL..77.1210C,2000ApJ...530L..69B,2005MNRAS.362L...4P,2002A&A...387..725O}. The strange quark stars (QSs) could have very thin hadronic crust~\cite[the mass of crust $M_{\rm crust}\sim 10^{-5} M_{\odot}$, or even smaller, $10^{-6} M_{\odot}$,][]{1992ApJ...400..647G, 1997A&A...325..189H}, and the total amount of energy liberated in the conversion is $E_{\rm conv}\sim 10^{53}$ erg~\citep{2000ApJ...530L..69B} for a neutron star (NS) of a typical mass ($\sim1.4M_\odot$). 

For a long time, the long-standing debate has mainly focused on if those co-called conventional pure QSs really exist, or the strange quark matter is most stable. Some works have investigated the deconfined quark matter within the framework of the Nambu-Jona-Lasino (NJL) model ~\cite[e.g.,][]{PhysRev.122.345,PhysRev.124.246,2005PhR...407..205B,2005PhRvD..72f5020B,2005PhRvD..72c4004R, 2024arXiv240915811S}, treating dynamically generated quark masses self-consistently. Note that even if three-flavor color superconducting condensates are considered, the energy per baryon $E/A>930$ MeV; the idea of absolutely stable strange QM is not supported in the NJL model if the vacuum properties of the model are kept at least qualitatively unchanged. 

There are several scenarios for a combustion from a neutron star to a conventional strange QS in the Bodmer-Witten hypothesis. One proposed in ~\cite{1996PhRvL..77.1210C} is that NSs in low-mass X-ray binaries can accrete sufficient mass to undergo a phase transition; in this case, baryon contamination is mainly from the crust of strange QS and a high-entropy clean fireball is produced. In addition, the collapse events and binary merger events could both produce QSs if the density reaches that of quark deconfinement~\cite[e.g.,][]{2006ApJ...639..382L,2009JCAP...09..007C,2019PhRvL.122f1101M,2022PhRvD.106j3030Z}. However, for the latter two cases, the number
of baryons loaded with the fireball is unlikely to be small~\citep{1992ApJ...397..570M,1996PhRvL..77.1210C,2003ApJ...588..931B}.
In this paper, $M_{\rm crust}$ of the QS is calculated within the framework of different theoretical models (MIT bag model and NJL model), for comparison with the observation of the GRBs.

The paper is organized as follows: in Section~\ref{sec:paracons}, a special subset of GRBs detected in the past 16 years are collected and used to estimate $M_{\rm crust}$ under the assumption of QSs as their central engines; in Sections~\ref{sec:MITBAG} and ~\ref{sec:NJLmodel}, compact stars (strange stars or hybrid stars) are constructed within the MIT bag model and NJL model, respectively; the maximum $M_{\rm crust}$ is extracted for comparison with the GRB phenomenology; the results are discussed and summarized in Section~\ref{sec:discussion}.

\section{The Constraints from GRB phenomenology }\label{sec:paracons}
In this paper, we pay specific attention to these GRBs whose prompt emissions are quasi-thermal-dominated \footnote{The low energy photon index in time-averaged spectrum should be well above the synchrotron death line ($>-2/3$) within one standard deviation, or at least in the duration during which more that half of the radiated energy are released in the prompt emission phase.}. This is because the problem of baryon contamination could be avoided in many other mechanisms~\citep[e.g.,][]{1992ApJ...392L...9D,1994MNRAS.270..480T,1996ApJ...473L..79B}, where power is extracted from rapidly spinning neutron stars or black holes with strong magnetic fields by Poynting flux; typically, the outflow in those GRBs should be much cleaner~\citep[e.g.,][]{2013ApJ...765..125L}\footnote{However, even if including those Poynting-flux-dominated GRBs, e.g. GRB 130427A and the B.O.A.T GRB 221009A, the final results will not be affected. In addition, if the Poynting flux is dissipated below the photosphere, the spectrum could have a quasi-thermal component; however, this does not affect the final result much, so the origins of the thermal prompt emission are not strictly required. }. Note that this does not mean that GRBs with QSs as the central engines (QS-GRBs) can not be Poynting-flux-dominated, but rather that we can not distinguish between these two central engines only from the baryon loading of the outflow. Moreover, they are required to have well-measured (or well-estimated) redshifts ($z$) because we find that $z$ is vital for the estimation of the energies of the outflows of GRBs. 

The durations of QS-GRBs are not restricted by theoretical predictions. The previous works on the combustion of a neutron star to a QS~\citep[e.g.,][]{1987PhLB..192...71O,1988PhLB..213..516H,2013PhRvD..87j3007P,2011PhRvD..84h3002H,2010PhRvC..82f2801N} show that the conversion occurs in a very short time in the range of 1 ms$-$ 1 s; thus some works assume a detonation mode ~\citep[][]{1996PhRvL..77.1210C,2000ApJ...530L..69B}, while others proposed a long-term conversion~\citep[e.g.,][]{2002A&A...387..725O,2015PhRvC..92d5801D} or a process that is absolutely unstable with no well-defined burn front.~\citep{1994PhLB..326..111C}.
This does not mean that the produced GRBs must have a short or long duration, because the following processes in the successive forming of a fireball would proceed during a time: cooling by the emission of neutrinos and antineutrinos ($\nu_{\rm e}$ and $\overline{\nu}_{\rm e}$), energy deposition via absorption of $\nu_{\rm e}$ and $\overline{\nu}_{\rm e}$ by nucleons, and the final formation of a fireball via pair production of photons. In particular, the durations of cooling processes are increased by thermal relaxation times affected by the crust~\citep[e.g.,][]{2022A&A...663A..19Z}, which will cause different durations of QS-GRBs. Therefore, the samples selected here should include both short ($<2$ s) and long-duration ($\gtrsim2$ s) GRBs. Approximately 60 GRBs have been selected in the past 16 years, as listed in Figure~\ref{tab:samples}, of which a small fraction ($\sim$ 10\%) has short durations.
\begin{figure}
\begin{center}
\includegraphics[width=0.5\textwidth]{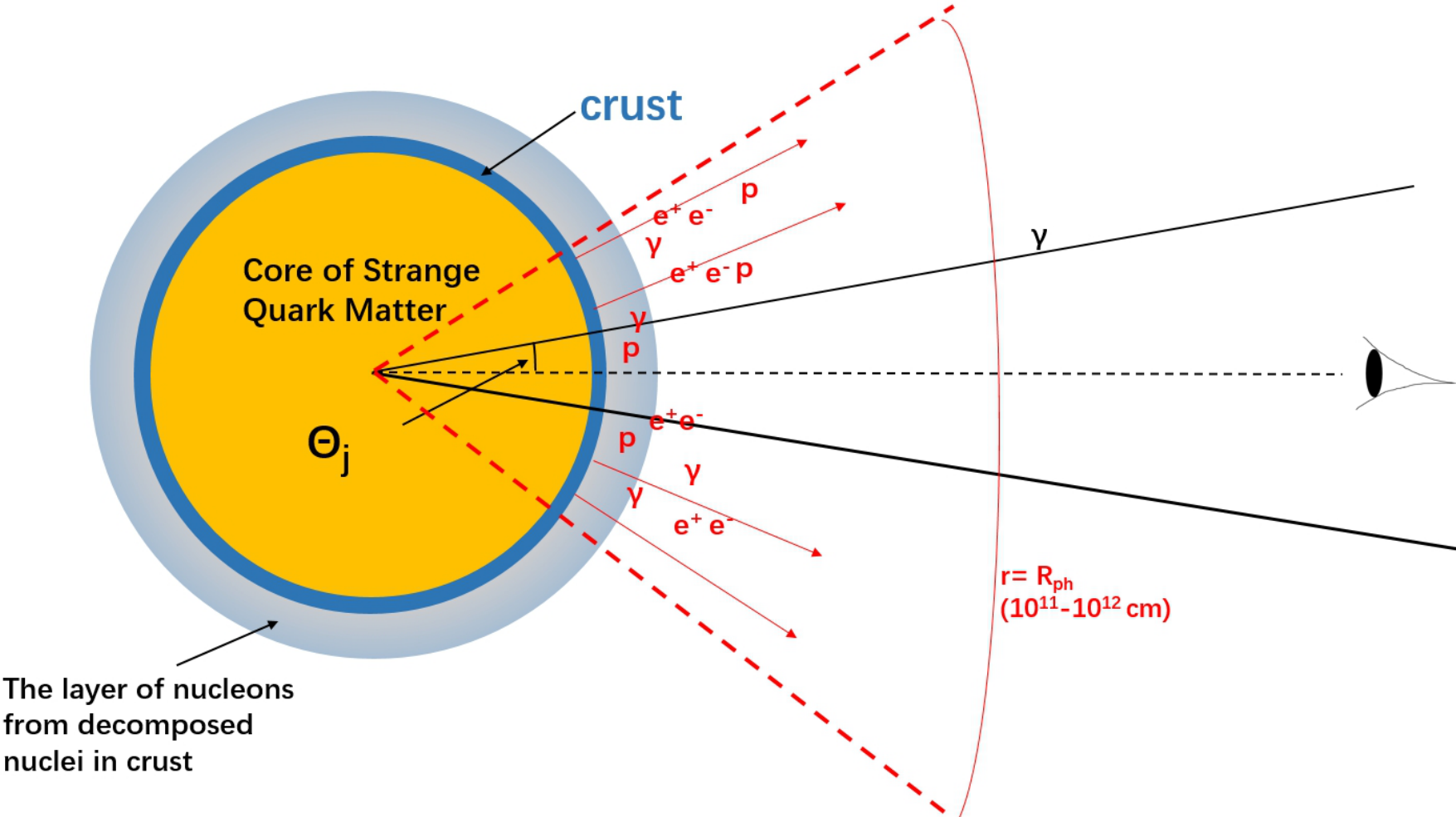}\put(-80,120){(a) }\\
\includegraphics[width=0.42\textwidth]{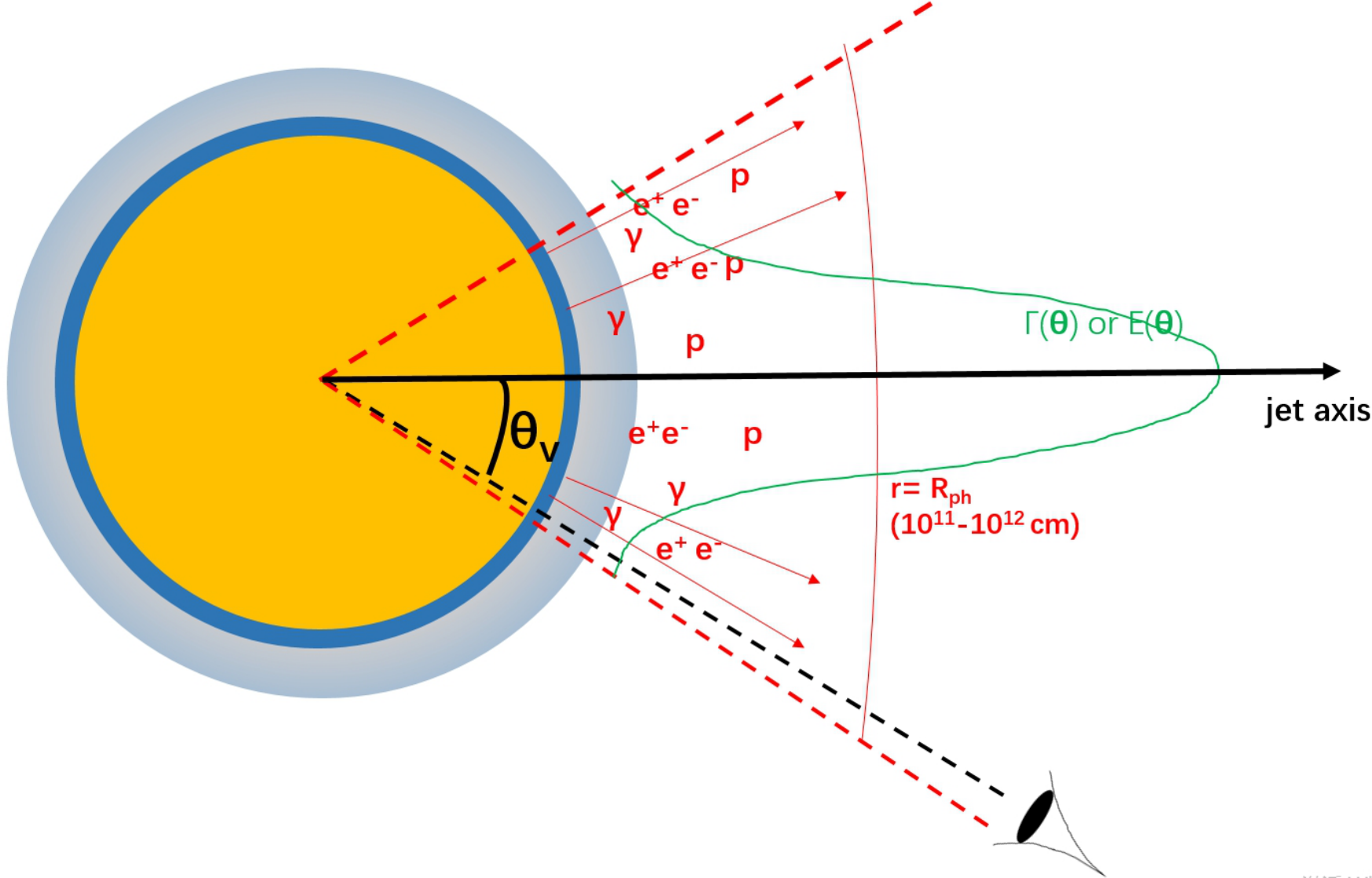}\put(-70,120){(b) }\\
\caption{\label{fig:theta_crust} The sketch of geometric relation (head-on observation in (a) and off-axis in (b), see the details in the text). The dimensions in this figure are distorted for clarity. }
 \end{center}
\end{figure}

After the combustion process, the crust would be heated and the nuclei of this crust may decompose into nucleons~\citep{1996PhRvL..77.1210C}. These nucleons contribute to the baryon contamination of the GRB outflows. The QS-GRB has a low baryon loading ($M_{\rm BL}$) that must be less than $M_{\rm crust}$ in the scenario proposed in \cite{1996PhRvL..77.1210C}. This might be a rough estimate. It is unknown if there appears to be an additional mechanism that causes nucleons at the solid angle of $4\pi$ from the whole star to enter the beam-like outflow; even if there is a strong mechanism for collimation, the nucleons are much heavier than electrons and positrons. First, we assume that there exists an additional strong mechanism for collimation, although it may not be very clear to us.  To simplify, we focus on these QS-GRBs that correspond to NSs with $1.4M_\odot$ and $E_{\rm conv}\sim10^{53}$ erg. Note that NSs with smaller masses could have smaller $E_{\rm conv}$, but there is no reason why QS-GRBs from NSs of larger masses do not exist. If almost all of $E_{\rm conv}$($\sim10^{53}$ erg) is used to generate a narrow jet, the inferred isotropic energy $E_{\rm iso}$ must be extremely large ($\sim10^{55}$ erg given the opening angle $\lesssim0.1$).  Only attention should be paid to the extremely bright bursts with $E_{\rm iso}\gtrsim10^{55}$ erg detected. However, there is almost no GRB in the samples that satisfies this criterion\footnote{
Furthermore, even if we consider those GRBs of which outflows are Poynting-flux dominated, e.g., the B.O.A.T GRB 221009A (even if their outflows are Poynting-flux dominated), their bulk Lorentz vector ($\Gamma$) is not large enough and corresponds to a baryon contamination that is much larger than the allowed maximum $M_{\rm crust}$. }.

Thus, we consider that there is no strong mechanism for collimation in this scenario; the fireball could be launched from a considerable region within a solid angle and the outflow is approximately uniform, as shown between the two dashed red lines in Figure~\ref{fig:theta_crust} (a); this means that a smaller $E_{\rm iso}$ could be allowed. The observed flux is the average in a small solid angle, which corresponds to the head-on observation, and is not affected by the outflows outside of the cone of $\theta_{\rm j}$ in other directions far from the line of sight; note that measured $\theta_{\rm j}$ should be smaller than that of the fireball-launching region. 
Since there is no strong mechanism for collimation, the decomposed nucleons at the other angles in $4\pi$ can hardly enter the solid angle of the region in which the fireball is launched and $M_{\rm BL}\lesssim M_{\rm crust}f_{\rm b}$. Given $M_{\rm BL}=E_{\rm iso, \gamma} f_{\rm b}/\eta_{\gamma}\Gamma$ (where $\eta_\gamma$ is the radiative efficiency in $\gamma$-rays and $f_{\rm b}=1-\rm{cos}\theta_{\rm j}$ is the beaming factor), $M_{\rm crust}\gtrsim E_{\rm iso, \gamma} /\eta_{\gamma}\Gamma$. Note that $E_{\rm iso, \gamma} /\eta_{\gamma}\gtrsim10^{53}$ erg if $E_{\rm conv}\sim10^{53} $ erg. In addition, the effect of rotating could not affect the result obtained here much, since $E_{\rm iso,\gamma}$ and $M_{\rm BL}$ are both time-integrated values. Moreover, rapidly rotating stars are not considered, as mentioned in the beginning of this section.

 For the structured jet (for example, the distributions of $\Gamma$ and energy are power law functions of the angular distance from the center as proposed in~\cite{2002MNRAS.332..945R}), the case is similar to the above discussion if $\theta_{\rm v}\ll \theta_{\rm j}$ ($\theta_{\rm v}$ is the viewing angle as proposed in ~\cite{2002ApJ...571..876Z} and \cite{2002MNRAS.332..945R}, while here $\theta_{\rm j}$ is the opening angle of the jet). For off-axis observation ($\theta_{\rm v}\sim\theta_{\rm j}$) as shown in Figure~\ref{fig:theta_crust} (b), the total energy of the outflow ($E_{\rm tot}$) is underestimated if it is estimated by $E_{\rm iso, \gamma}f_{\rm b}/\eta_{\gamma}$. If $\Gamma$ is estimated by the jet break time~\citep{2001ApJ...562L..55F}, it should be larger than that corresponding to $\theta_{\rm v}$ since $\Gamma(\theta)$ decreases with $\theta$. As a consequence, $M_{\rm BL}$ could be underestimated with the same method as that in head-on observation\footnote{If the other method based on the spectrum is used, the estimated $\Gamma$ could be corresponds to the off-axis $\Gamma$ and $M_{\rm crust}$ is not underestimated.} and $E_{\rm iso, \gamma}/\eta_\gamma$ could be smaller than $10^{53}$ erg if $E_{\rm conv}\sim10^{53} $ erg. In summary, $M_{\rm crust}\gtrsim E_{\rm iso, \gamma} /\eta_{\gamma}\Gamma$, which relates $M_{\rm crust}$ and the observed quantities; the lower limits (L.L.) of $M_{\rm crust}$ are extracted as shown in Table~\ref{tab:samples}.

 For the first part from 090902B to 110731A labeled `A' in the column of comments in Table~\ref{tab:samples}, $\Gamma$ values are from \cite{2015ApJ...813..127P} and estimated by the method proposed for the pure hot fireball \citep{2007ApJ...664L...1P,2015ApJ...813..127P} \footnote{$\Gamma=[(1.06)(1+z)^{2}d_{\rm L}\frac{Y\sigma_{\rm T}F^{\rm ob}}{2m_{p}c^{3}\mathcal{R}}]^{1/4}$, which where $d_{\rm L}$ is the luminosity distance, $\sigma_{\rm T}$ is the Thomson scattering cross section, and $F^{\rm ob}$ is the observed flux.$\mathcal{R}=(\frac{F^{\rm ob}_{\rm thermal}}{\sigma T^{4}_{\rm max}})^{1/2}$ where $F^{\rm ob}_{\rm thermal}$ is the thermal emission flux. $\sigma$ is Stefan–Boltzmann constant. Note that in this method, the emission is assumed to be from the saturated regime, thus $\Gamma$ reaches the value of $\eta$. $Y$ is the ratio between the total outflow energy and the energy emitted in the gamma rays, and $Y\geq1$. $Y$ is taken to be 2 and $\eta_{\gamma}$=0.5.}.  Note that for photospheric emission from the hybrid outflow,  $\Gamma$ (or $\eta$) should be diagnosed by `top-down' approach~\citep{2015ApJ...801..103G,2024ApJ...961..137S}. The data of the GRB samples in the second part (labeled `B') are mainly from published references (see the notes below the table). In fact, the uncertainties of the estimated $\Gamma$ do not have much impact on the order of magnitude of $M_{\rm crust}$ compared with those of $E_{\rm iso, \gamma}$. For the third part that begins from GRB 081118 (labeled `C'), $\Gamma$ (or $\eta$) are not determined in any published references and are estimated with the methods mentioned above.

Among the samples, some GRBs appear to be thermally dominated in the prompt emission phase in the $\gamma$-ray energy band but are followed by X-ray afterglows, for example, 101219B~\citep{2015ApJ...800L..34L}. For 101219B, the kinematic energy ($E_{\rm k}$) is about 20 times greater than $E_{\rm iso, \gamma}$. These cases are labeled as `*' in the first column of Table~\ref{tab:samples}. $E_{\rm k}$ could be derived from data of \textit{Swift} with the method in, e.g.~\cite{2007ApJ...655..989Z}. However, there are some uncertainties in the model of afterglow~\citep[e.g.][]{2006MNRAS.369..197F,2007ApJ...655..989Z} as well as in estimating $\theta_{\rm j}$, which bring considerable uncertainty to $\eta_{\gamma}$. Their prompt emissions are not luminous and X-ray afterglows are observed, which indicate small $\eta_{\gamma}$ and are inconsistent with the photosphere model of a pure hot fireball or the internal-collision-induced magnetic reconnection and turbulence (ICMART) model~\citep{2011ApJ...726...90Z}. The possible model is the internal shock model, with $\eta_{\gamma}\lesssim10\%$, which is used to estimate the lower limits of $M_{\rm crust}$ and $E_{\rm iso,\gamma}/\eta_{\gamma}$.

\begin{table*}
\caption{The GRB sample for the search of QS-GRBs.\label{tab:samples}}
\begin{ruledtabular}
\begin{tabular}{cccccccc}
GRB ID
& Type
&z
&Fluence from about 8 to 1000 keV
&$E_{\gamma,\rm iso}$
&$\Gamma$(or $\eta$)
&Estimated $M_{\rm crust}$
&Comments
\\
&
&
&(10$^{-6}$ erg cm$^{-2}$)
&($10^{52}$ erg)
&
&($10^{-7} M_{\odot}$)
&\\
\hline
090902B &II &1.82 &$436.00\pm6.00$ &$373.69\pm5.14$  &$995\pm75$ &$41963.18\pm3215.34$ &\\
090926B &II &1.24 &$145.00\pm4.00$ &$60.74\pm1.68$   &$110\pm10$ &$61699.40\pm5861.59$ &\\
101219B$^{*}$ &II &0.55 &$5.50\pm0.40$ &$0.45\pm0.03$   &$138\pm8$ &$3680.61\pm1487.68$ &A\footnote{ $\Gamma$ values are from \cite{2015ApJ...813..127P} and estimated by the method proposed for the pure hot fireball \citep{2007ApJ...664L...1P,2015ApJ...813..127P}.}\\
100724B &II &1.00 &$244.00\pm0.60$ &$67.34\pm0.17$   &$325\pm100$ &$23149.17\pm7123.05$\\
110731A &II &2.83 &$22.18\pm0.06$ &$40.83\pm0.11$   &$765\pm200$ &$5963.84\pm1559.26$\\\hline
141207A &II &10.00 &$74.70\pm3.00$ &$915.37\pm36.76$   &$\sim1000$ &$102276.52\pm4107.49$ &\\
190109A &II &1.50 &$7.60\pm0.60$ &$4.56\pm0.36$   &$\sim150$ &$3397.20\pm268.20$ &\\
210121A &II &0.37 &$123.00\pm8.00$ &$4.48\pm0.29$   &$\sim200$ &$2504.83\pm162.92$&\\
210610B &II &1.13 &$17.30\pm0.30$ &$6.10\pm0.11$   &$\sim400$ &$1704.41\pm29.56$&B\footnote{141207A: The redshift and $\Gamma$ are from ~\cite{2016ApJ...833..139A}, where $z$ is estimated with the Yonetoku relation~\citep{2004ApJ...609..935Y}.  190109A: $\Gamma$ is from ~\cite{2022ApJ...932...69L}.
210121A:$z$ and $\Gamma$ are from ~\cite{2022ApJ...931..112S}, which are estimated by fitting the data to an intermediate photospheric model from a structured jet.
210610B and 221022B: $\Gamma$ is from ~\cite{2024ApJ...961..137S}, which is estimated by the “top-down” approach~\citep{2015ApJ...801..103G} proposed by Gao and Zhang, with a characteristic temperature and flux. 
220426A: $\Gamma$ is from ~\cite{2022MNRAS.517.2088S}.
230307A: $\Gamma$ is estimated using the same method as 210610B and 221022B.
231129C: $\Gamma$ is from ~\cite{2024ApJ...972..132C}.} \\
220426A &II &1.40 &$101.00\pm1.00$ &$53.27\pm0.53$   &$\sim500$ &$11903.57\pm117.86$ &\\
221022B &II &0.61 &$71.40\pm0.70$ &$7.30\pm0.07$   &$\sim300$ &$2717.51\pm26.64$ &\\
230307A &I &0.07 &$4200.00\pm80.00$ &$4.29\pm0.08$   &$\sim400$ &$1199.66\pm22.85$&\\
231129C &II &0.50 &$84.10\pm0.40$ &$5.71\pm0.03$   &$\sim300$ &$2127.86\pm10.12$ &\\\hline
081118$^{*}$~\cite{2008GCN..8550....1B,2008GCN..8531....1D} &II &2.58 &$0.11\pm0.06$ &$0.18\pm0.10$ &$113\pm9$ &$\gtrsim872$&\\
081221~\cite{2008GCN..8700....1P,2008GCN..8704....1W} &II &0.70 &$37.00\pm1.00$ &$5.00\pm0.14$ &$135\pm1$ &$4135.22\pm115.79$&\\
081222~\cite{2008GCN..8715....1B,2008GCN..8718....1G} &II &2.77 &$13.20\pm0.40$ &$23.45\pm0.71$ &$369\pm12$ &$7100.05\pm322.28$&\\
090424~\cite{2009GCN..9230....1C,2009GCN..9243....1C} &II &0.54 &$52.00\pm5.00$ &$4.20\pm0.40$ &$183\pm3$ &$2568.82\pm249.87$&\\
091020~\cite{2009GCN.10095....1C,2009GCN.10053....1X} &II &1.71 &$10.00\pm2.00$ &$7.64\pm1.53$ &$152\pm12$ &$5617.14\pm1206.28$&\\
100414A~\citep{2010GCN.10595....1F,2010GCN.10606....1C} &II &1.37 &$129.00\pm2.00$ &$65.14\pm1.01$ &$648\pm13$ &$11231.36\pm277.98$&\\
100814A~\cite{2010GCN.11099....1V,2010GCN.11091....1F} &II &1.44 &$19.80\pm0.60$ &$11.01\pm0.33$ &$187\pm12$ &$6594.77\pm475.52$&\\
100728A~\cite{2010GCN.11021....1G,2013GCN.14500....1K} &II &2.11 &$195.00\pm35.00$ &$216.34\pm38.83$ &$567\pm14$ &$42600.55\pm7716.31$&\\
120712A~\cite{2012GCN.13469....1G,2012GCN.13458....1T} &II &4.15 &$4.43\pm0.05$ &$15.12\pm0.17$ &$422\pm44$ &$4002.84\pm422.12$&\\
120922A~\cite{2012GCN.13809....1Y,2012GCN.13810....1K} &II &3.10 &$6.50\pm0.40$ &$13.91\pm0.86$ &$154\pm7$ &$10066.15\pm779.78$&\\
121211A$^{*}$~\cite{2012GCN.14078....1Y,2012GCN.14059....1P} &II &1.02 &$0.49\pm0.05$ &$0.14\pm0.01$ &$138\pm9$ &$\gtrsim568$&\\
130408A~\cite{2013GCN.14368....1G} &II &3.76 &$12.00\pm2.00$ &$35.05\pm5.84$ &$614\pm44$ &$6376.93\pm1157.27$&\\
130609B~\cite{2013GCN.14869....1P,2013GCN.14888....1R} &II &1.30 &$60.20\pm0.70$ &$27.60\pm0.32$ &$420\pm9$ &$7342.70\pm172.53$&\\
140206A~\cite{2014GCN.15796....1V,2014GCN.15802....1D} &II &2.73 &$14.70\pm0.30$ &$25.48\pm0.52$ &$355\pm9$ &$8017.29\pm259.54$&\\
140419A~\cite{2014GCN.16134....1G} &II &3.96 &$4.90\pm1.90$ &$15.52\pm6.02$ &$544\pm82$ &$3188.42\pm1327.13$&\\
140423A~\cite{2014GCN.16152....1V,2014GCN.16150....1T} &II &3.26 &$21.00\pm1.00$ &$48.81\pm2.32$ &$365\pm13$ &$14941.33\pm896.36$&\\
140801A~\cite{2014GCN.16660....1G} &II &1.32 &$12.20\pm0.10$ &$5.75\pm0.05$ &$261\pm3$ &$2456.87\pm36.39$&\\
141028A~\cite{2014GCN.16971....1R,2014GCN.16983....1X} &II &2.30 &$34.78\pm0.09$ &$45.00\pm0.12$ &$500\pm13$ &$10057.97\pm254.92$&\\
141225A~\cite{2014GCN.17241....1J,2014GCN.17234....1G} &II &0.92 &$6.50\pm0.30$ &$1.51\pm0.07$ &$187\pm15$ &$897.87\pm81.19$&\\
150206A~\cite{2015GCN.17427....1G} &II &2.09 &$55.20\pm0.64$ &$60.27\pm0.70$ &$438\pm34$ &$15364.87\pm1192.91$&\textbf{C\footnote{see the description in the text. The data (e.g. $z$, fluence) from 081118 to 241107A are from the public data; data that support the findings of this article are openly available on the web of General Coordinates Network (GCN, https://gcn.nasa.gov/circulars/).
 }}\\
150314A~\cite{2015GCN.17587....1G} &II &1.76 &$91.00\pm4.00$ &$73.11\pm3.21$ &$606\pm6$ &$13473.50\pm606.60$&\\
160521B~\cite{2016GCN.19443....1Y,2016GCN.19456....1R} &II &2.50 &$13.20\pm1.60$ &$19.71\pm2.39$ &$499\pm11$ &$4417.52\pm544.31$&\\
180314A~\cite{2018GCN.22513....1T} &II &1.45 &$14.70\pm0.60$ &$8.23\pm0.34$ &$219\pm3$ &$4189.21\pm182.60$&\\
180620B~\cite{2018GCN.22825....1S} &II &1.12 &$7.70\pm0.04$ &$2.64\pm0.01$ &$165\pm12$ &$1789.88\pm126.99$&\\
180914B~\cite{2018GCN.23240....1F,2018GCN.23246....1D} &II &1.10 &$1150.00\pm50.00$ &$379.65\pm16.51$ &$511\pm15$ &$83057.22\pm4362.80$&\\
190114C~\cite{2019GCN.23737....1F} &II &0.42 &$483.00\pm1.00$ &$23.40\pm0.05$ &$463\pm3$ &$5649.12\pm35.66$&\\
191004B~\cite{2019GCN.25974....1S} &II &1.26 &$4.13\pm0.40$ &$1.78\pm0.17$ &$272\pm17$ &$732.81\pm84.36$&\\
200826A$^{*}$~\cite{2020GCN.28287....1M,2020GCN.28301....1S} &II &0.75 &$4.80\pm0.10$ &$0.74\pm0.02$ &$183\pm4$ &$\gtrsim2271$&\\
201020B~\cite{2020GCN.28710....1M,2020GCN.28765....1K} &II &0.80 &$39.29\pm0.40$ &$7.03\pm0.07$ &$219\pm3$ &$3587.06\pm57.78$&\\
210731A~\cite{2021GCN.30583....1K,2021GCN.30573....1L} &II &1.25 &$4.90\pm0.20$ &$2.09\pm0.09$ &$230\pm7$ &$1015.93\pm52.59$&\\
210822A~\cite{2021GCN.30694....1F} &II &1.74 &$120.00\pm11.00$ &$94.23\pm8.64$ &$694\pm27$ &$15161.66\pm1505.62$&\\
220101A~\cite{2022GCN.31433....1T} &II &4.62 &$4.00\pm0.07$ &$16.10\pm0.28$ &$591\pm14$ &$3043.13\pm87.54$&\\
220527A~\cite{2022GCN.32152....1L} &II &0.86 &$59.80\pm3.10$ &$12.16\pm0.63$ &$265\pm3$ &$5129.16\pm271.83$&\\
221226B$^{*}$~\cite{2022GCN.33112....1L,2022GCN.33110....1X} &II &2.69 &$0.78\pm0.05$ &$1.32\pm0.08$ &$279\pm11$ &$\gtrsim2648$&\\
230812B~\cite{2023GCN.34409....1D,2023GCN.34403....1F} &II &0.36 &$327.00\pm7.00$ &$11.26\pm0.24$ &$339\pm2$ &$3714.03\pm82.68$&\\
231210B~\cite{2023GCN.35359....1F} &II &3.13 &$4.02\pm0.56$ &$8.74\pm1.22$ &$457\pm36$ &$2135.76\pm342.58$&\\
231215A~\cite{2023GCN.35373....1T,2023GCN.35377....1F} &II &2.31 &$102.00\pm9.00$ &$132.48\pm11.69$ &$923\pm49$ &$16045.43\pm1648.57$&\\
240825A~\cite{2024GCN.37302....1F} &II &0.65 &$166.00\pm8.00$ &$19.31\pm0.93$ &$502\pm5$ &$4298.34\pm212.35$&\\
100206A$^{*}$~\cite{2010GCN.10381....1V,2010GCN.10410....1B} &I &0.41 &$0.93\pm0.04$ &$0.04\pm0.00$ &$323\pm27$ &$\gtrsim72$&\\
150424A$^{*}$~\cite{2015GCN.17758....1C,2015GCN.17752....1G} &I &0.30 &$18.10\pm1.10$ &$0.43\pm0.03$ &$516\pm22$ &$\gtrsim462$&\\
150906B~\cite{2015GCN.18296....1R} &I &0.12 &$28.00\pm2.00$ &$0.10\pm0.01$ &$322\pm17$ &$34.62\pm3.06$&\\
201227A$^{*}$~\cite{2020GCN.29196....1S,2021GCN.29281....1B} &I &0.05 &$3.60\pm0.10$ &$0.0022\pm0.0001$ &$300\pm13$ &$\gtrsim4.02$&\\
210704A$^{*}$~\cite{2021GCN.30388....1R,2021GCN.30452....1M} &I &0.11 &$19.50\pm0.20$ &$0.06\pm0.01$ &$150\pm2$ &$\gtrsim43$&\\
240615A$^{*}$~\cite{2024GCN.36777....1V,2024GCN.36685....1R} &I &4.50 &$1.56\pm0.06$ &$6.04\pm0.23$ &$\lesssim2100$ &$\gtrsim1588$&\\
241107A~\cite{2024GCN.38205....1F,2024GCN.38205....1F,2024GCN.38187....1M} &I &0.52 &$1.43\pm0.30$ &$0.11\pm0.02$ &$427\pm23$ &$27.91\pm6.04$&\\
\end{tabular}
\end{ruledtabular}
\end{table*}

\section{ Conventional strange stars constructed in the MIT bag model }\label{sec:MITBAG}
The bulk properties of quark matter could be described with the phenomenological
MIT bag model~\citep[e.g.,][]{1976PhLB...62..241B, 1984PhRvD..30.2379F,1986A&A...160..121H}. The thermodynamic potential density is a function of the mass of the strange quark ($M_s$) and the strong
interaction coupling constant ($\alpha_{\rm c}$) by allowing for transformations mediated by weak interactions between quarks and leptons. To the first order of $\alpha_{\rm c}$, it is determined to be (see details in \cite{1978PhRvD..17.1109F,1984PhRvD..30.2379F}): 
\begin{eqnarray}
    \Omega_{\rm NQ} &=& - (1-\frac{2\alpha_{\rm c}}{\pi})\frac{3 \mu^4}{4\pi^2}
+ \frac{(3-\frac{2\alpha_{\rm c}}{\pi}) M_s^2 \mu^2}{4\pi^2}+ O(\frac{M_s^4}{10^2})\nonumber\\
&+&O(\frac{M_s^6}{10^4\mu^2}) +B,
\label{eq:NQ}
\end{eqnarray}
where $\mu$ is the quark chemical potential and $B$ is the bag constant.
Due to the slight deficit of $s$ quarks relative to $u$ and $d$, a few electrons will appear in chemical equilibrium in strange quark matter and the electrons bound by the Coulomb force can extend several hundred fermis beyond the quark surface. The large outward-directed electric field is capable of supporting some normal material, which gives birth to a thin hadronic crust \citep{1986ApJ...310..261A,1992ApJ...400..647G}. The density at the base of the nuclear crust ($\rho_{\rm crust}$) has an upper limit of the neutron drip density ($\rho_{\rm drip}$). Some works~\citep[e.g.,][]{1997A&A...325..189H} revised the value of $\rho_{\rm crust}$ by solving the Poisson's equation around the gap width of $Z_{\rm g}$ between the crust and the QM core as below ($\alpha$ is the fine-structure constant; $V$ the electrical potential and the subscripts $q$ and $c$ denote those of the quark core and the crust, respectively; $Z_{\rm g}\gtrsim200$ nm, it is taken to be 200 nm to obtain the upper limit of $M_{\rm crust}$),
\begin{equation}
\frac{d^2V}{dz^2}=\left\{
\begin{array}{c}
\frac{4\alpha}{3\pi}(V^3-V_{\rm q}^3),\text{  }%
z\leq0, \\
\frac{4\alpha}{3\pi}V^3,\text{ 
}%
0<z \leq Z_{\rm g}, \\
\frac{4\alpha}{3\pi}(V^3-V^3_{\rm c}),\text{ 
}z>Z_{\rm g}.%
\end{array}%
\right.
\end{equation}
$\rho_{\rm crust}$ is determined more accurately and is affected by various $\alpha_{\rm c}$, $B$, and $M_s$. For the static configuration, the mass of the crust ($M^{\rm stat}_{\rm crust}$) and other properties of quark stars are obtained by solving the well-known Tolman-Oppenheimer-Volkoff (TOV) equation for the hydrostatic equilibrium of self-gravitating matter~\citep{1939PhRv...55..374O}.

\begin{figure*}
\begin{center}
\includegraphics[width=0.9\textwidth]{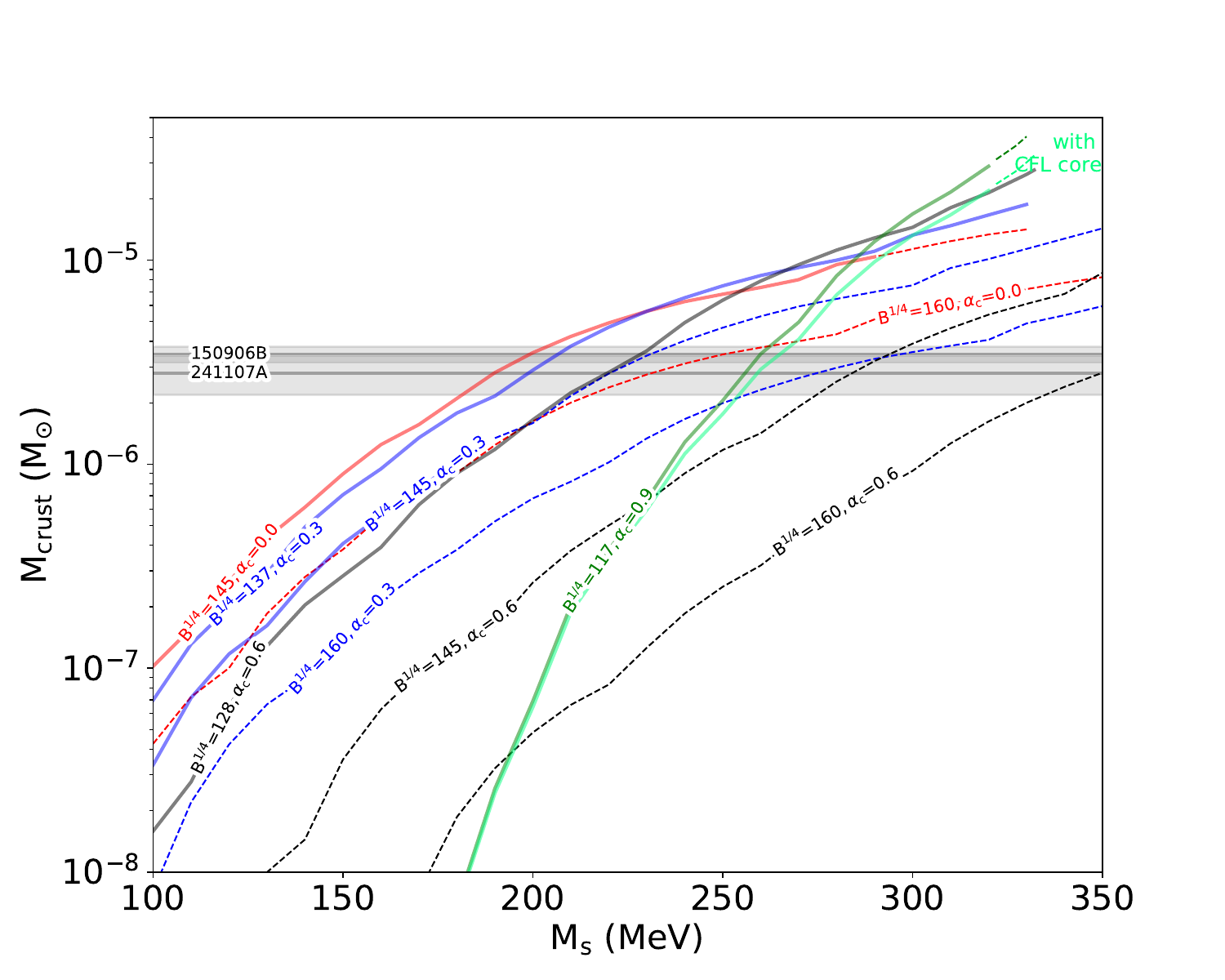}\put(-130,90){(a) }\\
\includegraphics[width=0.45\textwidth]{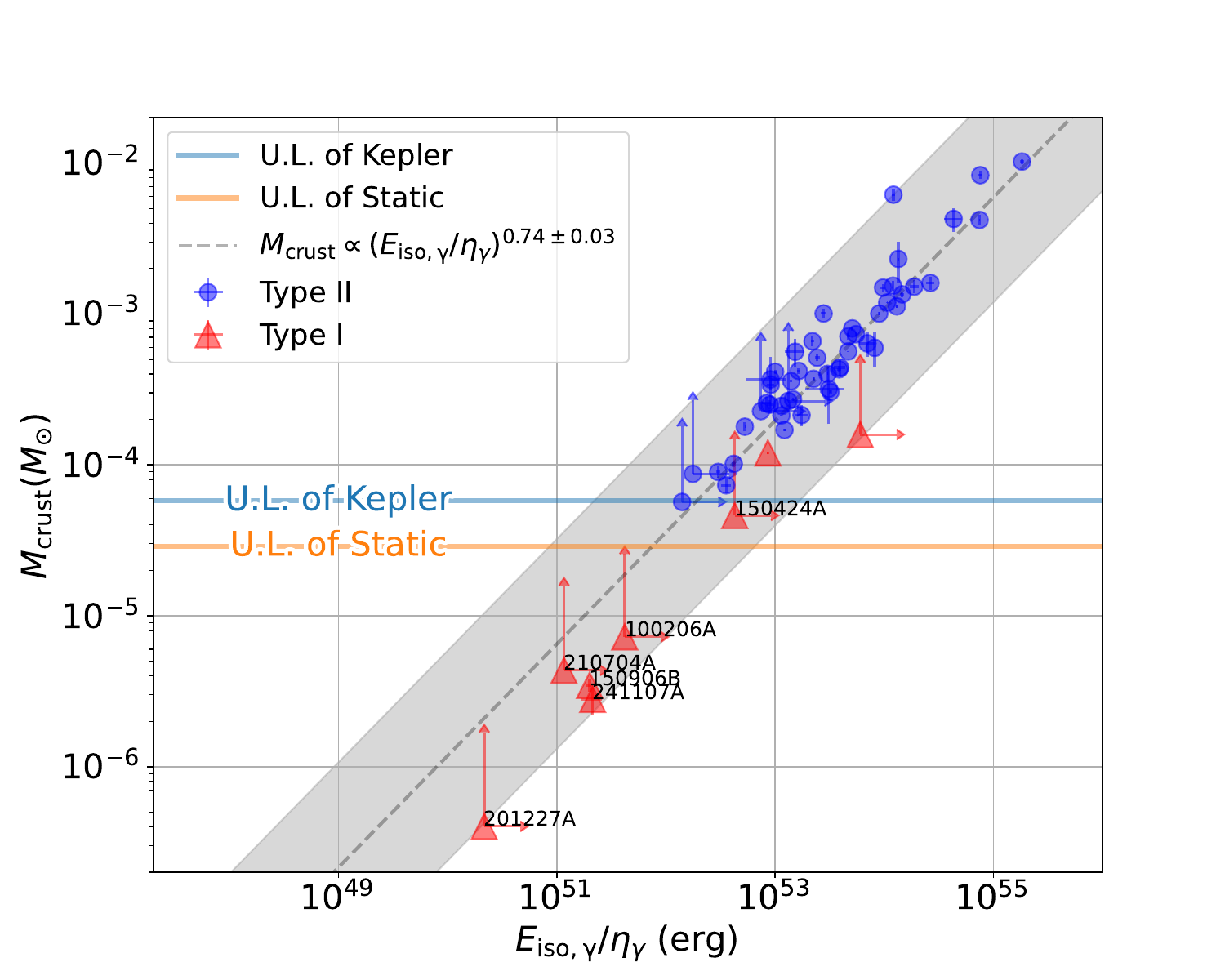}\put(-130,120){(b) }
\includegraphics[width=0.45\textwidth]{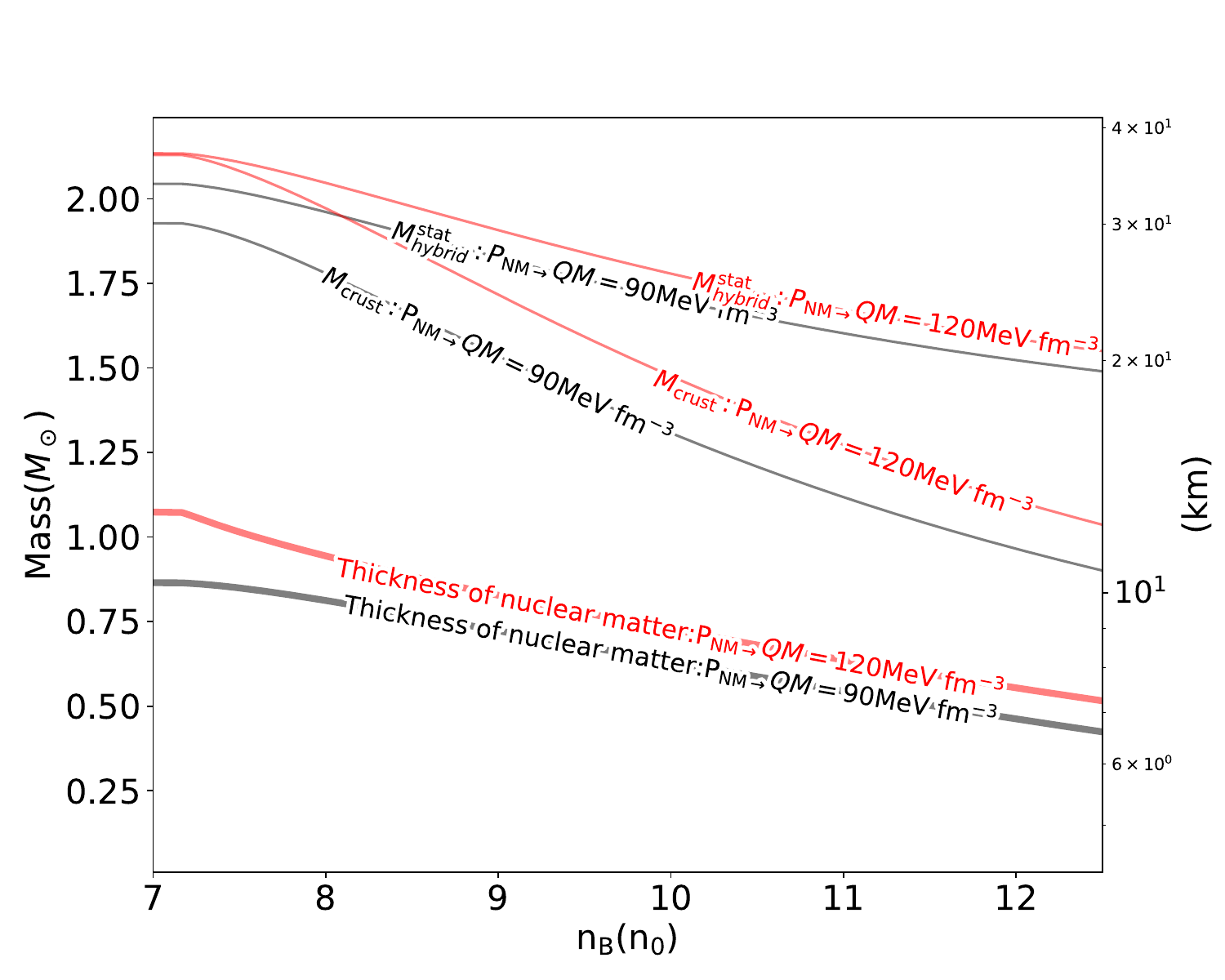}\put(-130,120){(c) }\\
\caption{ (a) The estimated maximum $M_{\rm crust}$ in MIT bag model with various parameters. The dashed lines or dashed parts in lines denote those with $E/A>930$ MeV. The light green line labeled `with CFL core' denotes $M_{\rm crust}$ with CFL condensate being considered for $\alpha=0.9$ and $B^{1/4}=117$ MeV. (b) $M_{\rm crust}$ versus $E_{\rm iso,\gamma}/\eta_\gamma$. The gray dashed line shows the correlation between L.L. of $M_{\rm crust}$ and $E_{\gamma, \rm iso}/\eta_{\gamma}$ above the upper limit of $M_{\rm crust}$ in MIT bag model (the blue horizontal line), $M_{\rm crust}=10^{-42.91\pm0.70} (E_{\gamma, \rm iso}/\eta_{\gamma})^{0.74\pm0.03}$ ; the gray shadow denotes the region within the one standard deviation of ratio of $M_{\rm crust}/(E_{\gamma, \rm iso}/\eta_{\gamma})^{0.74}$. (c) In the hybrid stars in NJL model: $M_{\rm crust}$, masses (the left y-axis) and thicknesses of nuclear matter (the right y-axis) as functions of the central density $n_{B}$ in the unit of $n_0$ where $n_0=0.16$ fm$^{-3}$ is the nuclear saturation density.\label{fig:Mcrust}}
 \end{center}
\end{figure*}

The bag constant should be large enough, so that nuclei with high atomic numbers would be unstable against decay into non-strange two-flavor quark matter~\citep{1984PhRvD..30.2379F}. For $\alpha_{\rm c}=0$, $B^{1/4}$ should be greater than 145 MeV and a smaller $B^{1/4}$ is allowed for $\alpha_{\rm c}>0$ ($B^{1/4}\gtrsim$137, 128 and 117 MeV for $\alpha=0.3$, $0.6$ and 0.9). In the numerical results, it is found that for the same $\alpha_{\rm c}$ and ranges of $M_{s}$, a smaller $B^{1/4}$ usually corresponds to a larger $M^{\rm stat}_{\rm crust}$.  
The maximum $M^{\rm stat}_{\rm crust}$ for various parameters is shown in Figure~\ref{fig:Mcrust} (a). The largest $M^{\rm stat}_{\rm crust}$ is about $3\times10^{-5} M_{\odot}$ at $M_s\simeq320$ MeV at $\alpha_{\rm c}$=0.9. A higher $M_s$ will not be considered, because $E/A$ becomes larger than 930 MeV, or the pressure would not vanish on the surface of the QS.
For a rotating QS, $M_{\rm crust}$ could be about two times larger and below $10^{-4}M_{\odot}$. There are two GRBs (150906B and 241107A) with estimated $M_{\rm crust}$ below this upper limit, as shown in Figure~\ref{fig:Mcrust} (b).

Let us discuss the impact of the formation of color superconducting condensates on $M_{\rm crust}$. In the MIT bag model, the spin-zero two-flavor superconducting (2SC) phase may not exist because its thermodynamic potential is higher than that of unpaired quarks~\citep{2002JHEP...06..031A}. Thus, the color-flavor-locked (CFL) phase seems to be the most favored. However, QM in the CFL phase is rigorously electrically
neutral and no electrons are required~\citep{2001PhRvL..86.3492R}. Therefore, a thin crust cannot be suspended with a gap from the quark core where all QM is in the CFL phase. Note that the criterion for the stability of the CFL phase (with the gap parameter $\Delta_{\rm CFL}$) is $\Delta_{\rm CFL}>\frac{M_s^2}{4\mu}$ \citep[e.g.][]{2001PhRvD..64g4017A,2002JHEP...06..031A}. where $\Delta_{\rm CFL}$ ranges around $10-100$ MeV~\citep{2001PhRvL..86.3492R, 2004PhRvL..92v2001A,2005PhRvD..71e4009A}. There exists a probability that the configuration that the CFL core is surrounded by an unpaired QM could be assumed\footnote{As discussed in \cite{2024arXiv240915811S}, it is possible that the CFL core is surrounded by QM in the CSL phase if the temperature falls below a few MeV. The difference between the equations of the state for unpaired QM and CSL QM is very small; therefore, the numerical results obtained for unpaired QM could be approximately taken as those for CSL QM.}. Given $\alpha_{\rm c}=0.9$, $\Delta_{\rm CFL}$ ranging from 20 to 80 MeV and $160$ MeV$\gtrsim M_s\gtrsim320$ MeV, the maximum $M_{\rm crust}$ considering the CFL condensate is shown in Figure~\ref{fig:Mcrust}(a) (in light green line), which is smaller than that of the unpaired QM.

There are some other possibilities for the existence of a subnuclear crust if the formation of color superconducting condensates is considered, e.g., color-spin-locked QM (CSL)~\cite[e.g.,][]{2003PhRvL..91x2301S, 2005PhRvD..71e4016S, 2005PhRvD..72c4008A,2024arXiv240915811S}, or gapless CFL QM~\citep[e.g.,][]{2004PhRvL..92v2001A,2005PhRvD..71e4009A}. However, regardless of the types of color superconducting condensates, they would not lead to a much larger gravitational mass as well as $M_{\rm crust}$.\footnote{As shown in \cite{2011ApJ...740L..14W} and \cite{2024arXiv240915811S}, pure CFL QSs could have larger gravitational masses. However, for QSs with CFL core surrounded by unpaired QM (or CSL QM), the masses could be slightly smaller than those of pure unpaired QM QS as shown by the numerical results in Figure 2 (d) in \cite{2024arXiv240915811S}. } In summary, the upper limit of $M_{\rm crust}$ would not be larger even if color superconducting condensates formed in the quark core.

 

\section{Hybrid stars constructed by NJL model}\label{sec:NJLmodel}
In the framework of NJL model, the thermodynamic potential density is given by,
\begin{eqnarray}
\Omega_{\rm NJL} &=& \Omega_{e}-\Omega_{ \rm Vac} +
\Omega_{\rm diquark,\Delta}
+2 G_S \sum_{\alpha=1}^{3} \sigma_\alpha^2 \nonumber\\
&-& 4 K \sigma_u \sigma_d \sigma_s
-T \sum_{n} \int_{|\fettu k|<\Lambda} \frac{d^3 k}{(2\pi)^3} \frac{1}{2} \Tr \ln  \frac{S^{-1}(i\omega_n, \fettu{k})}{T}, \nonumber \\
\label{eq:Omega_NJL}
\end{eqnarray}
where $\Omega_e$ is the thermodynamic potential of ultrarelativistic
electrons; $\Omega_{\rm Vac}$ is contribution from the vacuum at $T=\mu=0$, where $T$ is the temperature and the unit is MeV with the Boltzmann constant $k_{\rm B}=1$ in natural units;
$\Omega_{\rm diquark,\Delta}$, $2 G_S \sum_{\alpha=1}^{3} \sigma_\alpha^2 \nonumber$ and $4 K \sigma_u \sigma_d \sigma_s$ denote the contributions from diquark condensates, quark-antiquark condensates and the `t Hooft interaction, respectively; the last term denotes the summation of thermodynamic potential of all (quasi-)particles. In real QCD the ultraviolet modes decouple because of asymptotic freedom, but in the NJL model this feature is added by hand, through a UV momentum cutoff $\Lambda$ in the momentum ($\fettu{k}$) integrals. $S^{-1}$ is the inverse full quark propagator 
in the Nambu-Gorkov representation (e.g. see the details in \cite{2005PhR...407..205B,2005PhRvD..72f5020B,2005PhRvD..72c4004R}).

As concluded in previous works~\citep[e.g.,][]{2005PhR...407..205B, 2024ApJ...966....3Y,2024arXiv240915811S}, the NJL model does not support the idea of absolutely stable quark matter if one must keep the vacuum properties of the model at least qualitatively unchanged. The bag constant is determined to be $B^{1/4}=218$ MeV ($B=292$ MeV fm$^{-3}$) with parameters in \cite{1996PhRvC..53..410R} in the NJL model and $E/A$ always greater than 930 MeV. Therefore, a hybrid configuration with a quark core is predicted.

There is no difference from the case of neutron stars as the central engines for GRBs in the baryon contamination. For example, in the regime of weak and intermediate diquark coupling strength, a hybrid star (with the critical pressure of neutron star matter to quark matter required to be $P_{\rm NM\rightarrow QM}\gtrsim 90$ MeV fm$^{-3}$) is constructed. The masses of layers made of subnuclear and nuclear matter are shown in Figure~\ref{fig:Mcrust} (c)\footnote{Above this critical pressure, the maximum static gravitational mass will be above $2M_\odot$ although the hybrid configuration above $2M_\odot$ is not stable. However, we just used it as a sample to indicate that the baryon contamination of hybrid stars is the same as that in neutron stars.}.
In this case, in addition to the outermost subnuclear matter, the nuclear matter surrounding the quark core could also contribute to baryon loading in the outflow of the GRB, which could cause a larger baryon contamination comparable to observation.

\section{discussion and summary}\label{sec:discussion}
 In this paper, a search is performed for clean fireballs with very small baryon contamination. In Figures~\ref{fig:Mcrust} (a) and (b), there are GRBs (241107A and 150906B) with estimated $M_{\rm crust}$ below the upper limits predicted by the MIT bag model. However, this cannot provide positive support for the hypothesis of QS-GRBs for two reasons:
 \begin{itemize}
   \item 
   some previous works~\cite{2010ApJ...725.2209L,2012ApJ...751...49L} discovered a positive correlation between $\Gamma$ and $E_{\gamma, \rm iso}$ of $\Gamma\simeq91 E_{\gamma, \rm iso,52}^{0.29}$. Therefore, it could be inferred that there could be a correlation between L.L. of $M_{\rm crust}(\simeq E_{\gamma, \rm iso}/\eta_{\gamma}\Gamma$) and $E_{\gamma, \rm iso}/\eta_{\gamma}$. As shown in Figure~\ref{fig:Mcrust} (b), a correlation of $M_{\rm crust}\propto (E_{\gamma, \rm iso}/\eta_{\gamma})^{0.74\pm0.03}$ is extracted with the samples above the upper limit of the predicted $M_{\rm crust}$ in the MIT bag model (denoted by the blue horizontal line). The relations of $M_{\rm crust}$ versus $E_{\rm iso, \gamma}/\eta_\gamma$ of 241107A and 150906B appear to be consistent with this correlation within one standard deviation, indicating that there does not appear to be a particular scenario in baryon loading in their outflows compared to those samples above the upper limit of the predicted $M_{\rm crust}$ in the MIT bag model; otherwise, if they were QS-GRBs, their relations of $M_{\rm crust}$ versus $E_{\rm iso, \gamma}/\eta_\gamma$ should be well below the correlation. 
   
   \item the observed total energies carried by their outflows ($E_{\rm tot}\sim10^{49}$ erg if $\theta_{\rm j}\sim0.1$) are much smaller than the typical total amount of energy liberated in the NS-QS conversion ($\sim 10^{53}$ erg). If they were QS-GRBs, they should be from QSs with small masses, or the line of sight of the observer may be near the edge of the structured jet, as discussed in Section~\ref{sec:paracons}.  However, for those GRBs with $E_{\rm iso, \gamma}/\eta_\gamma>10^{53}$ erg, their $M_{\rm crust}$ are greater than $10^{-4}M_\odot$ and well above the upper limit of $M_{\rm crust}$, indicating that the predicted QS-GRB corresponding to a NS with typical mass is not found.
 \end{itemize}
   In the final analysis, the predicted QS-GRBs with very low baryon contamination may not exist. If the strange quark matter is the most stable matter as predicted, there seems no reason that such a low-mass X-ray binary scenario cannot work. The phenomenology of GRB appears to challenge the Bodmer-Witten hypothesis, while the NJL model in which the existence of hybrid stars is allowed is more consistent with the observation. However, note that the GRBs are downstream products of compact stars, in which mechanisms (e.g., emission sites and collimation) are not very clear to us; also, although there are several simulations as mentioned in Section~\ref{sec:paracons}, the details of NS-QS conversion are unknown, which may affect the production of fireball; thus, a more conservative conclusion is that GRBs have so far failed to provide positive support for this hypothesis. The other observational methods, such as the mass-radius relation measurement (via gravitational waves or X-ray pulse profile modeling) and thermal evolution of compact stars, may be much more direct on the confirmation of QSs.


\acknowledgements
The author thanks the support of the National Natural Science Foundation of China (grant No. 12303052). 
 The author is very grateful for the public GRB data of Fermi/GBM, HXMT, swift, konus-Wind and GECAM-B data.
 X.Y. Song is very grateful for the comments and suggestions from the anonymous referees and suggestions from Prof. Kin-Wang Ng on the combustion from NS to QS.
\bibliographystyle{apsrev4-1}
\bibliography{hybridCSL}

\begin{thebibliography}{143}%
\makeatletter
\providecommand \@ifxundefined [1]{%
 \@ifx{#1\undefined}
}%
\providecommand \@ifnum [1]{%
 \ifnum #1\expandafter \@firstoftwo
 \else \expandafter \@secondoftwo
 \fi
}%
\providecommand \@ifx [1]{%
 \ifx #1\expandafter \@firstoftwo
 \else \expandafter \@secondoftwo
 \fi
}%
\providecommand \natexlab [1]{#1}%
\providecommand \enquote  [1]{``#1''}%
\providecommand \bibnamefont  [1]{#1}%
\providecommand \bibfnamefont [1]{#1}%
\providecommand \citenamefont [1]{#1}%
\providecommand \href@noop [0]{\@secondoftwo}%
\providecommand \href [0]{\begingroup \@sanitize@url \@href}%
\providecommand \@href[1]{\@@startlink{#1}\@@href}%
\providecommand \@@href[1]{\endgroup#1\@@endlink}%
\providecommand \@sanitize@url [0]{\catcode `\\12\catcode `\$12\catcode `\&12\catcode `\#12\catcode `\^12\catcode `\_12\catcode `\%12\relax}%
\providecommand \@@startlink[1]{}%
\providecommand \@@endlink[0]{}%
\providecommand \url  [0]{\begingroup\@sanitize@url \@url }%
\providecommand \@url [1]{\endgroup\@href {#1}{\urlprefix }}%
\providecommand \urlprefix  [0]{URL }%
\providecommand \Eprint [0]{\href }%
\providecommand \doibase [0]{http://dx.doi.org/}%
\providecommand \selectlanguage [0]{\@gobble}%
\providecommand \bibinfo  [0]{\@secondoftwo}%
\providecommand \bibfield  [0]{\@secondoftwo}%
\providecommand \translation [1]{[#1]}%
\providecommand \BibitemOpen [0]{}%
\providecommand \bibitemStop [0]{}%
\providecommand \bibitemNoStop [0]{.\EOS\space}%
\providecommand \EOS [0]{\spacefactor3000\relax}%
\providecommand \BibitemShut  [1]{\csname bibitem#1\endcsname}%
\let\auto@bib@innerbib\@empty
\bibitem [{\citenamefont {{Bodmer}}(1971)}]{1971PhRvD...4.1601B}%
  \BibitemOpen
  \bibfield  {author} {\bibinfo {author} {\bibfnamefont {A.~R.}\ \bibnamefont {{Bodmer}}},\ }\href {\doibase 10.1103/PhysRevD.4.1601} {\bibfield  {journal} {\bibinfo  {journal} {\prd}\ }\textbf {\bibinfo {volume} {4}},\ \bibinfo {pages} {1601} (\bibinfo {year} {1971})}\BibitemShut {NoStop}%
\bibitem [{\citenamefont {{Baym}}\ and\ \citenamefont {{Chin}}(1976)}]{1976PhLB...62..241B}%
  \BibitemOpen
  \bibfield  {author} {\bibinfo {author} {\bibfnamefont {G.}~\bibnamefont {{Baym}}}\ and\ \bibinfo {author} {\bibfnamefont {S.~A.}\ \bibnamefont {{Chin}}},\ }\href {\doibase 10.1016/0370-2693(76)90517-7} {\bibfield  {journal} {\bibinfo  {journal} {Physics Letters B}\ }\textbf {\bibinfo {volume} {62}},\ \bibinfo {pages} {241} (\bibinfo {year} {1976})}\BibitemShut {NoStop}%
\bibitem [{\citenamefont {{Witten}}(1984)}]{1984PhRvD..30..272W}%
  \BibitemOpen
  \bibfield  {author} {\bibinfo {author} {\bibfnamefont {E.}~\bibnamefont {{Witten}}},\ }\href {\doibase 10.1103/PhysRevD.30.272} {\bibfield  {journal} {\bibinfo  {journal} {\prd}\ }\textbf {\bibinfo {volume} {30}},\ \bibinfo {pages} {272} (\bibinfo {year} {1984})}\BibitemShut {NoStop}%
\bibitem [{\citenamefont {{Farhi}}\ and\ \citenamefont {{Jaffe}}(1984)}]{1984PhRvD..30.2379F}%
  \BibitemOpen
  \bibfield  {author} {\bibinfo {author} {\bibfnamefont {E.}~\bibnamefont {{Farhi}}}\ and\ \bibinfo {author} {\bibfnamefont {R.~L.}\ \bibnamefont {{Jaffe}}},\ }\href {\doibase 10.1103/PhysRevD.30.2379} {\bibfield  {journal} {\bibinfo  {journal} {\prd}\ }\textbf {\bibinfo {volume} {30}},\ \bibinfo {pages} {2379} (\bibinfo {year} {1984})}\BibitemShut {NoStop}%
\bibitem [{\citenamefont {{Alcock}}\ \emph {et~al.}(1986)\citenamefont {{Alcock}}, \citenamefont {{Farhi}},\ and\ \citenamefont {{Olinto}}}]{1986ApJ...310..261A}%
  \BibitemOpen
  \bibfield  {author} {\bibinfo {author} {\bibfnamefont {C.}~\bibnamefont {{Alcock}}}, \bibinfo {author} {\bibfnamefont {E.}~\bibnamefont {{Farhi}}}, \ and\ \bibinfo {author} {\bibfnamefont {A.}~\bibnamefont {{Olinto}}},\ }\href {\doibase 10.1086/164679} {\bibfield  {journal} {\bibinfo  {journal} {\apj}\ }\textbf {\bibinfo {volume} {310}},\ \bibinfo {pages} {261} (\bibinfo {year} {1986})}\BibitemShut {NoStop}%
\bibitem [{\citenamefont {{Cheng}}\ and\ \citenamefont {{Dai}}(1996)}]{1996PhRvL..77.1210C}%
  \BibitemOpen
  \bibfield  {author} {\bibinfo {author} {\bibfnamefont {K.~S.}\ \bibnamefont {{Cheng}}}\ and\ \bibinfo {author} {\bibfnamefont {Z.~G.}\ \bibnamefont {{Dai}}},\ }\href {\doibase 10.1103/PhysRevLett.77.1210} {\bibfield  {journal} {\bibinfo  {journal} {\prl}\ }\textbf {\bibinfo {volume} {77}},\ \bibinfo {pages} {1210} (\bibinfo {year} {1996})},\ \Eprint {http://arxiv.org/abs/astro-ph/9510073} {arXiv:astro-ph/9510073 [astro-ph]} \BibitemShut {NoStop}%
\bibitem [{\citenamefont {{Bombaci}}\ and\ \citenamefont {{Datta}}(2000)}]{2000ApJ...530L..69B}%
  \BibitemOpen
  \bibfield  {author} {\bibinfo {author} {\bibfnamefont {I.}~\bibnamefont {{Bombaci}}}\ and\ \bibinfo {author} {\bibfnamefont {B.}~\bibnamefont {{Datta}}},\ }\href {\doibase 10.1086/312497} {\bibfield  {journal} {\bibinfo  {journal} {\apjl}\ }\textbf {\bibinfo {volume} {530}},\ \bibinfo {pages} {L69} (\bibinfo {year} {2000})},\ \Eprint {http://arxiv.org/abs/astro-ph/0001478} {arXiv:astro-ph/0001478 [astro-ph]} \BibitemShut {NoStop}%
\bibitem [{\citenamefont {{Paczy{\'n}ski}}\ and\ \citenamefont {{Haensel}}(2005)}]{2005MNRAS.362L...4P}%
  \BibitemOpen
  \bibfield  {author} {\bibinfo {author} {\bibfnamefont {B.}~\bibnamefont {{Paczy{\'n}ski}}}\ and\ \bibinfo {author} {\bibfnamefont {P.}~\bibnamefont {{Haensel}}},\ }\href {\doibase 10.1111/j.1745-3933.2005.00059.x} {\bibfield  {journal} {\bibinfo  {journal} {\mnras{362}{L4}{2005}}\ }\textbf {\bibinfo {volume} {362}},\ \bibinfo {pages} {L4} (\bibinfo {year} {2005})},\ \Eprint {http://arxiv.org/abs/astro-ph/0502297} {arXiv:astro-ph/0502297 [astro-ph]} \BibitemShut {NoStop}%
\bibitem [{\citenamefont {{Ouyed}}\ and\ \citenamefont {{Sannino}}(2002)}]{2002A&A...387..725O}%
  \BibitemOpen
  \bibfield  {author} {\bibinfo {author} {\bibfnamefont {R.}~\bibnamefont {{Ouyed}}}\ and\ \bibinfo {author} {\bibfnamefont {F.}~\bibnamefont {{Sannino}}},\ }\href {\doibase 10.1051/0004-6361:20020409} {\bibfield  {journal} {\bibinfo  {journal} {\ASAS}\ }\textbf {\bibinfo {volume} {387}},\ \bibinfo {pages} {725} (\bibinfo {year} {2002})},\ \Eprint {http://arxiv.org/abs/astro-ph/0103022} {arXiv:astro-ph/0103022 [astro-ph]} \BibitemShut {NoStop}%
\bibitem [{\citenamefont {{Glendenning}}\ and\ \citenamefont {{Weber}}(1992)}]{1992ApJ...400..647G}%
  \BibitemOpen
  \bibfield  {author} {\bibinfo {author} {\bibfnamefont {N.~K.}\ \bibnamefont {{Glendenning}}}\ and\ \bibinfo {author} {\bibfnamefont {F.}~\bibnamefont {{Weber}}},\ }\href {\doibase 10.1086/172026} {\bibfield  {journal} {\bibinfo  {journal} {\apj}\ }\textbf {\bibinfo {volume} {400}},\ \bibinfo {pages} {647} (\bibinfo {year} {1992})}\BibitemShut {NoStop}%
\bibitem [{\citenamefont {{Huang}}\ and\ \citenamefont {{Lu}}(1997)}]{1997A&A...325..189H}%
  \BibitemOpen
  \bibfield  {author} {\bibinfo {author} {\bibfnamefont {Y.~F.}\ \bibnamefont {{Huang}}}\ and\ \bibinfo {author} {\bibfnamefont {T.}~\bibnamefont {{Lu}}},\ }\href@noop {} {\bibfield  {journal} {\bibinfo  {journal} {\ASAS}\ }\textbf {\bibinfo {volume} {325}},\ \bibinfo {pages} {189} (\bibinfo {year} {1997})}\BibitemShut {NoStop}%
\bibitem [{\citenamefont {Nambu}\ and\ \citenamefont {Jona-Lasinio}(1961{\natexlab{a}})}]{PhysRev.122.345}%
  \BibitemOpen
  \bibfield  {author} {\bibinfo {author} {\bibfnamefont {Y.}~\bibnamefont {Nambu}}\ and\ \bibinfo {author} {\bibfnamefont {G.}~\bibnamefont {Jona-Lasinio}},\ }\href {\doibase 10.1103/PhysRev.122.345} {\bibfield  {journal} {\bibinfo  {journal} {Phys. Rev.}\ }\textbf {\bibinfo {volume} {122}},\ \bibinfo {pages} {345} (\bibinfo {year} {1961}{\natexlab{a}})}\BibitemShut {NoStop}%
\bibitem [{\citenamefont {Nambu}\ and\ \citenamefont {Jona-Lasinio}(1961{\natexlab{b}})}]{PhysRev.124.246}%
  \BibitemOpen
  \bibfield  {author} {\bibinfo {author} {\bibfnamefont {Y.}~\bibnamefont {Nambu}}\ and\ \bibinfo {author} {\bibfnamefont {G.}~\bibnamefont {Jona-Lasinio}},\ }\href {\doibase 10.1103/PhysRev.124.246} {\bibfield  {journal} {\bibinfo  {journal} {Phys. Rev.}\ }\textbf {\bibinfo {volume} {124}},\ \bibinfo {pages} {246} (\bibinfo {year} {1961}{\natexlab{b}})}\BibitemShut {NoStop}%
\bibitem [{\citenamefont {{Buballa}}(2005)}]{2005PhR...407..205B}%
  \BibitemOpen
  \bibfield  {author} {\bibinfo {author} {\bibfnamefont {M.}~\bibnamefont {{Buballa}}},\ }\href {\doibase 10.1016/j.physrep.2004.11.004} {\bibfield  {journal} {\bibinfo  {journal} {\PR}\ }\textbf {\bibinfo {volume} {407}},\ \bibinfo {pages} {205} (\bibinfo {year} {2005})},\ \Eprint {http://arxiv.org/abs/hep-ph/0402234} {arXiv:hep-ph/0402234 [hep-ph]} \BibitemShut {NoStop}%
\bibitem [{\citenamefont {{Blaschke}}\ \emph {et~al.}(2005)\citenamefont {{Blaschke}}, \citenamefont {{Fredriksson}}, \citenamefont {{Grigorian}}, \citenamefont {{{\"O}zta{\c{s}}}},\ and\ \citenamefont {{Sandin}}}]{2005PhRvD..72f5020B}%
  \BibitemOpen
  \bibfield  {author} {\bibinfo {author} {\bibfnamefont {D.}~\bibnamefont {{Blaschke}}}, \bibinfo {author} {\bibfnamefont {S.}~\bibnamefont {{Fredriksson}}}, \bibinfo {author} {\bibfnamefont {H.}~\bibnamefont {{Grigorian}}}, \bibinfo {author} {\bibfnamefont {A.~M.}\ \bibnamefont {{{\"O}zta{\c{s}}}}}, \ and\ \bibinfo {author} {\bibfnamefont {F.}~\bibnamefont {{Sandin}}},\ }\href {\doibase 10.1103/PhysRevD.72.065020} {\bibfield  {journal} {\bibinfo  {journal} {\prd}\ }\textbf {\bibinfo {volume} {72}},\ \bibinfo {eid} {065020} (\bibinfo {year} {2005})},\ \Eprint {http://arxiv.org/abs/hep-ph/0503194} {arXiv:hep-ph/0503194 [hep-ph]} \BibitemShut {NoStop}%
\bibitem [{\citenamefont {{R{\"u}ster}}\ \emph {et~al.}(2005)\citenamefont {{R{\"u}ster}}, \citenamefont {{Werth}}, \citenamefont {{Buballa}}, \citenamefont {{Shovkovy}},\ and\ \citenamefont {{Rischke}}}]{2005PhRvD..72c4004R}%
  \BibitemOpen
  \bibfield  {author} {\bibinfo {author} {\bibfnamefont {S.~B.}\ \bibnamefont {{R{\"u}ster}}}, \bibinfo {author} {\bibfnamefont {V.}~\bibnamefont {{Werth}}}, \bibinfo {author} {\bibfnamefont {M.}~\bibnamefont {{Buballa}}}, \bibinfo {author} {\bibfnamefont {I.~A.}\ \bibnamefont {{Shovkovy}}}, \ and\ \bibinfo {author} {\bibfnamefont {D.~H.}\ \bibnamefont {{Rischke}}},\ }\href {\doibase 10.1103/PhysRevD.72.034004} {\bibfield  {journal} {\bibinfo  {journal} {\prd}\ }\textbf {\bibinfo {volume} {72}},\ \bibinfo {eid} {034004} (\bibinfo {year} {2005})},\ \Eprint {http://arxiv.org/abs/hep-ph/0503184} {arXiv:hep-ph/0503184 [hep-ph]} \BibitemShut {NoStop}%
\bibitem [{\citenamefont {{Song}}(2024)}]{2024arXiv240915811S}%
  \BibitemOpen
  \bibfield  {author} {\bibinfo {author} {\bibfnamefont {X.-Y.}\ \bibnamefont {{Song}}},\ }\href {\doibase 10.48550/arXiv.2409.15811} {\bibfield  {journal} {\bibinfo  {journal} {arXiv e-prints}\ ,\ \bibinfo {eid} {arXiv:2409.15811}} (\bibinfo {year} {2024})},\ \Eprint {http://arxiv.org/abs/2409.15811} {arXiv:2409.15811 [hep-ph]} \BibitemShut {NoStop}%
\bibitem [{\citenamefont {{Lin}}\ \emph {et~al.}(2006)\citenamefont {{Lin}}, \citenamefont {{Cheng}}, \citenamefont {{Chu}},\ and\ \citenamefont {{Suen}}}]{2006ApJ...639..382L}%
  \BibitemOpen
  \bibfield  {author} {\bibinfo {author} {\bibfnamefont {L.~M.}\ \bibnamefont {{Lin}}}, \bibinfo {author} {\bibfnamefont {K.~S.}\ \bibnamefont {{Cheng}}}, \bibinfo {author} {\bibfnamefont {M.~C.}\ \bibnamefont {{Chu}}}, \ and\ \bibinfo {author} {\bibfnamefont {W.~M.}\ \bibnamefont {{Suen}}},\ }\href {\doibase 10.1086/499202} {\bibfield  {journal} {\bibinfo  {journal} {\apj}\ }\textbf {\bibinfo {volume} {639}},\ \bibinfo {pages} {382} (\bibinfo {year} {2006})},\ \Eprint {http://arxiv.org/abs/astro-ph/0509447} {arXiv:astro-ph/0509447 [astro-ph]} \BibitemShut {NoStop}%
\bibitem [{\citenamefont {{Cheng}}\ \emph {et~al.}(2009)\citenamefont {{Cheng}}, \citenamefont {{Harko}}, \citenamefont {{Huang}}, \citenamefont {{Lin}}, \citenamefont {{Suen}},\ and\ \citenamefont {{Tian}}}]{2009JCAP...09..007C}%
  \BibitemOpen
  \bibfield  {author} {\bibinfo {author} {\bibfnamefont {K.~S.}\ \bibnamefont {{Cheng}}}, \bibinfo {author} {\bibfnamefont {T.}~\bibnamefont {{Harko}}}, \bibinfo {author} {\bibfnamefont {Y.~F.}\ \bibnamefont {{Huang}}}, \bibinfo {author} {\bibfnamefont {L.~M.}\ \bibnamefont {{Lin}}}, \bibinfo {author} {\bibfnamefont {W.~M.}\ \bibnamefont {{Suen}}}, \ and\ \bibinfo {author} {\bibfnamefont {X.~L.}\ \bibnamefont {{Tian}}},\ }\href {\doibase 10.1088/1475-7516/2009/09/007} {\bibfield  {journal} {\bibinfo  {journal} {Journal of Cosmology and Astroparticle Physics}\ }\textbf {\bibinfo {volume} {2009}},\ \bibinfo {eid} {007} (\bibinfo {year} {2009})},\ \Eprint {http://arxiv.org/abs/0908.1834} {arXiv:0908.1834 [astro-ph.HE]} \BibitemShut {NoStop}%
\bibitem [{\citenamefont {{Most}}\ \emph {et~al.}(2019)\citenamefont {{Most}}, \citenamefont {{Papenfort}}, \citenamefont {{Dexheimer}}, \citenamefont {{Hanauske}}, \citenamefont {{Schramm}}, \citenamefont {{St{\"o}cker}},\ and\ \citenamefont {{Rezzolla}}}]{2019PhRvL.122f1101M}%
  \BibitemOpen
  \bibfield  {author} {\bibinfo {author} {\bibfnamefont {E.~R.}\ \bibnamefont {{Most}}}, \bibinfo {author} {\bibfnamefont {L.~J.}\ \bibnamefont {{Papenfort}}}, \bibinfo {author} {\bibfnamefont {V.}~\bibnamefont {{Dexheimer}}}, \bibinfo {author} {\bibfnamefont {M.}~\bibnamefont {{Hanauske}}}, \bibinfo {author} {\bibfnamefont {S.}~\bibnamefont {{Schramm}}}, \bibinfo {author} {\bibfnamefont {H.}~\bibnamefont {{St{\"o}cker}}}, \ and\ \bibinfo {author} {\bibfnamefont {L.}~\bibnamefont {{Rezzolla}}},\ }\href {\doibase 10.1103/PhysRevLett.122.061101} {\bibfield  {journal} {\bibinfo  {journal} {\prl}\ }\textbf {\bibinfo {volume} {122}},\ \bibinfo {eid} {061101} (\bibinfo {year} {2019})},\ \Eprint {http://arxiv.org/abs/1807.03684} {arXiv:1807.03684 [astro-ph.HE]} \BibitemShut {NoStop}%
\bibitem [{\citenamefont {{Zhou}}\ \emph {et~al.}(2022)\citenamefont {{Zhou}}, \citenamefont {{Kiuchi}}, \citenamefont {{Shibata}}, \citenamefont {{Tsokaros}},\ and\ \citenamefont {{Ury{\r{A}}, K{\={o}}ji}}}]{2022PhRvD.106j3030Z}%
  \BibitemOpen
  \bibfield  {author} {\bibinfo {author} {\bibfnamefont {E.}~\bibnamefont {{Zhou}}}, \bibinfo {author} {\bibfnamefont {K.}~\bibnamefont {{Kiuchi}}}, \bibinfo {author} {\bibfnamefont {M.}~\bibnamefont {{Shibata}}}, \bibinfo {author} {\bibfnamefont {A.}~\bibnamefont {{Tsokaros}}}, \ and\ \bibinfo {author} {\bibnamefont {{Ury{\r{A}}, K{\={o}}ji}}},\ }\href {\doibase 10.1103/PhysRevD.106.103030} {\bibfield  {journal} {\bibinfo  {journal} {\prd}\ }\textbf {\bibinfo {volume} {106}},\ \bibinfo {eid} {103030} (\bibinfo {year} {2022})},\ \Eprint {http://arxiv.org/abs/2111.00958} {arXiv:2111.00958 [astro-ph.HE]} \BibitemShut {NoStop}%
\bibitem [{\citenamefont {{Meszaros}}\ and\ \citenamefont {{Rees}}(1992)}]{1992ApJ...397..570M}%
  \BibitemOpen
  \bibfield  {author} {\bibinfo {author} {\bibfnamefont {P.}~\bibnamefont {{Meszaros}}}\ and\ \bibinfo {author} {\bibfnamefont {M.~J.}\ \bibnamefont {{Rees}}},\ }\href {\doibase 10.1086/171813} {\bibfield  {journal} {\bibinfo  {journal} {\apj}\ }\textbf {\bibinfo {volume} {397}},\ \bibinfo {pages} {570} (\bibinfo {year} {1992})}\BibitemShut {NoStop}%
\bibitem [{\citenamefont {{Beloborodov}}(2003)}]{2003ApJ...588..931B}%
  \BibitemOpen
  \bibfield  {author} {\bibinfo {author} {\bibfnamefont {A.~M.}\ \bibnamefont {{Beloborodov}}},\ }\href {\doibase 10.1086/374217} {\bibfield  {journal} {\bibinfo  {journal} {\apj}\ }\textbf {\bibinfo {volume} {588}},\ \bibinfo {pages} {931} (\bibinfo {year} {2003})},\ \Eprint {http://arxiv.org/abs/astro-ph/0210522} {arXiv:astro-ph/0210522 [astro-ph]} \BibitemShut {NoStop}%
\bibitem [{\citenamefont {{Duncan}}\ and\ \citenamefont {{Thompson}}(1992)}]{1992ApJ...392L...9D}%
  \BibitemOpen
  \bibfield  {author} {\bibinfo {author} {\bibfnamefont {R.~C.}\ \bibnamefont {{Duncan}}}\ and\ \bibinfo {author} {\bibfnamefont {C.}~\bibnamefont {{Thompson}}},\ }\href {\doibase 10.1086/186413} {\bibfield  {journal} {\bibinfo  {journal} {\apjl}\ }\textbf {\bibinfo {volume} {392}},\ \bibinfo {pages} {L9} (\bibinfo {year} {1992})}\BibitemShut {NoStop}%
\bibitem [{\citenamefont {{Thompson}}(1994)}]{1994MNRAS.270..480T}%
  \BibitemOpen
  \bibfield  {author} {\bibinfo {author} {\bibfnamefont {C.}~\bibnamefont {{Thompson}}},\ }\href {\doibase 10.1093/mnras/270.3.480} {\bibfield  {journal} {\bibinfo  {journal} {\mnras{270}{480}{1994}}\ }\textbf {\bibinfo {volume} {270}},\ \bibinfo {pages} {480} (\bibinfo {year} {1994})}\BibitemShut {NoStop}%
\bibitem [{\citenamefont {{Blackman}}\ \emph {et~al.}(1996)\citenamefont {{Blackman}}, \citenamefont {{Yi}},\ and\ \citenamefont {{Field}}}]{1996ApJ...473L..79B}%
  \BibitemOpen
  \bibfield  {author} {\bibinfo {author} {\bibfnamefont {E.~G.}\ \bibnamefont {{Blackman}}}, \bibinfo {author} {\bibfnamefont {I.}~\bibnamefont {{Yi}}}, \ and\ \bibinfo {author} {\bibfnamefont {G.~B.}\ \bibnamefont {{Field}}},\ }\href {\doibase 10.1086/310403} {\bibfield  {journal} {\bibinfo  {journal} {\apjl}\ }\textbf {\bibinfo {volume} {473}},\ \bibinfo {pages} {L79} (\bibinfo {year} {1996})},\ \Eprint {http://arxiv.org/abs/astro-ph/9609116} {arXiv:astro-ph/9609116 [astro-ph]} \BibitemShut {NoStop}%
\bibitem [{\citenamefont {{Lei}}\ \emph {et~al.}(2013)\citenamefont {{Lei}}, \citenamefont {{Zhang}},\ and\ \citenamefont {{Liang}}}]{2013ApJ...765..125L}%
  \BibitemOpen
  \bibfield  {author} {\bibinfo {author} {\bibfnamefont {W.-H.}\ \bibnamefont {{Lei}}}, \bibinfo {author} {\bibfnamefont {B.}~\bibnamefont {{Zhang}}}, \ and\ \bibinfo {author} {\bibfnamefont {E.-W.}\ \bibnamefont {{Liang}}},\ }\href {\doibase 10.1088/0004-637X/765/2/125} {\bibfield  {journal} {\bibinfo  {journal} {\apj}\ }\textbf {\bibinfo {volume} {765}},\ \bibinfo {eid} {125} (\bibinfo {year} {2013})},\ \Eprint {http://arxiv.org/abs/1209.4427} {arXiv:1209.4427 [astro-ph.HE]} \BibitemShut {NoStop}%
\bibitem [{\citenamefont {{Olinto}}(1987)}]{1987PhLB..192...71O}%
  \BibitemOpen
  \bibfield  {author} {\bibinfo {author} {\bibfnamefont {A.~V.}\ \bibnamefont {{Olinto}}},\ }\href {\doibase 10.1016/0370-2693(87)91144-0} {\bibfield  {journal} {\bibinfo  {journal} {Physics Letters B}\ }\textbf {\bibinfo {volume} {192}},\ \bibinfo {pages} {71} (\bibinfo {year} {1987})}\BibitemShut {NoStop}%
\bibitem [{\citenamefont {{Horvath}}\ and\ \citenamefont {{Benvenuto}}(1988)}]{1988PhLB..213..516H}%
  \BibitemOpen
  \bibfield  {author} {\bibinfo {author} {\bibfnamefont {J.~E.}\ \bibnamefont {{Horvath}}}\ and\ \bibinfo {author} {\bibfnamefont {O.~G.}\ \bibnamefont {{Benvenuto}}},\ }\href {\doibase 10.1016/0370-2693(88)91302-0} {\bibfield  {journal} {\bibinfo  {journal} {Physics Letters B}\ }\textbf {\bibinfo {volume} {213}},\ \bibinfo {pages} {516} (\bibinfo {year} {1988})}\BibitemShut {NoStop}%
\bibitem [{\citenamefont {{Pagliara}}\ \emph {et~al.}(2013)\citenamefont {{Pagliara}}, \citenamefont {{Herzog}},\ and\ \citenamefont {{R{\"o}pke}}}]{2013PhRvD..87j3007P}%
  \BibitemOpen
  \bibfield  {author} {\bibinfo {author} {\bibfnamefont {G.}~\bibnamefont {{Pagliara}}}, \bibinfo {author} {\bibfnamefont {M.}~\bibnamefont {{Herzog}}}, \ and\ \bibinfo {author} {\bibfnamefont {F.~K.}\ \bibnamefont {{R{\"o}pke}}},\ }\href {\doibase 10.1103/PhysRevD.87.103007} {\bibfield  {journal} {\bibinfo  {journal} {\prd}\ }\textbf {\bibinfo {volume} {87}},\ \bibinfo {eid} {103007} (\bibinfo {year} {2013})},\ \Eprint {http://arxiv.org/abs/1304.6884} {arXiv:1304.6884 [astro-ph.HE]} \BibitemShut {NoStop}%
\bibitem [{\citenamefont {{Herzog}}\ and\ \citenamefont {{R{\"o}pke}}(2011)}]{2011PhRvD..84h3002H}%
  \BibitemOpen
  \bibfield  {author} {\bibinfo {author} {\bibfnamefont {M.}~\bibnamefont {{Herzog}}}\ and\ \bibinfo {author} {\bibfnamefont {F.~K.}\ \bibnamefont {{R{\"o}pke}}},\ }\href {\doibase 10.1103/PhysRevD.84.083002} {\bibfield  {journal} {\bibinfo  {journal} {\prd}\ }\textbf {\bibinfo {volume} {84}},\ \bibinfo {eid} {083002} (\bibinfo {year} {2011})},\ \Eprint {http://arxiv.org/abs/1109.0539} {arXiv:1109.0539 [astro-ph.HE]} \BibitemShut {NoStop}%
\bibitem [{\citenamefont {{Niebergal}}\ \emph {et~al.}(2010)\citenamefont {{Niebergal}}, \citenamefont {{Ouyed}},\ and\ \citenamefont {{Jaikumar}}}]{2010PhRvC..82f2801N}%
  \BibitemOpen
  \bibfield  {author} {\bibinfo {author} {\bibfnamefont {B.}~\bibnamefont {{Niebergal}}}, \bibinfo {author} {\bibfnamefont {R.}~\bibnamefont {{Ouyed}}}, \ and\ \bibinfo {author} {\bibfnamefont {P.}~\bibnamefont {{Jaikumar}}},\ }\href {\doibase 10.1103/PhysRevC.82.062801} {\bibfield  {journal} {\bibinfo  {journal} {\prc}\ }\textbf {\bibinfo {volume} {82}},\ \bibinfo {eid} {062801} (\bibinfo {year} {2010})},\ \Eprint {http://arxiv.org/abs/1008.4806} {arXiv:1008.4806 [nucl-th]} \BibitemShut {NoStop}%
\bibitem [{\citenamefont {{Drago}}\ and\ \citenamefont {{Pagliara}}(2015)}]{2015PhRvC..92d5801D}%
  \BibitemOpen
  \bibfield  {author} {\bibinfo {author} {\bibfnamefont {A.}~\bibnamefont {{Drago}}}\ and\ \bibinfo {author} {\bibfnamefont {G.}~\bibnamefont {{Pagliara}}},\ }\href {\doibase 10.1103/PhysRevC.92.045801} {\bibfield  {journal} {\bibinfo  {journal} {\prc}\ }\textbf {\bibinfo {volume} {92}},\ \bibinfo {eid} {045801} (\bibinfo {year} {2015})},\ \Eprint {http://arxiv.org/abs/1506.08337} {arXiv:1506.08337 [nucl-th]} \BibitemShut {NoStop}%
\bibitem [{\citenamefont {{Cho}}\ \emph {et~al.}(1994)\citenamefont {{Cho}}, \citenamefont {{Ng}},\ and\ \citenamefont {{Speliotopoulos}}}]{1994PhLB..326..111C}%
  \BibitemOpen
  \bibfield  {author} {\bibinfo {author} {\bibfnamefont {H.~T.}\ \bibnamefont {{Cho}}}, \bibinfo {author} {\bibfnamefont {K.~W.}\ \bibnamefont {{Ng}}}, \ and\ \bibinfo {author} {\bibfnamefont {A.~D.}\ \bibnamefont {{Speliotopoulos}}},\ }\href {\doibase 10.1016/0370-2693(94)91201-7} {\bibfield  {journal} {\bibinfo  {journal} {Physics Letters B}\ }\textbf {\bibinfo {volume} {326}},\ \bibinfo {pages} {111} (\bibinfo {year} {1994})},\ \Eprint {http://arxiv.org/abs/astro-ph/9305006} {arXiv:astro-ph/9305006 [astro-ph]} \BibitemShut {NoStop}%
\bibitem [{\citenamefont {{Zapata}}\ \emph {et~al.}(2022)\citenamefont {{Zapata}}, \citenamefont {{Sales}}, \citenamefont {{Jaikumar}},\ and\ \citenamefont {{Negreiros}}}]{2022A&A...663A..19Z}%
  \BibitemOpen
  \bibfield  {author} {\bibinfo {author} {\bibfnamefont {J.}~\bibnamefont {{Zapata}}}, \bibinfo {author} {\bibfnamefont {T.}~\bibnamefont {{Sales}}}, \bibinfo {author} {\bibfnamefont {P.}~\bibnamefont {{Jaikumar}}}, \ and\ \bibinfo {author} {\bibfnamefont {R.}~\bibnamefont {{Negreiros}}},\ }\href {\doibase 10.1051/0004-6361/202243148} {\bibfield  {journal} {\bibinfo  {journal} {\ASAS}\ }\textbf {\bibinfo {volume} {663}},\ \bibinfo {eid} {A19} (\bibinfo {year} {2022})},\ \Eprint {http://arxiv.org/abs/2201.06928} {arXiv:2201.06928 [hep-ph]} \BibitemShut {NoStop}%
\bibitem [{\citenamefont {{Rossi}}\ \emph {et~al.}(2002)\citenamefont {{Rossi}}, \citenamefont {{Lazzati}},\ and\ \citenamefont {{Rees}}}]{2002MNRAS.332..945R}%
  \BibitemOpen
  \bibfield  {author} {\bibinfo {author} {\bibfnamefont {E.}~\bibnamefont {{Rossi}}}, \bibinfo {author} {\bibfnamefont {D.}~\bibnamefont {{Lazzati}}}, \ and\ \bibinfo {author} {\bibfnamefont {M.~J.}\ \bibnamefont {{Rees}}},\ }\href {\doibase 10.1046/j.1365-8711.2002.05363.x} {\bibfield  {journal} {\bibinfo  {journal} {\mnras{332}{945}{2002}}\ }\textbf {\bibinfo {volume} {332}},\ \bibinfo {pages} {945} (\bibinfo {year} {2002})},\ \Eprint {http://arxiv.org/abs/astro-ph/0112083} {arXiv:astro-ph/0112083 [astro-ph]} \BibitemShut {NoStop}%
\bibitem [{\citenamefont {{Zhang}}\ and\ \citenamefont {{M{\'e}sz{\'a}ros}}(2002)}]{2002ApJ...571..876Z}%
  \BibitemOpen
  \bibfield  {author} {\bibinfo {author} {\bibfnamefont {B.}~\bibnamefont {{Zhang}}}\ and\ \bibinfo {author} {\bibfnamefont {P.}~\bibnamefont {{M{\'e}sz{\'a}ros}}},\ }\href {\doibase 10.1086/339981} {\bibfield  {journal} {\bibinfo  {journal} {\apj}\ }\textbf {\bibinfo {volume} {571}},\ \bibinfo {pages} {876} (\bibinfo {year} {2002})},\ \Eprint {http://arxiv.org/abs/astro-ph/0112118} {arXiv:astro-ph/0112118 [astro-ph]} \BibitemShut {NoStop}%
\bibitem [{\citenamefont {{Frail}}\ \emph {et~al.}(2001)\citenamefont {{Frail}}, \citenamefont {{Kulkarni}}, \citenamefont {{Sari}}, \citenamefont {{Djorgovski}}, \citenamefont {{Bloom}}, \citenamefont {{Galama}}, \citenamefont {{Reichart}}, \citenamefont {{Berger}}, \citenamefont {{Harrison}}, \citenamefont {{Price}}, \citenamefont {{Yost}}, \citenamefont {{Diercks}}, \citenamefont {{Goodrich}},\ and\ \citenamefont {{Chaffee}}}]{2001ApJ...562L..55F}%
  \BibitemOpen
  \bibfield  {author} {\bibinfo {author} {\bibfnamefont {D.~A.}\ \bibnamefont {{Frail}}}, \bibinfo {author} {\bibfnamefont {S.~R.}\ \bibnamefont {{Kulkarni}}}, \bibinfo {author} {\bibfnamefont {R.}~\bibnamefont {{Sari}}}, \bibinfo {author} {\bibfnamefont {S.~G.}\ \bibnamefont {{Djorgovski}}}, \bibinfo {author} {\bibfnamefont {J.~S.}\ \bibnamefont {{Bloom}}}, \bibinfo {author} {\bibfnamefont {T.~J.}\ \bibnamefont {{Galama}}}, \bibinfo {author} {\bibfnamefont {D.~E.}\ \bibnamefont {{Reichart}}}, \bibinfo {author} {\bibfnamefont {E.}~\bibnamefont {{Berger}}}, \bibinfo {author} {\bibfnamefont {F.~A.}\ \bibnamefont {{Harrison}}}, \bibinfo {author} {\bibfnamefont {P.~A.}\ \bibnamefont {{Price}}}, \bibinfo {author} {\bibfnamefont {S.~A.}\ \bibnamefont {{Yost}}}, \bibinfo {author} {\bibfnamefont {A.}~\bibnamefont {{Diercks}}}, \bibinfo {author} {\bibfnamefont {R.~W.}\ \bibnamefont {{Goodrich}}}, \ and\ \bibinfo {author} {\bibfnamefont {F.}~\bibnamefont {{Chaffee}}},\ }\href {\doibase 10.1086/338119} {\bibfield
  {journal} {\bibinfo  {journal} {\apjl}\ }\textbf {\bibinfo {volume} {562}},\ \bibinfo {pages} {L55} (\bibinfo {year} {2001})},\ \Eprint {http://arxiv.org/abs/astro-ph/0102282} {arXiv:astro-ph/0102282 [astro-ph]} \BibitemShut {NoStop}%
\bibitem [{\citenamefont {{Pe'er}}\ \emph {et~al.}(2015)\citenamefont {{Pe'er}}, \citenamefont {{Barlow}}, \citenamefont {{O'Mahony}}, \citenamefont {{Margutti}}, \citenamefont {{Ryde}}, \citenamefont {{Larsson}}, \citenamefont {{Lazzati}},\ and\ \citenamefont {{Livio}}}]{2015ApJ...813..127P}%
  \BibitemOpen
  \bibfield  {author} {\bibinfo {author} {\bibfnamefont {A.}~\bibnamefont {{Pe'er}}}, \bibinfo {author} {\bibfnamefont {H.}~\bibnamefont {{Barlow}}}, \bibinfo {author} {\bibfnamefont {S.}~\bibnamefont {{O'Mahony}}}, \bibinfo {author} {\bibfnamefont {R.}~\bibnamefont {{Margutti}}}, \bibinfo {author} {\bibfnamefont {F.}~\bibnamefont {{Ryde}}}, \bibinfo {author} {\bibfnamefont {J.}~\bibnamefont {{Larsson}}}, \bibinfo {author} {\bibfnamefont {D.}~\bibnamefont {{Lazzati}}}, \ and\ \bibinfo {author} {\bibfnamefont {M.}~\bibnamefont {{Livio}}},\ }\href {\doibase 10.1088/0004-637X/813/2/127} {\bibfield  {journal} {\bibinfo  {journal} {\apj}\ }\textbf {\bibinfo {volume} {813}},\ \bibinfo {eid} {127} (\bibinfo {year} {2015})},\ \Eprint {http://arxiv.org/abs/1507.00873} {arXiv:1507.00873 [astro-ph.HE]} \BibitemShut {NoStop}%
\bibitem [{\citenamefont {{Pe'er}}\ \emph {et~al.}(2007)\citenamefont {{Pe'er}}, \citenamefont {{Ryde}}, \citenamefont {{Wijers}}, \citenamefont {{M{\'e}sz{\'a}ros}},\ and\ \citenamefont {{Rees}}}]{2007ApJ...664L...1P}%
  \BibitemOpen
  \bibfield  {author} {\bibinfo {author} {\bibfnamefont {A.}~\bibnamefont {{Pe'er}}}, \bibinfo {author} {\bibfnamefont {F.}~\bibnamefont {{Ryde}}}, \bibinfo {author} {\bibfnamefont {R.~A.~M.~J.}\ \bibnamefont {{Wijers}}}, \bibinfo {author} {\bibfnamefont {P.}~\bibnamefont {{M{\'e}sz{\'a}ros}}}, \ and\ \bibinfo {author} {\bibfnamefont {M.~J.}\ \bibnamefont {{Rees}}},\ }\href {\doibase 10.1086/520534} {\bibfield  {journal} {\bibinfo  {journal} {\apjl}\ }\textbf {\bibinfo {volume} {664}},\ \bibinfo {pages} {L1} (\bibinfo {year} {2007})},\ \Eprint {http://arxiv.org/abs/astro-ph/0703734} {arXiv:astro-ph/0703734 [astro-ph]} \BibitemShut {NoStop}%
\bibitem [{\citenamefont {{Gao}}\ and\ \citenamefont {{Zhang}}(2015)}]{2015ApJ...801..103G}%
  \BibitemOpen
  \bibfield  {author} {\bibinfo {author} {\bibfnamefont {H.}~\bibnamefont {{Gao}}}\ and\ \bibinfo {author} {\bibfnamefont {B.}~\bibnamefont {{Zhang}}},\ }\href {\doibase 10.1088/0004-637X/801/2/103} {\bibfield  {journal} {\bibinfo  {journal} {\apj}\ }\textbf {\bibinfo {volume} {801}},\ \bibinfo {eid} {103} (\bibinfo {year} {2015})},\ \Eprint {http://arxiv.org/abs/1409.3584} {arXiv:1409.3584 [astro-ph.HE]} \BibitemShut {NoStop}%
\bibitem [{\citenamefont {{Song}}\ \emph {et~al.}(2024)\citenamefont {{Song}}, \citenamefont {{Wang}},\ and\ \citenamefont {{Zhang}}}]{2024ApJ...961..137S}%
  \BibitemOpen
  \bibfield  {author} {\bibinfo {author} {\bibfnamefont {X.-Y.}\ \bibnamefont {{Song}}}, \bibinfo {author} {\bibfnamefont {L.-J.}\ \bibnamefont {{Wang}}}, \ and\ \bibinfo {author} {\bibfnamefont {S.}~\bibnamefont {{Zhang}}},\ }\href {\doibase 10.3847/1538-4357/ad0df3} {\bibfield  {journal} {\bibinfo  {journal} {\apj}\ }\textbf {\bibinfo {volume} {961}},\ \bibinfo {eid} {137} (\bibinfo {year} {2024})},\ \Eprint {http://arxiv.org/abs/2306.02248} {arXiv:2306.02248 [astro-ph.HE]} \BibitemShut {NoStop}%
\bibitem [{\citenamefont {{Larsson}}\ \emph {et~al.}(2015)\citenamefont {{Larsson}}, \citenamefont {{Racusin}},\ and\ \citenamefont {{Burgess}}}]{2015ApJ...800L..34L}%
  \BibitemOpen
  \bibfield  {author} {\bibinfo {author} {\bibfnamefont {J.}~\bibnamefont {{Larsson}}}, \bibinfo {author} {\bibfnamefont {J.~L.}\ \bibnamefont {{Racusin}}}, \ and\ \bibinfo {author} {\bibfnamefont {J.~M.}\ \bibnamefont {{Burgess}}},\ }\href {\doibase 10.1088/2041-8205/800/2/L34} {\bibfield  {journal} {\bibinfo  {journal} {\apjl}\ }\textbf {\bibinfo {volume} {800}},\ \bibinfo {eid} {L34} (\bibinfo {year} {2015})},\ \Eprint {http://arxiv.org/abs/1502.00645} {arXiv:1502.00645 [astro-ph.HE]} \BibitemShut {NoStop}%
\bibitem [{\citenamefont {{Zhang}}\ \emph {et~al.}(2007)\citenamefont {{Zhang}}, \citenamefont {{Liang}}, \citenamefont {{Page}}, \citenamefont {{Grupe}}, \citenamefont {{Zhang}}, \citenamefont {{Barthelmy}}, \citenamefont {{Burrows}}, \citenamefont {{Campana}}, \citenamefont {{Chincarini}}, \citenamefont {{Gehrels}}, \citenamefont {{Kobayashi}}, \citenamefont {{M{\'e}sz{\'a}ros}}, \citenamefont {{Moretti}}, \citenamefont {{Nousek}}, \citenamefont {{O'Brien}}, \citenamefont {{Osborne}}, \citenamefont {{Roming}}, \citenamefont {{Sakamoto}}, \citenamefont {{Schady}},\ and\ \citenamefont {{Willingale}}}]{2007ApJ...655..989Z}%
  \BibitemOpen
  \bibfield  {author} {\bibinfo {author} {\bibfnamefont {B.}~\bibnamefont {{Zhang}}}, \bibinfo {author} {\bibfnamefont {E.}~\bibnamefont {{Liang}}}, \bibinfo {author} {\bibfnamefont {K.~L.}\ \bibnamefont {{Page}}}, \bibinfo {author} {\bibfnamefont {D.}~\bibnamefont {{Grupe}}}, \bibinfo {author} {\bibfnamefont {B.-B.}\ \bibnamefont {{Zhang}}}, \bibinfo {author} {\bibfnamefont {S.~D.}\ \bibnamefont {{Barthelmy}}}, \bibinfo {author} {\bibfnamefont {D.~N.}\ \bibnamefont {{Burrows}}}, \bibinfo {author} {\bibfnamefont {S.}~\bibnamefont {{Campana}}}, \bibinfo {author} {\bibfnamefont {G.}~\bibnamefont {{Chincarini}}}, \bibinfo {author} {\bibfnamefont {N.}~\bibnamefont {{Gehrels}}}, \bibinfo {author} {\bibfnamefont {S.}~\bibnamefont {{Kobayashi}}}, \bibinfo {author} {\bibfnamefont {P.}~\bibnamefont {{M{\'e}sz{\'a}ros}}}, \bibinfo {author} {\bibfnamefont {A.}~\bibnamefont {{Moretti}}}, \bibinfo {author} {\bibfnamefont {J.~A.}\ \bibnamefont {{Nousek}}}, \bibinfo {author} {\bibfnamefont {P.~T.}\ \bibnamefont
  {{O'Brien}}}, \bibinfo {author} {\bibfnamefont {J.~P.}\ \bibnamefont {{Osborne}}}, \bibinfo {author} {\bibfnamefont {P.~W.~A.}\ \bibnamefont {{Roming}}}, \bibinfo {author} {\bibfnamefont {T.}~\bibnamefont {{Sakamoto}}}, \bibinfo {author} {\bibfnamefont {P.}~\bibnamefont {{Schady}}}, \ and\ \bibinfo {author} {\bibfnamefont {R.}~\bibnamefont {{Willingale}}},\ }\href {\doibase 10.1086/510110} {\bibfield  {journal} {\bibinfo  {journal} {\apj}\ }\textbf {\bibinfo {volume} {655}},\ \bibinfo {pages} {989} (\bibinfo {year} {2007})},\ \Eprint {http://arxiv.org/abs/astro-ph/0610177} {arXiv:astro-ph/0610177 [astro-ph]} \BibitemShut {NoStop}%
\bibitem [{\citenamefont {{Fan}}\ and\ \citenamefont {{Piran}}(2006)}]{2006MNRAS.369..197F}%
  \BibitemOpen
  \bibfield  {author} {\bibinfo {author} {\bibfnamefont {Y.}~\bibnamefont {{Fan}}}\ and\ \bibinfo {author} {\bibfnamefont {T.}~\bibnamefont {{Piran}}},\ }\href {\doibase 10.1111/j.1365-2966.2006.10280.x} {\bibfield  {journal} {\bibinfo  {journal} {\mnras{369}{197}{2006}}\ }\textbf {\bibinfo {volume} {369}},\ \bibinfo {pages} {197} (\bibinfo {year} {2006})},\ \Eprint {http://arxiv.org/abs/astro-ph/0601054} {arXiv:astro-ph/0601054 [astro-ph]} \BibitemShut {NoStop}%
\bibitem [{\citenamefont {{Zhang}}\ and\ \citenamefont {{Yan}}(2011)}]{2011ApJ...726...90Z}%
  \BibitemOpen
  \bibfield  {author} {\bibinfo {author} {\bibfnamefont {B.}~\bibnamefont {{Zhang}}}\ and\ \bibinfo {author} {\bibfnamefont {H.}~\bibnamefont {{Yan}}},\ }\href {\doibase 10.1088/0004-637X/726/2/90} {\bibfield  {journal} {\bibinfo  {journal} {\apj}\ }\textbf {\bibinfo {volume} {726}},\ \bibinfo {eid} {90} (\bibinfo {year} {2011})},\ \Eprint {http://arxiv.org/abs/1011.1197} {arXiv:1011.1197 [astro-ph.HE]} \BibitemShut {NoStop}%
\bibitem [{\citenamefont {{Arimoto}}\ \emph {et~al.}(2016)\citenamefont {{Arimoto}}, \citenamefont {{Asano}}, \citenamefont {{Ohno}}, \citenamefont {{Veres}}, \citenamefont {{Axelsson}}, \citenamefont {{Bissaldi}}, \citenamefont {{Tachibana}},\ and\ \citenamefont {{Kawai}}}]{2016ApJ...833..139A}%
  \BibitemOpen
  \bibfield  {author} {\bibinfo {author} {\bibfnamefont {M.}~\bibnamefont {{Arimoto}}}, \bibinfo {author} {\bibfnamefont {K.}~\bibnamefont {{Asano}}}, \bibinfo {author} {\bibfnamefont {M.}~\bibnamefont {{Ohno}}}, \bibinfo {author} {\bibfnamefont {P.}~\bibnamefont {{Veres}}}, \bibinfo {author} {\bibfnamefont {M.}~\bibnamefont {{Axelsson}}}, \bibinfo {author} {\bibfnamefont {E.}~\bibnamefont {{Bissaldi}}}, \bibinfo {author} {\bibfnamefont {Y.}~\bibnamefont {{Tachibana}}}, \ and\ \bibinfo {author} {\bibfnamefont {N.}~\bibnamefont {{Kawai}}},\ }\href {\doibase 10.3847/1538-4357/833/2/139} {\bibfield  {journal} {\bibinfo  {journal} {\apj}\ }\textbf {\bibinfo {volume} {833}},\ \bibinfo {eid} {139} (\bibinfo {year} {2016})},\ \Eprint {http://arxiv.org/abs/1610.04867} {arXiv:1610.04867 [astro-ph.HE]} \BibitemShut {NoStop}%
\bibitem [{\citenamefont {{Yonetoku}}\ \emph {et~al.}(2004)\citenamefont {{Yonetoku}}, \citenamefont {{Murakami}}, \citenamefont {{Nakamura}}, \citenamefont {{Yamazaki}}, \citenamefont {{Inoue}},\ and\ \citenamefont {{Ioka}}}]{2004ApJ...609..935Y}%
  \BibitemOpen
  \bibfield  {author} {\bibinfo {author} {\bibfnamefont {D.}~\bibnamefont {{Yonetoku}}}, \bibinfo {author} {\bibfnamefont {T.}~\bibnamefont {{Murakami}}}, \bibinfo {author} {\bibfnamefont {T.}~\bibnamefont {{Nakamura}}}, \bibinfo {author} {\bibfnamefont {R.}~\bibnamefont {{Yamazaki}}}, \bibinfo {author} {\bibfnamefont {A.~K.}\ \bibnamefont {{Inoue}}}, \ and\ \bibinfo {author} {\bibfnamefont {K.}~\bibnamefont {{Ioka}}},\ }\href {\doibase 10.1086/421285} {\bibfield  {journal} {\bibinfo  {journal} {\apj}\ }\textbf {\bibinfo {volume} {609}},\ \bibinfo {pages} {935} (\bibinfo {year} {2004})},\ \Eprint {http://arxiv.org/abs/astro-ph/0309217} {arXiv:astro-ph/0309217 [astro-ph]} \BibitemShut {NoStop}%
\bibitem [{\citenamefont {{Li}}\ \emph {et~al.}(2022)\citenamefont {{Li}}, \citenamefont {{Wang}}, \citenamefont {{Jiang}}, \citenamefont {{Du}}, \citenamefont {{Liu}}, \citenamefont {{Gan}}, \citenamefont {{Zhou}}, \citenamefont {{Lin}},\ and\ \citenamefont {{Liang}}}]{2022ApJ...932...69L}%
  \BibitemOpen
  \bibfield  {author} {\bibinfo {author} {\bibfnamefont {R.-Q.}\ \bibnamefont {{Li}}}, \bibinfo {author} {\bibfnamefont {X.-G.}\ \bibnamefont {{Wang}}}, \bibinfo {author} {\bibfnamefont {L.-Y.}\ \bibnamefont {{Jiang}}}, \bibinfo {author} {\bibfnamefont {S.-S.}\ \bibnamefont {{Du}}}, \bibinfo {author} {\bibfnamefont {H.-Y.}\ \bibnamefont {{Liu}}}, \bibinfo {author} {\bibfnamefont {Y.-Y.}\ \bibnamefont {{Gan}}}, \bibinfo {author} {\bibfnamefont {Z.-M.}\ \bibnamefont {{Zhou}}}, \bibinfo {author} {\bibfnamefont {D.-B.}\ \bibnamefont {{Lin}}}, \ and\ \bibinfo {author} {\bibfnamefont {E.-W.}\ \bibnamefont {{Liang}}},\ }\href {\doibase 10.3847/1538-4357/ac6d5d} {\bibfield  {journal} {\bibinfo  {journal} {\apj}\ }\textbf {\bibinfo {volume} {932}},\ \bibinfo {eid} {69} (\bibinfo {year} {2022})}\BibitemShut {NoStop}%
\bibitem [{\citenamefont {{Song}}\ \emph {et~al.}(2022{\natexlab{a}})\citenamefont {{Song}}, \citenamefont {{Zhang}}, \citenamefont {{Zhang}}, \citenamefont {{Xiong}},\ and\ \citenamefont {{Song}}}]{2022ApJ...931..112S}%
  \BibitemOpen
  \bibfield  {author} {\bibinfo {author} {\bibfnamefont {X.-Y.}\ \bibnamefont {{Song}}}, \bibinfo {author} {\bibfnamefont {S.-N.}\ \bibnamefont {{Zhang}}}, \bibinfo {author} {\bibfnamefont {S.}~\bibnamefont {{Zhang}}}, \bibinfo {author} {\bibfnamefont {S.-L.}\ \bibnamefont {{Xiong}}}, \ and\ \bibinfo {author} {\bibfnamefont {L.-M.}\ \bibnamefont {{Song}}},\ }\href {\doibase 10.3847/1538-4357/ac6b33} {\bibfield  {journal} {\bibinfo  {journal} {\apj}\ }\textbf {\bibinfo {volume} {931}},\ \bibinfo {eid} {112} (\bibinfo {year} {2022}{\natexlab{a}})},\ \Eprint {http://arxiv.org/abs/2204.09430} {arXiv:2204.09430 [astro-ph.HE]} \BibitemShut {NoStop}%
\bibitem [{\citenamefont {{Song}}\ \emph {et~al.}(2022{\natexlab{b}})\citenamefont {{Song}}, \citenamefont {{Zhang}}, \citenamefont {{Ge}},\ and\ \citenamefont {{Zhang}}}]{2022MNRAS.517.2088S}%
  \BibitemOpen
  \bibfield  {author} {\bibinfo {author} {\bibfnamefont {X.-Y.}\ \bibnamefont {{Song}}}, \bibinfo {author} {\bibfnamefont {S.-N.}\ \bibnamefont {{Zhang}}}, \bibinfo {author} {\bibfnamefont {M.-Y.}\ \bibnamefont {{Ge}}}, \ and\ \bibinfo {author} {\bibfnamefont {S.}~\bibnamefont {{Zhang}}},\ }\href {\doibase 10.1093/mnras/stac2764} {\bibfield  {journal} {\bibinfo  {journal} {\mnras{517}{2088}{2022}}\ }\textbf {\bibinfo {volume} {517}},\ \bibinfo {pages} {2088} (\bibinfo {year} {2022}{\natexlab{b}})},\ \Eprint {http://arxiv.org/abs/2209.10832} {arXiv:2209.10832 [astro-ph.HE]} \BibitemShut {NoStop}%
\bibitem [{\citenamefont {{Chen}}\ \emph {et~al.}(2024)\citenamefont {{Chen}}, \citenamefont {{Zhu}}, \citenamefont {{Peng}},\ and\ \citenamefont {{Zhang}}}]{2024ApJ...972..132C}%
  \BibitemOpen
  \bibfield  {author} {\bibinfo {author} {\bibfnamefont {J.-M.}\ \bibnamefont {{Chen}}}, \bibinfo {author} {\bibfnamefont {K.-R.}\ \bibnamefont {{Zhu}}}, \bibinfo {author} {\bibfnamefont {Z.-Y.}\ \bibnamefont {{Peng}}}, \ and\ \bibinfo {author} {\bibfnamefont {L.}~\bibnamefont {{Zhang}}},\ }\href {\doibase 10.3847/1538-4357/ad5f93} {\bibfield  {journal} {\bibinfo  {journal} {\apj}\ }\textbf {\bibinfo {volume} {972}},\ \bibinfo {eid} {132} (\bibinfo {year} {2024})}\BibitemShut {NoStop}%
\bibitem [{\citenamefont {{Bhat}}\ \emph {et~al.}(2008)\citenamefont {{Bhat}}, \citenamefont {{Preece}},\ and\ \citenamefont {{van der Horst}}}]{2008GCN..8550....1B}%
  \BibitemOpen
  \bibfield  {author} {\bibinfo {author} {\bibfnamefont {P.~N.}\ \bibnamefont {{Bhat}}}, \bibinfo {author} {\bibfnamefont {R.~D.}\ \bibnamefont {{Preece}}}, \ and\ \bibinfo {author} {\bibfnamefont {A.~J.}\ \bibnamefont {{van der Horst}}},\ }\href@noop {} {\bibfield  {journal} {\bibinfo  {journal} {GRB Coordinates Network}\ }\textbf {\bibinfo {volume} {8550}},\ \bibinfo {pages} {1} (\bibinfo {year} {2008})}\BibitemShut {NoStop}%
\bibitem [{\citenamefont {{D'Elia}}\ \emph {et~al.}(2008)\citenamefont {{D'Elia}}, \citenamefont {{Thoene}}, \citenamefont {{de Ugarte Postigo}}, \citenamefont {{D'Avanzo}}, \citenamefont {{Covino}}, \citenamefont {{Piranomonte}}, \citenamefont {{Salvaterra}},\ and\ \citenamefont {{Chincarini}}}]{2008GCN..8531....1D}%
  \BibitemOpen
  \bibfield  {author} {\bibinfo {author} {\bibfnamefont {V.}~\bibnamefont {{D'Elia}}}, \bibinfo {author} {\bibfnamefont {C.~C.}\ \bibnamefont {{Thoene}}}, \bibinfo {author} {\bibfnamefont {A.}~\bibnamefont {{de Ugarte Postigo}}}, \bibinfo {author} {\bibfnamefont {P.}~\bibnamefont {{D'Avanzo}}}, \bibinfo {author} {\bibfnamefont {S.}~\bibnamefont {{Covino}}}, \bibinfo {author} {\bibfnamefont {S.}~\bibnamefont {{Piranomonte}}}, \bibinfo {author} {\bibfnamefont {R.}~\bibnamefont {{Salvaterra}}}, \ and\ \bibinfo {author} {\bibfnamefont {G.}~\bibnamefont {{Chincarini}}},\ }\href@noop {} {\bibfield  {journal} {\bibinfo  {journal} {GRB Coordinates Network}\ }\textbf {\bibinfo {volume} {8531}},\ \bibinfo {pages} {1} (\bibinfo {year} {2008})}\BibitemShut {NoStop}%
\bibitem [{\citenamefont {{Pelangeon}}\ and\ \citenamefont {{Atteia}}(2008)}]{2008GCN..8700....1P}%
  \BibitemOpen
  \bibfield  {author} {\bibinfo {author} {\bibfnamefont {A.}~\bibnamefont {{Pelangeon}}}\ and\ \bibinfo {author} {\bibfnamefont {J.~L.}\ \bibnamefont {{Atteia}}},\ }\href@noop {} {\bibfield  {journal} {\bibinfo  {journal} {GRB Coordinates Network}\ }\textbf {\bibinfo {volume} {8700}},\ \bibinfo {pages} {1} (\bibinfo {year} {2008})}\BibitemShut {NoStop}%
\bibitem [{\citenamefont {{Wilson-Hodge}}(2008)}]{2008GCN..8704....1W}%
  \BibitemOpen
  \bibfield  {author} {\bibinfo {author} {\bibfnamefont {C.~A.}\ \bibnamefont {{Wilson-Hodge}}},\ }\href@noop {} {\bibfield  {journal} {\bibinfo  {journal} {GRB Coordinates Network}\ }\textbf {\bibinfo {volume} {8704}},\ \bibinfo {pages} {1} (\bibinfo {year} {2008})}\BibitemShut {NoStop}%
\bibitem [{\citenamefont {{Bissaldi}}\ and\ \citenamefont {{McBreen}}(2008)}]{2008GCN..8715....1B}%
  \BibitemOpen
  \bibfield  {author} {\bibinfo {author} {\bibfnamefont {E.}~\bibnamefont {{Bissaldi}}}\ and\ \bibinfo {author} {\bibfnamefont {S.}~\bibnamefont {{McBreen}}},\ }\href@noop {} {\bibfield  {journal} {\bibinfo  {journal} {GRB Coordinates Network}\ }\textbf {\bibinfo {volume} {8715}},\ \bibinfo {pages} {1} (\bibinfo {year} {2008})}\BibitemShut {NoStop}%
\bibitem [{\citenamefont {{Graham}}\ \emph {et~al.}(2008)\citenamefont {{Graham}}, \citenamefont {{Tanvir}}, \citenamefont {{Fruchter}}, \citenamefont {{Wiersema}},\ and\ \citenamefont {{Levan}}}]{2008GCN..8718....1G}%
  \BibitemOpen
  \bibfield  {author} {\bibinfo {author} {\bibfnamefont {J.~F.}\ \bibnamefont {{Graham}}}, \bibinfo {author} {\bibfnamefont {N.~R.}\ \bibnamefont {{Tanvir}}}, \bibinfo {author} {\bibfnamefont {A.~S.}\ \bibnamefont {{Fruchter}}}, \bibinfo {author} {\bibfnamefont {K.}~\bibnamefont {{Wiersema}}}, \ and\ \bibinfo {author} {\bibfnamefont {A.~J.}\ \bibnamefont {{Levan}}},\ }\href@noop {} {\bibfield  {journal} {\bibinfo  {journal} {GRB Coordinates Network}\ }\textbf {\bibinfo {volume} {8718}},\ \bibinfo {pages} {1} (\bibinfo {year} {2008})}\BibitemShut {NoStop}%
\bibitem [{\citenamefont {{Connaughton}}(2009)}]{2009GCN..9230....1C}%
  \BibitemOpen
  \bibfield  {author} {\bibinfo {author} {\bibfnamefont {V.}~\bibnamefont {{Connaughton}}},\ }\href@noop {} {\bibfield  {journal} {\bibinfo  {journal} {GRB Coordinates Network}\ }\textbf {\bibinfo {volume} {9230}},\ \bibinfo {pages} {1} (\bibinfo {year} {2009})}\BibitemShut {NoStop}%
\bibitem [{\citenamefont {{Chornock}}\ \emph {et~al.}(2009)\citenamefont {{Chornock}}, \citenamefont {{Perley}}, \citenamefont {{Cenko}},\ and\ \citenamefont {{Bloom}}}]{2009GCN..9243....1C}%
  \BibitemOpen
  \bibfield  {author} {\bibinfo {author} {\bibfnamefont {R.}~\bibnamefont {{Chornock}}}, \bibinfo {author} {\bibfnamefont {D.~A.}\ \bibnamefont {{Perley}}}, \bibinfo {author} {\bibfnamefont {S.~B.}\ \bibnamefont {{Cenko}}}, \ and\ \bibinfo {author} {\bibfnamefont {J.~S.}\ \bibnamefont {{Bloom}}},\ }\href@noop {} {\bibfield  {journal} {\bibinfo  {journal} {GRB Coordinates Network}\ }\textbf {\bibinfo {volume} {9243}},\ \bibinfo {pages} {1} (\bibinfo {year} {2009})}\BibitemShut {NoStop}%
\bibitem [{\citenamefont {{Chaplin}}(2009)}]{2009GCN.10095....1C}%
  \BibitemOpen
  \bibfield  {author} {\bibinfo {author} {\bibfnamefont {V.}~\bibnamefont {{Chaplin}}},\ }\href@noop {} {\bibfield  {journal} {\bibinfo  {journal} {GRB Coordinates Network}\ }\textbf {\bibinfo {volume} {10095}},\ \bibinfo {pages} {1} (\bibinfo {year} {2009})}\BibitemShut {NoStop}%
\bibitem [{\citenamefont {{Xu}}\ \emph {et~al.}(2009)\citenamefont {{Xu}}, \citenamefont {{Fynbo}}, \citenamefont {{Tanvir}}, \citenamefont {{Hjorth}}, \citenamefont {{Leloudas}}, \citenamefont {{Malesani}}, \citenamefont {{Jakobsson}}, \citenamefont {{Wilson}},\ and\ \citenamefont {{Andersen}}}]{2009GCN.10053....1X}%
  \BibitemOpen
  \bibfield  {author} {\bibinfo {author} {\bibfnamefont {D.}~\bibnamefont {{Xu}}}, \bibinfo {author} {\bibfnamefont {J.~P.~U.}\ \bibnamefont {{Fynbo}}}, \bibinfo {author} {\bibfnamefont {N.~R.}\ \bibnamefont {{Tanvir}}}, \bibinfo {author} {\bibfnamefont {J.}~\bibnamefont {{Hjorth}}}, \bibinfo {author} {\bibfnamefont {G.}~\bibnamefont {{Leloudas}}}, \bibinfo {author} {\bibfnamefont {D.}~\bibnamefont {{Malesani}}}, \bibinfo {author} {\bibfnamefont {P.}~\bibnamefont {{Jakobsson}}}, \bibinfo {author} {\bibfnamefont {P.~A.}\ \bibnamefont {{Wilson}}}, \ and\ \bibinfo {author} {\bibfnamefont {J.}~\bibnamefont {{Andersen}}},\ }\href@noop {} {\bibfield  {journal} {\bibinfo  {journal} {GRB Coordinates Network}\ }\textbf {\bibinfo {volume} {10053}},\ \bibinfo {pages} {1} (\bibinfo {year} {2009})}\BibitemShut {NoStop}%
\bibitem [{\citenamefont {{Foley}}(2010)}]{2010GCN.10595....1F}%
  \BibitemOpen
  \bibfield  {author} {\bibinfo {author} {\bibfnamefont {S.}~\bibnamefont {{Foley}}},\ }\href@noop {} {\bibfield  {journal} {\bibinfo  {journal} {GRB Coordinates Network}\ }\textbf {\bibinfo {volume} {10595}},\ \bibinfo {pages} {1} (\bibinfo {year} {2010})}\BibitemShut {NoStop}%
\bibitem [{\citenamefont {{Cucchiara}}\ and\ \citenamefont {{Fox}}(2010)}]{2010GCN.10606....1C}%
  \BibitemOpen
  \bibfield  {author} {\bibinfo {author} {\bibfnamefont {A.}~\bibnamefont {{Cucchiara}}}\ and\ \bibinfo {author} {\bibfnamefont {D.~B.}\ \bibnamefont {{Fox}}},\ }\href@noop {} {\bibfield  {journal} {\bibinfo  {journal} {GRB Coordinates Network}\ }\textbf {\bibinfo {volume} {10606}},\ \bibinfo {pages} {1} (\bibinfo {year} {2010})}\BibitemShut {NoStop}%
\bibitem [{\citenamefont {{von Kienlin}}(2010{\natexlab{a}})}]{2010GCN.11099....1V}%
  \BibitemOpen
  \bibfield  {author} {\bibinfo {author} {\bibfnamefont {A.}~\bibnamefont {{von Kienlin}}},\ }\href@noop {} {\bibfield  {journal} {\bibinfo  {journal} {GRB Coordinates Network}\ }\textbf {\bibinfo {volume} {11099}},\ \bibinfo {pages} {1} (\bibinfo {year} {2010}{\natexlab{a}})}\BibitemShut {NoStop}%
\bibitem [{\citenamefont {{Filgas}}\ \emph {et~al.}(2010)\citenamefont {{Filgas}}, \citenamefont {{Schady}},\ and\ \citenamefont {{Greiner}}}]{2010GCN.11091....1F}%
  \BibitemOpen
  \bibfield  {author} {\bibinfo {author} {\bibfnamefont {R.}~\bibnamefont {{Filgas}}}, \bibinfo {author} {\bibfnamefont {P.}~\bibnamefont {{Schady}}}, \ and\ \bibinfo {author} {\bibfnamefont {J.}~\bibnamefont {{Greiner}}},\ }\href@noop {} {\bibfield  {journal} {\bibinfo  {journal} {GRB Coordinates Network}\ }\textbf {\bibinfo {volume} {11091}},\ \bibinfo {pages} {1} (\bibinfo {year} {2010})}\BibitemShut {NoStop}%
\bibitem [{\citenamefont {{Golenetskii}}\ \emph {et~al.}(2010)\citenamefont {{Golenetskii}}, \citenamefont {{Aptekar}}, \citenamefont {{Frederiks}}, \citenamefont {{Mazets}}, \citenamefont {{Pal'Shin}}, \citenamefont {{Oleynik}}, \citenamefont {{Ulanov}}, \citenamefont {{Svinkin}},\ and\ \citenamefont {{Cline}}}]{2010GCN.11021....1G}%
  \BibitemOpen
  \bibfield  {author} {\bibinfo {author} {\bibfnamefont {S.}~\bibnamefont {{Golenetskii}}}, \bibinfo {author} {\bibfnamefont {R.}~\bibnamefont {{Aptekar}}}, \bibinfo {author} {\bibfnamefont {D.}~\bibnamefont {{Frederiks}}}, \bibinfo {author} {\bibfnamefont {E.}~\bibnamefont {{Mazets}}}, \bibinfo {author} {\bibfnamefont {V.}~\bibnamefont {{Pal'Shin}}}, \bibinfo {author} {\bibfnamefont {P.}~\bibnamefont {{Oleynik}}}, \bibinfo {author} {\bibfnamefont {M.}~\bibnamefont {{Ulanov}}}, \bibinfo {author} {\bibfnamefont {D.}~\bibnamefont {{Svinkin}}}, \ and\ \bibinfo {author} {\bibfnamefont {T.}~\bibnamefont {{Cline}}},\ }\href@noop {} {\bibfield  {journal} {\bibinfo  {journal} {GRB Coordinates Network}\ }\textbf {\bibinfo {volume} {11021}},\ \bibinfo {pages} {1} (\bibinfo {year} {2010})}\BibitemShut {NoStop}%
\bibitem [{\citenamefont {{Kruehler}}\ \emph {et~al.}(2013)\citenamefont {{Kruehler}}, \citenamefont {{Greiner}},\ and\ \citenamefont {{Kann}}}]{2013GCN.14500....1K}%
  \BibitemOpen
  \bibfield  {author} {\bibinfo {author} {\bibfnamefont {T.}~\bibnamefont {{Kruehler}}}, \bibinfo {author} {\bibfnamefont {J.}~\bibnamefont {{Greiner}}}, \ and\ \bibinfo {author} {\bibfnamefont {D.~A.}\ \bibnamefont {{Kann}}},\ }\href@noop {} {\bibfield  {journal} {\bibinfo  {journal} {GRB Coordinates Network}\ }\textbf {\bibinfo {volume} {14500}},\ \bibinfo {pages} {1} (\bibinfo {year} {2013})}\BibitemShut {NoStop}%
\bibitem [{\citenamefont {{Gruber}}(2012)}]{2012GCN.13469....1G}%
  \BibitemOpen
  \bibfield  {author} {\bibinfo {author} {\bibfnamefont {D.}~\bibnamefont {{Gruber}}},\ }\href@noop {} {\bibfield  {journal} {\bibinfo  {journal} {GRB Coordinates Network}\ }\textbf {\bibinfo {volume} {13469}},\ \bibinfo {pages} {1} (\bibinfo {year} {2012})}\BibitemShut {NoStop}%
\bibitem [{\citenamefont {{Tanvir}}\ \emph {et~al.}(2012)\citenamefont {{Tanvir}}, \citenamefont {{Levan}},\ and\ \citenamefont {{Krogsrud}}}]{2012GCN.13458....1T}%
  \BibitemOpen
  \bibfield  {author} {\bibinfo {author} {\bibfnamefont {N.~R.}\ \bibnamefont {{Tanvir}}}, \bibinfo {author} {\bibfnamefont {A.~J.}\ \bibnamefont {{Levan}}}, \ and\ \bibinfo {author} {\bibfnamefont {D.}~\bibnamefont {{Krogsrud}}},\ }\href@noop {} {\bibfield  {journal} {\bibinfo  {journal} {GRB Coordinates Network}\ }\textbf {\bibinfo {volume} {13458}},\ \bibinfo {pages} {1} (\bibinfo {year} {2012})}\BibitemShut {NoStop}%
\bibitem [{\citenamefont {{Younes}}(2012)}]{2012GCN.13809....1Y}%
  \BibitemOpen
  \bibfield  {author} {\bibinfo {author} {\bibfnamefont {G.}~\bibnamefont {{Younes}}},\ }\href@noop {} {\bibfield  {journal} {\bibinfo  {journal} {GRB Coordinates Network}\ }\textbf {\bibinfo {volume} {13809}},\ \bibinfo {pages} {1} (\bibinfo {year} {2012})}\BibitemShut {NoStop}%
\bibitem [{\citenamefont {{Knust}}\ \emph {et~al.}(2012)\citenamefont {{Knust}}, \citenamefont {{Kruehler}}, \citenamefont {{Klose}},\ and\ \citenamefont {{Greiner}}}]{2012GCN.13810....1K}%
  \BibitemOpen
  \bibfield  {author} {\bibinfo {author} {\bibfnamefont {F.}~\bibnamefont {{Knust}}}, \bibinfo {author} {\bibfnamefont {T.}~\bibnamefont {{Kruehler}}}, \bibinfo {author} {\bibfnamefont {S.}~\bibnamefont {{Klose}}}, \ and\ \bibinfo {author} {\bibfnamefont {J.}~\bibnamefont {{Greiner}}},\ }\href@noop {} {\bibfield  {journal} {\bibinfo  {journal} {GRB Coordinates Network}\ }\textbf {\bibinfo {volume} {13810}},\ \bibinfo {pages} {1} (\bibinfo {year} {2012})}\BibitemShut {NoStop}%
\bibitem [{\citenamefont {{Yu}}(2012)}]{2012GCN.14078....1Y}%
  \BibitemOpen
  \bibfield  {author} {\bibinfo {author} {\bibfnamefont {D.}~\bibnamefont {{Yu}}},\ }\href@noop {} {\bibfield  {journal} {\bibinfo  {journal} {GRB Coordinates Network}\ }\textbf {\bibinfo {volume} {14078}},\ \bibinfo {pages} {1} (\bibinfo {year} {2012})}\BibitemShut {NoStop}%
\bibitem [{\citenamefont {{Perley}}\ \emph {et~al.}(2012)\citenamefont {{Perley}}, \citenamefont {{Prochaska}},\ and\ \citenamefont {{Morgan}}}]{2012GCN.14059....1P}%
  \BibitemOpen
  \bibfield  {author} {\bibinfo {author} {\bibfnamefont {D.~A.}\ \bibnamefont {{Perley}}}, \bibinfo {author} {\bibfnamefont {J.~X.}\ \bibnamefont {{Prochaska}}}, \ and\ \bibinfo {author} {\bibfnamefont {A.~N.}\ \bibnamefont {{Morgan}}},\ }\href@noop {} {\bibfield  {journal} {\bibinfo  {journal} {GRB Coordinates Network}\ }\textbf {\bibinfo {volume} {14059}},\ \bibinfo {pages} {1} (\bibinfo {year} {2012})}\BibitemShut {NoStop}%
\bibitem [{\citenamefont {{Golenetskii}}\ \emph {et~al.}(2013)\citenamefont {{Golenetskii}}, \citenamefont {{Aptekar}}, \citenamefont {{Frederiks}}, \citenamefont {{Mazets}}, \citenamefont {{Pal'Shin}}, \citenamefont {{Oleynik}}, \citenamefont {{Ulanov}}, \citenamefont {{Svinkin}},\ and\ \citenamefont {{Cline}}}]{2013GCN.14368....1G}%
  \BibitemOpen
  \bibfield  {author} {\bibinfo {author} {\bibfnamefont {S.}~\bibnamefont {{Golenetskii}}}, \bibinfo {author} {\bibfnamefont {R.}~\bibnamefont {{Aptekar}}}, \bibinfo {author} {\bibfnamefont {D.}~\bibnamefont {{Frederiks}}}, \bibinfo {author} {\bibfnamefont {E.}~\bibnamefont {{Mazets}}}, \bibinfo {author} {\bibfnamefont {V.}~\bibnamefont {{Pal'Shin}}}, \bibinfo {author} {\bibfnamefont {P.}~\bibnamefont {{Oleynik}}}, \bibinfo {author} {\bibfnamefont {M.}~\bibnamefont {{Ulanov}}}, \bibinfo {author} {\bibfnamefont {D.}~\bibnamefont {{Svinkin}}}, \ and\ \bibinfo {author} {\bibfnamefont {T.}~\bibnamefont {{Cline}}},\ }\href@noop {} {\bibfield  {journal} {\bibinfo  {journal} {GRB Coordinates Network}\ }\textbf {\bibinfo {volume} {14368}},\ \bibinfo {pages} {1} (\bibinfo {year} {2013})}\BibitemShut {NoStop}%
\bibitem [{\citenamefont {{Pelassa}}(2013)}]{2013GCN.14869....1P}%
  \BibitemOpen
  \bibfield  {author} {\bibinfo {author} {\bibfnamefont {V.}~\bibnamefont {{Pelassa}}},\ }\href@noop {} {\bibfield  {journal} {\bibinfo  {journal} {GRB Coordinates Network}\ }\textbf {\bibinfo {volume} {14869}},\ \bibinfo {pages} {1} (\bibinfo {year} {2013})}\BibitemShut {NoStop}%
\bibitem [{\citenamefont {{Ruffini}}\ \emph {et~al.}(2013)\citenamefont {{Ruffini}}, \citenamefont {{Bianco}}, \citenamefont {{Enderli}}, \citenamefont {{Muccino}}, \citenamefont {{Penacchioni}}, \citenamefont {{Pisani}}, \citenamefont {{Rueda}}, \citenamefont {{Sahakyan}}, \citenamefont {{Wang}},\ and\ \citenamefont {{Izzo}}}]{2013GCN.14888....1R}%
  \BibitemOpen
  \bibfield  {author} {\bibinfo {author} {\bibfnamefont {R.}~\bibnamefont {{Ruffini}}}, \bibinfo {author} {\bibfnamefont {C.~L.}\ \bibnamefont {{Bianco}}}, \bibinfo {author} {\bibfnamefont {M.}~\bibnamefont {{Enderli}}}, \bibinfo {author} {\bibfnamefont {M.}~\bibnamefont {{Muccino}}}, \bibinfo {author} {\bibfnamefont {A.~V.}\ \bibnamefont {{Penacchioni}}}, \bibinfo {author} {\bibfnamefont {G.~B.}\ \bibnamefont {{Pisani}}}, \bibinfo {author} {\bibfnamefont {J.~A.}\ \bibnamefont {{Rueda}}}, \bibinfo {author} {\bibfnamefont {N.}~\bibnamefont {{Sahakyan}}}, \bibinfo {author} {\bibfnamefont {Y.}~\bibnamefont {{Wang}}}, \ and\ \bibinfo {author} {\bibfnamefont {L.}~\bibnamefont {{Izzo}}},\ }\href@noop {} {\bibfield  {journal} {\bibinfo  {journal} {GRB Coordinates Network}\ }\textbf {\bibinfo {volume} {14888}},\ \bibinfo {pages} {1} (\bibinfo {year} {2013})}\BibitemShut {NoStop}%
\bibitem [{\citenamefont {{von Kienlin}}\ and\ \citenamefont {{Bhat}}(2014)}]{2014GCN.15796....1V}%
  \BibitemOpen
  \bibfield  {author} {\bibinfo {author} {\bibfnamefont {A.}~\bibnamefont {{von Kienlin}}}\ and\ \bibinfo {author} {\bibfnamefont {P.~N.}\ \bibnamefont {{Bhat}}},\ }\href@noop {} {\bibfield  {journal} {\bibinfo  {journal} {GRB Coordinates Network}\ }\textbf {\bibinfo {volume} {15796}},\ \bibinfo {pages} {1} (\bibinfo {year} {2014})}\BibitemShut {NoStop}%
\bibitem [{\citenamefont {{D'Elia}}\ \emph {et~al.}(2014)\citenamefont {{D'Elia}}, \citenamefont {{D'Avanzo}}, \citenamefont {{Covino}}, \citenamefont {{Melandri}}, \citenamefont {{Vergani}},\ and\ \citenamefont {{di Fabrizio}}}]{2014GCN.15802....1D}%
  \BibitemOpen
  \bibfield  {author} {\bibinfo {author} {\bibfnamefont {V.}~\bibnamefont {{D'Elia}}}, \bibinfo {author} {\bibfnamefont {P.}~\bibnamefont {{D'Avanzo}}}, \bibinfo {author} {\bibfnamefont {S.}~\bibnamefont {{Covino}}}, \bibinfo {author} {\bibfnamefont {A.}~\bibnamefont {{Melandri}}}, \bibinfo {author} {\bibfnamefont {S.~D.}\ \bibnamefont {{Vergani}}}, \ and\ \bibinfo {author} {\bibfnamefont {L.}~\bibnamefont {{di Fabrizio}}},\ }\href@noop {} {\bibfield  {journal} {\bibinfo  {journal} {GRB Coordinates Network}\ }\textbf {\bibinfo {volume} {15802}},\ \bibinfo {pages} {1} (\bibinfo {year} {2014})}\BibitemShut {NoStop}%
\bibitem [{\citenamefont {{Golenetskii}}\ \emph {et~al.}(2014{\natexlab{a}})\citenamefont {{Golenetskii}}, \citenamefont {{Aptekar}}, \citenamefont {{Frederiks}}, \citenamefont {{Pal'Shin}}, \citenamefont {{Oleynik}}, \citenamefont {{Ulanov}}, \citenamefont {{Svinkin}}, \citenamefont {{Tsvetkova}},\ and\ \citenamefont {{Cline}}}]{2014GCN.16134....1G}%
  \BibitemOpen
  \bibfield  {author} {\bibinfo {author} {\bibfnamefont {S.}~\bibnamefont {{Golenetskii}}}, \bibinfo {author} {\bibfnamefont {R.}~\bibnamefont {{Aptekar}}}, \bibinfo {author} {\bibfnamefont {D.}~\bibnamefont {{Frederiks}}}, \bibinfo {author} {\bibfnamefont {V.}~\bibnamefont {{Pal'Shin}}}, \bibinfo {author} {\bibfnamefont {P.}~\bibnamefont {{Oleynik}}}, \bibinfo {author} {\bibfnamefont {M.}~\bibnamefont {{Ulanov}}}, \bibinfo {author} {\bibfnamefont {D.}~\bibnamefont {{Svinkin}}}, \bibinfo {author} {\bibfnamefont {A.}~\bibnamefont {{Tsvetkova}}}, \ and\ \bibinfo {author} {\bibfnamefont {T.}~\bibnamefont {{Cline}}},\ }\href@noop {} {\bibfield  {journal} {\bibinfo  {journal} {GRB Coordinates Network}\ }\textbf {\bibinfo {volume} {16134}},\ \bibinfo {pages} {1} (\bibinfo {year} {2014}{\natexlab{a}})}\BibitemShut {NoStop}%
\bibitem [{\citenamefont {{von Kienlin}}(2014)}]{2014GCN.16152....1V}%
  \BibitemOpen
  \bibfield  {author} {\bibinfo {author} {\bibfnamefont {A.}~\bibnamefont {{von Kienlin}}},\ }\href@noop {} {\bibfield  {journal} {\bibinfo  {journal} {GRB Coordinates Network}\ }\textbf {\bibinfo {volume} {16152}},\ \bibinfo {pages} {1} (\bibinfo {year} {2014})}\BibitemShut {NoStop}%
\bibitem [{\citenamefont {{Tanvir}}\ \emph {et~al.}(2014)\citenamefont {{Tanvir}}, \citenamefont {{Levan}}, \citenamefont {{Wiersema}}, \citenamefont {{Petric}}, \citenamefont {{Chiboucas}},\ and\ \citenamefont {{Miller}}}]{2014GCN.16150....1T}%
  \BibitemOpen
  \bibfield  {author} {\bibinfo {author} {\bibfnamefont {N.~R.}\ \bibnamefont {{Tanvir}}}, \bibinfo {author} {\bibfnamefont {A.~J.}\ \bibnamefont {{Levan}}}, \bibinfo {author} {\bibfnamefont {K.}~\bibnamefont {{Wiersema}}}, \bibinfo {author} {\bibfnamefont {A.}~\bibnamefont {{Petric}}}, \bibinfo {author} {\bibfnamefont {K.}~\bibnamefont {{Chiboucas}}}, \ and\ \bibinfo {author} {\bibfnamefont {J.}~\bibnamefont {{Miller}}},\ }\href@noop {} {\bibfield  {journal} {\bibinfo  {journal} {GRB Coordinates Network}\ }\textbf {\bibinfo {volume} {16150}},\ \bibinfo {pages} {1} (\bibinfo {year} {2014})}\BibitemShut {NoStop}%
\bibitem [{\citenamefont {{Golenetskii}}\ \emph {et~al.}(2014{\natexlab{b}})\citenamefont {{Golenetskii}}, \citenamefont {{Aptekar}}, \citenamefont {{Frederiks}}, \citenamefont {{Pal'Shin}}, \citenamefont {{Oleynik}}, \citenamefont {{Ulanov}}, \citenamefont {{Svinkin}}, \citenamefont {{Tsvetkova}}, \citenamefont {{Lyssenko}},\ and\ \citenamefont {{Cline}}}]{2014GCN.16660....1G}%
  \BibitemOpen
  \bibfield  {author} {\bibinfo {author} {\bibfnamefont {S.}~\bibnamefont {{Golenetskii}}}, \bibinfo {author} {\bibfnamefont {R.}~\bibnamefont {{Aptekar}}}, \bibinfo {author} {\bibfnamefont {D.}~\bibnamefont {{Frederiks}}}, \bibinfo {author} {\bibfnamefont {V.}~\bibnamefont {{Pal'Shin}}}, \bibinfo {author} {\bibfnamefont {P.}~\bibnamefont {{Oleynik}}}, \bibinfo {author} {\bibfnamefont {M.}~\bibnamefont {{Ulanov}}}, \bibinfo {author} {\bibfnamefont {D.}~\bibnamefont {{Svinkin}}}, \bibinfo {author} {\bibfnamefont {A.}~\bibnamefont {{Tsvetkova}}}, \bibinfo {author} {\bibfnamefont {A.}~\bibnamefont {{Lyssenko}}}, \ and\ \bibinfo {author} {\bibfnamefont {T.}~\bibnamefont {{Cline}}},\ }\href@noop {} {\bibfield  {journal} {\bibinfo  {journal} {GRB Coordinates Network}\ }\textbf {\bibinfo {volume} {16660}},\ \bibinfo {pages} {1} (\bibinfo {year} {2014}{\natexlab{b}})}\BibitemShut {NoStop}%
\bibitem [{\citenamefont {{Roberts}}(2014)}]{2014GCN.16971....1R}%
  \BibitemOpen
  \bibfield  {author} {\bibinfo {author} {\bibfnamefont {O.~J.}\ \bibnamefont {{Roberts}}},\ }\href@noop {} {\bibfield  {journal} {\bibinfo  {journal} {GRB Coordinates Network}\ }\textbf {\bibinfo {volume} {16971}},\ \bibinfo {pages} {1} (\bibinfo {year} {2014})}\BibitemShut {NoStop}%
\bibitem [{\citenamefont {{Xu}}\ \emph {et~al.}(2014)\citenamefont {{Xu}}, \citenamefont {{Levan}}, \citenamefont {{Fynbo}}, \citenamefont {{Tanvir}}, \citenamefont {{D'Elia}},\ and\ \citenamefont {{Malesani}}}]{2014GCN.16983....1X}%
  \BibitemOpen
  \bibfield  {author} {\bibinfo {author} {\bibfnamefont {D.}~\bibnamefont {{Xu}}}, \bibinfo {author} {\bibfnamefont {A.~J.}\ \bibnamefont {{Levan}}}, \bibinfo {author} {\bibfnamefont {J.~P.~U.}\ \bibnamefont {{Fynbo}}}, \bibinfo {author} {\bibfnamefont {N.~R.}\ \bibnamefont {{Tanvir}}}, \bibinfo {author} {\bibfnamefont {V.}~\bibnamefont {{D'Elia}}}, \ and\ \bibinfo {author} {\bibfnamefont {D.}~\bibnamefont {{Malesani}}},\ }\href@noop {} {\bibfield  {journal} {\bibinfo  {journal} {GRB Coordinates Network}\ }\textbf {\bibinfo {volume} {16983}},\ \bibinfo {pages} {1} (\bibinfo {year} {2014})}\BibitemShut {NoStop}%
\bibitem [{\citenamefont {{Jenke}}(2014)}]{2014GCN.17241....1J}%
  \BibitemOpen
  \bibfield  {author} {\bibinfo {author} {\bibfnamefont {P.}~\bibnamefont {{Jenke}}},\ }\href@noop {} {\bibfield  {journal} {\bibinfo  {journal} {GRB Coordinates Network}\ }\textbf {\bibinfo {volume} {17241}},\ \bibinfo {pages} {1} (\bibinfo {year} {2014})}\BibitemShut {NoStop}%
\bibitem [{\citenamefont {{Gorosabel}}\ \emph {et~al.}(2014)\citenamefont {{Gorosabel}}, \citenamefont {{de Ugarte Postigo}}, \citenamefont {{Thoene}}, \citenamefont {{Tanvir}}, \citenamefont {{Fynbo}}, \citenamefont {{Garcia-Alvarez}},\ and\ \citenamefont {{Perez-Romero}}}]{2014GCN.17234....1G}%
  \BibitemOpen
  \bibfield  {author} {\bibinfo {author} {\bibfnamefont {J.}~\bibnamefont {{Gorosabel}}}, \bibinfo {author} {\bibfnamefont {A.}~\bibnamefont {{de Ugarte Postigo}}}, \bibinfo {author} {\bibfnamefont {C.~C.}\ \bibnamefont {{Thoene}}}, \bibinfo {author} {\bibfnamefont {N.}~\bibnamefont {{Tanvir}}}, \bibinfo {author} {\bibfnamefont {J.~P.~U.}\ \bibnamefont {{Fynbo}}}, \bibinfo {author} {\bibfnamefont {D.}~\bibnamefont {{Garcia-Alvarez}}}, \ and\ \bibinfo {author} {\bibfnamefont {A.}~\bibnamefont {{Perez-Romero}}},\ }\href@noop {} {\bibfield  {journal} {\bibinfo  {journal} {GRB Coordinates Network}\ }\textbf {\bibinfo {volume} {17234}},\ \bibinfo {pages} {1} (\bibinfo {year} {2014})}\BibitemShut {NoStop}%
\bibitem [{\citenamefont {{Golenetskii}}\ \emph {et~al.}(2015{\natexlab{a}})\citenamefont {{Golenetskii}}, \citenamefont {{Aptekar}}, \citenamefont {{Frederiks}}, \citenamefont {{Pal'Shin}}, \citenamefont {{Oleynik}}, \citenamefont {{Ulanov}}, \citenamefont {{Svinkin}}, \citenamefont {{Tsvetkova}}, \citenamefont {{Lysenko}},\ and\ \citenamefont {{Cline}}}]{2015GCN.17427....1G}%
  \BibitemOpen
  \bibfield  {author} {\bibinfo {author} {\bibfnamefont {S.}~\bibnamefont {{Golenetskii}}}, \bibinfo {author} {\bibfnamefont {R.}~\bibnamefont {{Aptekar}}}, \bibinfo {author} {\bibfnamefont {D.}~\bibnamefont {{Frederiks}}}, \bibinfo {author} {\bibfnamefont {V.}~\bibnamefont {{Pal'Shin}}}, \bibinfo {author} {\bibfnamefont {P.}~\bibnamefont {{Oleynik}}}, \bibinfo {author} {\bibfnamefont {M.}~\bibnamefont {{Ulanov}}}, \bibinfo {author} {\bibfnamefont {D.}~\bibnamefont {{Svinkin}}}, \bibinfo {author} {\bibfnamefont {A.}~\bibnamefont {{Tsvetkova}}}, \bibinfo {author} {\bibfnamefont {A.}~\bibnamefont {{Lysenko}}}, \ and\ \bibinfo {author} {\bibfnamefont {T.}~\bibnamefont {{Cline}}},\ }\href@noop {} {\bibfield  {journal} {\bibinfo  {journal} {GRB Coordinates Network}\ }\textbf {\bibinfo {volume} {17427}},\ \bibinfo {pages} {1} (\bibinfo {year} {2015}{\natexlab{a}})}\BibitemShut {NoStop}%
\bibitem [{\citenamefont {{Golenetskii}}\ \emph {et~al.}(2015{\natexlab{b}})\citenamefont {{Golenetskii}}, \citenamefont {{Aptekar}}, \citenamefont {{Frederiks}}, \citenamefont {{Pal'Shin}}, \citenamefont {{Oleynik}}, \citenamefont {{Ulanov}}, \citenamefont {{Svinkin}}, \citenamefont {{Tsvetkova}}, \citenamefont {{Lysenko}},\ and\ \citenamefont {{Cline}}}]{2015GCN.17587....1G}%
  \BibitemOpen
  \bibfield  {author} {\bibinfo {author} {\bibfnamefont {S.}~\bibnamefont {{Golenetskii}}}, \bibinfo {author} {\bibfnamefont {R.}~\bibnamefont {{Aptekar}}}, \bibinfo {author} {\bibfnamefont {D.}~\bibnamefont {{Frederiks}}}, \bibinfo {author} {\bibfnamefont {V.}~\bibnamefont {{Pal'Shin}}}, \bibinfo {author} {\bibfnamefont {P.}~\bibnamefont {{Oleynik}}}, \bibinfo {author} {\bibfnamefont {M.}~\bibnamefont {{Ulanov}}}, \bibinfo {author} {\bibfnamefont {D.}~\bibnamefont {{Svinkin}}}, \bibinfo {author} {\bibfnamefont {A.}~\bibnamefont {{Tsvetkova}}}, \bibinfo {author} {\bibfnamefont {A.}~\bibnamefont {{Lysenko}}}, \ and\ \bibinfo {author} {\bibfnamefont {T.}~\bibnamefont {{Cline}}},\ }\href@noop {} {\bibfield  {journal} {\bibinfo  {journal} {GRB Coordinates Network}\ }\textbf {\bibinfo {volume} {17587}},\ \bibinfo {pages} {1} (\bibinfo {year} {2015}{\natexlab{b}})}\BibitemShut {NoStop}%
\bibitem [{\citenamefont {{Yu}}\ and\ \citenamefont {{Veres}}(2016)}]{2016GCN.19443....1Y}%
  \BibitemOpen
  \bibfield  {author} {\bibinfo {author} {\bibfnamefont {H.~F.}\ \bibnamefont {{Yu}}}\ and\ \bibinfo {author} {\bibfnamefont {P.}~\bibnamefont {{Veres}}},\ }\href@noop {} {\bibfield  {journal} {\bibinfo  {journal} {GRB Coordinates Network}\ }\textbf {\bibinfo {volume} {19443}},\ \bibinfo {pages} {1} (\bibinfo {year} {2016})}\BibitemShut {NoStop}%
\bibitem [{\citenamefont {{Ruffini}}\ \emph {et~al.}(2016)\citenamefont {{Ruffini}}, \citenamefont {{Aimuratov}}, \citenamefont {{Becerra}}, \citenamefont {{Bianco}}, \citenamefont {{Kovacevic}}, \citenamefont {{Moradi}}, \citenamefont {{Muccino}}, \citenamefont {{Penacchioni}}, \citenamefont {{Pisani}}, \citenamefont {{Primorac}}, \citenamefont {{Rueda}},\ and\ \citenamefont {{Wang}}}]{2016GCN.19456....1R}%
  \BibitemOpen
  \bibfield  {author} {\bibinfo {author} {\bibfnamefont {R.}~\bibnamefont {{Ruffini}}}, \bibinfo {author} {\bibfnamefont {Y.}~\bibnamefont {{Aimuratov}}}, \bibinfo {author} {\bibfnamefont {L.}~\bibnamefont {{Becerra}}}, \bibinfo {author} {\bibfnamefont {C.~L.}\ \bibnamefont {{Bianco}}}, \bibinfo {author} {\bibfnamefont {M.}~\bibnamefont {{Kovacevic}}}, \bibinfo {author} {\bibfnamefont {R.}~\bibnamefont {{Moradi}}}, \bibinfo {author} {\bibfnamefont {M.}~\bibnamefont {{Muccino}}}, \bibinfo {author} {\bibfnamefont {A.~V.}\ \bibnamefont {{Penacchioni}}}, \bibinfo {author} {\bibfnamefont {G.~B.}\ \bibnamefont {{Pisani}}}, \bibinfo {author} {\bibfnamefont {D.}~\bibnamefont {{Primorac}}}, \bibinfo {author} {\bibfnamefont {J.}~\bibnamefont {{Rueda}}}, \ and\ \bibinfo {author} {\bibfnamefont {Y.}~\bibnamefont {{Wang}}},\ }\href@noop {} {\bibfield  {journal} {\bibinfo  {journal} {GRB Coordinates Network}\ }\textbf {\bibinfo {volume} {19456}},\ \bibinfo {pages} {1} (\bibinfo {year} {2016})}\BibitemShut {NoStop}%
\bibitem [{\citenamefont {{Tsvetkova}}\ \emph {et~al.}(2018)\citenamefont {{Tsvetkova}}, \citenamefont {{Golenetskii}}, \citenamefont {{Aptekar}}, \citenamefont {{Frederiks}}, \citenamefont {{Oleynik}}, \citenamefont {{Ulanov}}, \citenamefont {{Svinkin}}, \citenamefont {{Lysenko}}, \citenamefont {{Kozlova}},\ and\ \citenamefont {{Cline}}}]{2018GCN.22513....1T}%
  \BibitemOpen
  \bibfield  {author} {\bibinfo {author} {\bibfnamefont {A.}~\bibnamefont {{Tsvetkova}}}, \bibinfo {author} {\bibfnamefont {S.}~\bibnamefont {{Golenetskii}}}, \bibinfo {author} {\bibfnamefont {R.}~\bibnamefont {{Aptekar}}}, \bibinfo {author} {\bibfnamefont {D.}~\bibnamefont {{Frederiks}}}, \bibinfo {author} {\bibfnamefont {P.}~\bibnamefont {{Oleynik}}}, \bibinfo {author} {\bibfnamefont {M.}~\bibnamefont {{Ulanov}}}, \bibinfo {author} {\bibfnamefont {D.}~\bibnamefont {{Svinkin}}}, \bibinfo {author} {\bibfnamefont {A.}~\bibnamefont {{Lysenko}}}, \bibinfo {author} {\bibfnamefont {A.}~\bibnamefont {{Kozlova}}}, \ and\ \bibinfo {author} {\bibfnamefont {T.}~\bibnamefont {{Cline}}},\ }\href@noop {} {\bibfield  {journal} {\bibinfo  {journal} {GRB Coordinates Network}\ }\textbf {\bibinfo {volume} {22513}},\ \bibinfo {pages} {1} (\bibinfo {year} {2018})}\BibitemShut {NoStop}%
\bibitem [{\citenamefont {{Svinkin}}\ \emph {et~al.}(2018)\citenamefont {{Svinkin}}, \citenamefont {{Golenetskii}}, \citenamefont {{Aptekar}}, \citenamefont {{Frederiks}}, \citenamefont {{Ulanov}}, \citenamefont {{Tsvetkova}}, \citenamefont {{Lysenko}}, \citenamefont {{Kozlova}}, \citenamefont {{Cline}},\ and\ \citenamefont {{Konus-Wind Team}}}]{2018GCN.22825....1S}%
  \BibitemOpen
  \bibfield  {author} {\bibinfo {author} {\bibfnamefont {D.}~\bibnamefont {{Svinkin}}}, \bibinfo {author} {\bibfnamefont {S.}~\bibnamefont {{Golenetskii}}}, \bibinfo {author} {\bibfnamefont {R.}~\bibnamefont {{Aptekar}}}, \bibinfo {author} {\bibfnamefont {D.}~\bibnamefont {{Frederiks}}}, \bibinfo {author} {\bibfnamefont {M.}~\bibnamefont {{Ulanov}}}, \bibinfo {author} {\bibfnamefont {A.}~\bibnamefont {{Tsvetkova}}}, \bibinfo {author} {\bibfnamefont {A.}~\bibnamefont {{Lysenko}}}, \bibinfo {author} {\bibfnamefont {A.}~\bibnamefont {{Kozlova}}}, \bibinfo {author} {\bibfnamefont {T.}~\bibnamefont {{Cline}}}, \ and\ \bibinfo {author} {\bibnamefont {{Konus-Wind Team}}},\ }\href@noop {} {\bibfield  {journal} {\bibinfo  {journal} {GRB Coordinates Network}\ }\textbf {\bibinfo {volume} {22825}},\ \bibinfo {pages} {1} (\bibinfo {year} {2018})}\BibitemShut {NoStop}%
\bibitem [{\citenamefont {{Frederiks}}\ \emph {et~al.}(2018)\citenamefont {{Frederiks}}, \citenamefont {{Golenetskii}}, \citenamefont {{Aptekar}}, \citenamefont {{Kozlova}}, \citenamefont {{Lysenko}}, \citenamefont {{Svinkin}}, \citenamefont {{Tsvetkova}}, \citenamefont {{Ulanov}},\ and\ \citenamefont {{Cline}}}]{2018GCN.23240....1F}%
  \BibitemOpen
  \bibfield  {author} {\bibinfo {author} {\bibfnamefont {D.}~\bibnamefont {{Frederiks}}}, \bibinfo {author} {\bibfnamefont {S.}~\bibnamefont {{Golenetskii}}}, \bibinfo {author} {\bibfnamefont {R.}~\bibnamefont {{Aptekar}}}, \bibinfo {author} {\bibfnamefont {A.}~\bibnamefont {{Kozlova}}}, \bibinfo {author} {\bibfnamefont {A.}~\bibnamefont {{Lysenko}}}, \bibinfo {author} {\bibfnamefont {D.}~\bibnamefont {{Svinkin}}}, \bibinfo {author} {\bibfnamefont {A.}~\bibnamefont {{Tsvetkova}}}, \bibinfo {author} {\bibfnamefont {M.}~\bibnamefont {{Ulanov}}}, \ and\ \bibinfo {author} {\bibfnamefont {T.}~\bibnamefont {{Cline}}},\ }\href@noop {} {\bibfield  {journal} {\bibinfo  {journal} {GRB Coordinates Network}\ }\textbf {\bibinfo {volume} {23240}},\ \bibinfo {pages} {1} (\bibinfo {year} {2018})}\BibitemShut {NoStop}%
\bibitem [{\citenamefont {{D'Avanzo}}\ \emph {et~al.}(2018)\citenamefont {{D'Avanzo}}, \citenamefont {{Heintz}}, \citenamefont {{de Ugarte Postigo}}, \citenamefont {{Levan}}, \citenamefont {{Izzo}}, \citenamefont {{Malesani}}, \citenamefont {{Kann}},\ and\ \citenamefont {{Tanvir}}}]{2018GCN.23246....1D}%
  \BibitemOpen
  \bibfield  {author} {\bibinfo {author} {\bibfnamefont {P.}~\bibnamefont {{D'Avanzo}}}, \bibinfo {author} {\bibfnamefont {K.~E.}\ \bibnamefont {{Heintz}}}, \bibinfo {author} {\bibfnamefont {A.}~\bibnamefont {{de Ugarte Postigo}}}, \bibinfo {author} {\bibfnamefont {A.~J.}\ \bibnamefont {{Levan}}}, \bibinfo {author} {\bibfnamefont {L.}~\bibnamefont {{Izzo}}}, \bibinfo {author} {\bibfnamefont {D.~B.}\ \bibnamefont {{Malesani}}}, \bibinfo {author} {\bibfnamefont {D.~A.}\ \bibnamefont {{Kann}}}, \ and\ \bibinfo {author} {\bibfnamefont {N.~R.}\ \bibnamefont {{Tanvir}}},\ }\href@noop {} {\bibfield  {journal} {\bibinfo  {journal} {GRB Coordinates Network}\ }\textbf {\bibinfo {volume} {23246}},\ \bibinfo {pages} {1} (\bibinfo {year} {2018})}\BibitemShut {NoStop}%
\bibitem [{\citenamefont {{Frederiks}}\ \emph {et~al.}(2019)\citenamefont {{Frederiks}}, \citenamefont {{Golenetskii}}, \citenamefont {{Aptekar}}, \citenamefont {{Kozlova}}, \citenamefont {{Lysenko}}, \citenamefont {{Svinkin}}, \citenamefont {{Tsvetkova}}, \citenamefont {{Ulanov}},\ and\ \citenamefont {{Cline}}}]{2019GCN.23737....1F}%
  \BibitemOpen
  \bibfield  {author} {\bibinfo {author} {\bibfnamefont {D.}~\bibnamefont {{Frederiks}}}, \bibinfo {author} {\bibfnamefont {S.}~\bibnamefont {{Golenetskii}}}, \bibinfo {author} {\bibfnamefont {R.}~\bibnamefont {{Aptekar}}}, \bibinfo {author} {\bibfnamefont {A.}~\bibnamefont {{Kozlova}}}, \bibinfo {author} {\bibfnamefont {A.}~\bibnamefont {{Lysenko}}}, \bibinfo {author} {\bibfnamefont {D.}~\bibnamefont {{Svinkin}}}, \bibinfo {author} {\bibfnamefont {A.}~\bibnamefont {{Tsvetkova}}}, \bibinfo {author} {\bibfnamefont {M.}~\bibnamefont {{Ulanov}}}, \ and\ \bibinfo {author} {\bibfnamefont {T.}~\bibnamefont {{Cline}}},\ }\href@noop {} {\bibfield  {journal} {\bibinfo  {journal} {GRB Coordinates Network}\ }\textbf {\bibinfo {volume} {23737}},\ \bibinfo {pages} {1} (\bibinfo {year} {2019})}\BibitemShut {NoStop}%
\bibitem [{\citenamefont {{Svinkin}}\ \emph {et~al.}(2019)\citenamefont {{Svinkin}}, \citenamefont {{Golenetskii}}, \citenamefont {{Aptekar}}, \citenamefont {{Frederiks}}, \citenamefont {{Ulanov}}, \citenamefont {{Tsvetkova}}, \citenamefont {{Lysenko}}, \citenamefont {{Kozlova}}, \citenamefont {{Cline}},\ and\ \citenamefont {{Konus-Wind Team}}}]{2019GCN.25974....1S}%
  \BibitemOpen
  \bibfield  {author} {\bibinfo {author} {\bibfnamefont {D.}~\bibnamefont {{Svinkin}}}, \bibinfo {author} {\bibfnamefont {S.}~\bibnamefont {{Golenetskii}}}, \bibinfo {author} {\bibfnamefont {R.}~\bibnamefont {{Aptekar}}}, \bibinfo {author} {\bibfnamefont {D.}~\bibnamefont {{Frederiks}}}, \bibinfo {author} {\bibfnamefont {M.}~\bibnamefont {{Ulanov}}}, \bibinfo {author} {\bibfnamefont {A.}~\bibnamefont {{Tsvetkova}}}, \bibinfo {author} {\bibfnamefont {A.}~\bibnamefont {{Lysenko}}}, \bibinfo {author} {\bibfnamefont {A.}~\bibnamefont {{Kozlova}}}, \bibinfo {author} {\bibfnamefont {T.}~\bibnamefont {{Cline}}}, \ and\ \bibinfo {author} {\bibnamefont {{Konus-Wind Team}}},\ }\href@noop {} {\bibfield  {journal} {\bibinfo  {journal} {GRB Coordinates Network}\ }\textbf {\bibinfo {volume} {25974}},\ \bibinfo {pages} {1} (\bibinfo {year} {2019})}\BibitemShut {NoStop}%
\bibitem [{\citenamefont {{Mangan}}\ \emph {et~al.}(2020)\citenamefont {{Mangan}}, \citenamefont {{Dunwoody}}, \citenamefont {{Meegan}},\ and\ \citenamefont {{Fermi GBM Team}}}]{2020GCN.28287....1M}%
  \BibitemOpen
  \bibfield  {author} {\bibinfo {author} {\bibfnamefont {J.}~\bibnamefont {{Mangan}}}, \bibinfo {author} {\bibfnamefont {R.}~\bibnamefont {{Dunwoody}}}, \bibinfo {author} {\bibfnamefont {C.}~\bibnamefont {{Meegan}}}, \ and\ \bibinfo {author} {\bibnamefont {{Fermi GBM Team}}},\ }\href@noop {} {\bibfield  {journal} {\bibinfo  {journal} {GRB Coordinates Network}\ }\textbf {\bibinfo {volume} {28287}},\ \bibinfo {pages} {1} (\bibinfo {year} {2020})}\BibitemShut {NoStop}%
\bibitem [{\citenamefont {{Svinkin}}\ \emph {et~al.}(2020{\natexlab{a}})\citenamefont {{Svinkin}}, \citenamefont {{Frederiks}}, \citenamefont {{Ridnaia}}, \citenamefont {{Tsvetkova}},\ and\ \citenamefont {{Konus-Wind Team}}}]{2020GCN.28301....1S}%
  \BibitemOpen
  \bibfield  {author} {\bibinfo {author} {\bibfnamefont {D.}~\bibnamefont {{Svinkin}}}, \bibinfo {author} {\bibfnamefont {D.}~\bibnamefont {{Frederiks}}}, \bibinfo {author} {\bibfnamefont {A.}~\bibnamefont {{Ridnaia}}}, \bibinfo {author} {\bibfnamefont {A.}~\bibnamefont {{Tsvetkova}}}, \ and\ \bibinfo {author} {\bibnamefont {{Konus-Wind Team}}},\ }\href@noop {} {\bibfield  {journal} {\bibinfo  {journal} {GRB Coordinates Network}\ }\textbf {\bibinfo {volume} {28301}},\ \bibinfo {pages} {1} (\bibinfo {year} {2020}{\natexlab{a}})}\BibitemShut {NoStop}%
\bibitem [{\citenamefont {{Malacaria}}\ \emph {et~al.}(2020)\citenamefont {{Malacaria}}, \citenamefont {{Meegan}},\ and\ \citenamefont {{Fermi GBM Team}}}]{2020GCN.28710....1M}%
  \BibitemOpen
  \bibfield  {author} {\bibinfo {author} {\bibfnamefont {C.}~\bibnamefont {{Malacaria}}}, \bibinfo {author} {\bibfnamefont {C.}~\bibnamefont {{Meegan}}}, \ and\ \bibinfo {author} {\bibnamefont {{Fermi GBM Team}}},\ }\href@noop {} {\bibfield  {journal} {\bibinfo  {journal} {GRB Coordinates Network}\ }\textbf {\bibinfo {volume} {28710}},\ \bibinfo {pages} {1} (\bibinfo {year} {2020})}\BibitemShut {NoStop}%
\bibitem [{\citenamefont {{Kann}}\ \emph {et~al.}(2020)\citenamefont {{Kann}}, \citenamefont {{de Ugarte Postigo}}, \citenamefont {{Blazek}}, \citenamefont {{Agui Fernandez}}, \citenamefont {{Thoene}}, \citenamefont {{Geier}},\ and\ \citenamefont {{Rivero}}}]{2020GCN.28765....1K}%
  \BibitemOpen
  \bibfield  {author} {\bibinfo {author} {\bibfnamefont {D.~A.}\ \bibnamefont {{Kann}}}, \bibinfo {author} {\bibfnamefont {A.}~\bibnamefont {{de Ugarte Postigo}}}, \bibinfo {author} {\bibfnamefont {M.}~\bibnamefont {{Blazek}}}, \bibinfo {author} {\bibfnamefont {J.~F.}\ \bibnamefont {{Agui Fernandez}}}, \bibinfo {author} {\bibfnamefont {C.~C.}\ \bibnamefont {{Thoene}}}, \bibinfo {author} {\bibfnamefont {S.}~\bibnamefont {{Geier}}}, \ and\ \bibinfo {author} {\bibfnamefont {M.}~\bibnamefont {{Rivero}}},\ }\href@noop {} {\bibfield  {journal} {\bibinfo  {journal} {GRB Coordinates Network}\ }\textbf {\bibinfo {volume} {28765}},\ \bibinfo {pages} {1} (\bibinfo {year} {2020})}\BibitemShut {NoStop}%
\bibitem [{\citenamefont {{Kann}}\ \emph {et~al.}(2021)\citenamefont {{Kann}}, \citenamefont {{Izzo}}, \citenamefont {{Levan}}, \citenamefont {{Malesani}}, \citenamefont {{de Wet}},\ and\ \citenamefont {{Stargate Collaboration}}}]{2021GCN.30583....1K}%
  \BibitemOpen
  \bibfield  {author} {\bibinfo {author} {\bibfnamefont {D.~A.}\ \bibnamefont {{Kann}}}, \bibinfo {author} {\bibfnamefont {L.}~\bibnamefont {{Izzo}}}, \bibinfo {author} {\bibfnamefont {A.~J.}\ \bibnamefont {{Levan}}}, \bibinfo {author} {\bibfnamefont {D.~B.}\ \bibnamefont {{Malesani}}}, \bibinfo {author} {\bibfnamefont {S.}~\bibnamefont {{de Wet}}}, \ and\ \bibinfo {author} {\bibnamefont {{Stargate Collaboration}}},\ }\href@noop {} {\bibfield  {journal} {\bibinfo  {journal} {GRB Coordinates Network}\ }\textbf {\bibinfo {volume} {30583}},\ \bibinfo {pages} {1} (\bibinfo {year} {2021})}\BibitemShut {NoStop}%
\bibitem [{\citenamefont {{Lesage}}\ \emph {et~al.}(2021)\citenamefont {{Lesage}}, \citenamefont {{Meegan}},\ and\ \citenamefont {{Fermi Gamma-ray Burst Monitor Team}}}]{2021GCN.30573....1L}%
  \BibitemOpen
  \bibfield  {author} {\bibinfo {author} {\bibfnamefont {S.}~\bibnamefont {{Lesage}}}, \bibinfo {author} {\bibfnamefont {C.}~\bibnamefont {{Meegan}}}, \ and\ \bibinfo {author} {\bibnamefont {{Fermi Gamma-ray Burst Monitor Team}}},\ }\href@noop {} {\bibfield  {journal} {\bibinfo  {journal} {GRB Coordinates Network}\ }\textbf {\bibinfo {volume} {30573}},\ \bibinfo {pages} {1} (\bibinfo {year} {2021})}\BibitemShut {NoStop}%
\bibitem [{\citenamefont {{Frederiks}}\ \emph {et~al.}(2021)\citenamefont {{Frederiks}}, \citenamefont {{Golenetskii}}, \citenamefont {{Lysenko}}, \citenamefont {{Ridnaia}}, \citenamefont {{Svinkin}}, \citenamefont {{Tsvetkova}}, \citenamefont {{Ulanov}}, \citenamefont {{Cline}},\ and\ \citenamefont {{Konus-Wind Team}}}]{2021GCN.30694....1F}%
  \BibitemOpen
  \bibfield  {author} {\bibinfo {author} {\bibfnamefont {D.}~\bibnamefont {{Frederiks}}}, \bibinfo {author} {\bibfnamefont {S.}~\bibnamefont {{Golenetskii}}}, \bibinfo {author} {\bibfnamefont {A.}~\bibnamefont {{Lysenko}}}, \bibinfo {author} {\bibfnamefont {A.}~\bibnamefont {{Ridnaia}}}, \bibinfo {author} {\bibfnamefont {D.}~\bibnamefont {{Svinkin}}}, \bibinfo {author} {\bibfnamefont {A.}~\bibnamefont {{Tsvetkova}}}, \bibinfo {author} {\bibfnamefont {M.}~\bibnamefont {{Ulanov}}}, \bibinfo {author} {\bibfnamefont {T.}~\bibnamefont {{Cline}}}, \ and\ \bibinfo {author} {\bibnamefont {{Konus-Wind Team}}},\ }\href@noop {} {\bibfield  {journal} {\bibinfo  {journal} {GRB Coordinates Network}\ }\textbf {\bibinfo {volume} {30694}},\ \bibinfo {pages} {1} (\bibinfo {year} {2021})}\BibitemShut {NoStop}%
\bibitem [{\citenamefont {{Tsvetkova}}\ \emph {et~al.}(2022)\citenamefont {{Tsvetkova}}, \citenamefont {{Frederiks}}, \citenamefont {{Lysenko}}, \citenamefont {{Ridnaia}}, \citenamefont {{Svinkin}}, \citenamefont {{Ulanov}}, \citenamefont {{Cline}},\ and\ \citenamefont {{Konus-Wind Team}}}]{2022GCN.31433....1T}%
  \BibitemOpen
  \bibfield  {author} {\bibinfo {author} {\bibfnamefont {A.}~\bibnamefont {{Tsvetkova}}}, \bibinfo {author} {\bibfnamefont {D.}~\bibnamefont {{Frederiks}}}, \bibinfo {author} {\bibfnamefont {A.}~\bibnamefont {{Lysenko}}}, \bibinfo {author} {\bibfnamefont {A.}~\bibnamefont {{Ridnaia}}}, \bibinfo {author} {\bibfnamefont {D.}~\bibnamefont {{Svinkin}}}, \bibinfo {author} {\bibfnamefont {M.}~\bibnamefont {{Ulanov}}}, \bibinfo {author} {\bibfnamefont {T.}~\bibnamefont {{Cline}}}, \ and\ \bibinfo {author} {\bibnamefont {{Konus-Wind Team}}},\ }\href@noop {} {\bibfield  {journal} {\bibinfo  {journal} {GRB Coordinates Network}\ }\textbf {\bibinfo {volume} {31433}},\ \bibinfo {pages} {1} (\bibinfo {year} {2022})}\BibitemShut {NoStop}%
\bibitem [{\citenamefont {{Lysenko}}\ \emph {et~al.}(2022)\citenamefont {{Lysenko}}, \citenamefont {{Frederiks}}, \citenamefont {{Ridnaia}}, \citenamefont {{Svinkin}}, \citenamefont {{Tsvetkova}}, \citenamefont {{Ulanov}}, \citenamefont {{Cline}},\ and\ \citenamefont {{Konus-Wind Team}}}]{2022GCN.32152....1L}%
  \BibitemOpen
  \bibfield  {author} {\bibinfo {author} {\bibfnamefont {A.}~\bibnamefont {{Lysenko}}}, \bibinfo {author} {\bibfnamefont {D.}~\bibnamefont {{Frederiks}}}, \bibinfo {author} {\bibfnamefont {A.}~\bibnamefont {{Ridnaia}}}, \bibinfo {author} {\bibfnamefont {D.}~\bibnamefont {{Svinkin}}}, \bibinfo {author} {\bibfnamefont {A.}~\bibnamefont {{Tsvetkova}}}, \bibinfo {author} {\bibfnamefont {M.}~\bibnamefont {{Ulanov}}}, \bibinfo {author} {\bibfnamefont {T.}~\bibnamefont {{Cline}}}, \ and\ \bibinfo {author} {\bibnamefont {{Konus-Wind Team}}},\ }\href@noop {} {\bibfield  {journal} {\bibinfo  {journal} {GRB Coordinates Network}\ }\textbf {\bibinfo {volume} {32152}},\ \bibinfo {pages} {1} (\bibinfo {year} {2022})}\BibitemShut {NoStop}%
\bibitem [{\citenamefont {{Lesage}}\ \emph {et~al.}(2022)\citenamefont {{Lesage}}, \citenamefont {{Meegan}},\ and\ \citenamefont {{Fermi Gamma-ray Burst Monitor Team}}}]{2022GCN.33112....1L}%
  \BibitemOpen
  \bibfield  {author} {\bibinfo {author} {\bibfnamefont {S.}~\bibnamefont {{Lesage}}}, \bibinfo {author} {\bibfnamefont {C.}~\bibnamefont {{Meegan}}}, \ and\ \bibinfo {author} {\bibnamefont {{Fermi Gamma-ray Burst Monitor Team}}},\ }\href@noop {} {\bibfield  {journal} {\bibinfo  {journal} {GRB Coordinates Network}\ }\textbf {\bibinfo {volume} {33112}},\ \bibinfo {pages} {1} (\bibinfo {year} {2022})}\BibitemShut {NoStop}%
\bibitem [{\citenamefont {{Xu}}\ \emph {et~al.}(2022)\citenamefont {{Xu}}, \citenamefont {{Chrimes}}, \citenamefont {{Schneider}}, \citenamefont {{Izzo}}, \citenamefont {{Saccardi}}, \citenamefont {{Zhu}}, \citenamefont {{Tanvir}},\ and\ \citenamefont {{Stargate Collaboration}}}]{2022GCN.33110....1X}%
  \BibitemOpen
  \bibfield  {author} {\bibinfo {author} {\bibfnamefont {D.}~\bibnamefont {{Xu}}}, \bibinfo {author} {\bibfnamefont {A.}~\bibnamefont {{Chrimes}}}, \bibinfo {author} {\bibfnamefont {B.}~\bibnamefont {{Schneider}}}, \bibinfo {author} {\bibfnamefont {L.}~\bibnamefont {{Izzo}}}, \bibinfo {author} {\bibfnamefont {A.}~\bibnamefont {{Saccardi}}}, \bibinfo {author} {\bibfnamefont {Z.~P.}\ \bibnamefont {{Zhu}}}, \bibinfo {author} {\bibfnamefont {N.~R.}\ \bibnamefont {{Tanvir}}}, \ and\ \bibinfo {author} {\bibnamefont {{Stargate Collaboration}}},\ }\href@noop {} {\bibfield  {journal} {\bibinfo  {journal} {GRB Coordinates Network}\ }\textbf {\bibinfo {volume} {33110}},\ \bibinfo {pages} {1} (\bibinfo {year} {2022})}\BibitemShut {NoStop}%
\bibitem [{\citenamefont {{de Ugarte Postigo}}\ \emph {et~al.}(2023)\citenamefont {{de Ugarte Postigo}}, \citenamefont {{Agui Fernandez}}, \citenamefont {{Thoene}},\ and\ \citenamefont {{Izzo}}}]{2023GCN.34409....1D}%
  \BibitemOpen
  \bibfield  {author} {\bibinfo {author} {\bibfnamefont {A.}~\bibnamefont {{de Ugarte Postigo}}}, \bibinfo {author} {\bibfnamefont {J.~F.}\ \bibnamefont {{Agui Fernandez}}}, \bibinfo {author} {\bibfnamefont {C.~C.}\ \bibnamefont {{Thoene}}}, \ and\ \bibinfo {author} {\bibfnamefont {L.}~\bibnamefont {{Izzo}}},\ }\href@noop {} {\bibfield  {journal} {\bibinfo  {journal} {GRB Coordinates Network}\ }\textbf {\bibinfo {volume} {34409}},\ \bibinfo {pages} {1} (\bibinfo {year} {2023})}\BibitemShut {NoStop}%
\bibitem [{\citenamefont {{Frederiks}}\ \emph {et~al.}(2023{\natexlab{a}})\citenamefont {{Frederiks}}, \citenamefont {{Lysenko}}, \citenamefont {{Ridnaia}}, \citenamefont {{Svinkin}}, \citenamefont {{Tsvetkova}}, \citenamefont {{Ulanov}}, \citenamefont {{Cline}},\ and\ \citenamefont {{Konus-Wind Team}}}]{2023GCN.34403....1F}%
  \BibitemOpen
  \bibfield  {author} {\bibinfo {author} {\bibfnamefont {D.}~\bibnamefont {{Frederiks}}}, \bibinfo {author} {\bibfnamefont {A.}~\bibnamefont {{Lysenko}}}, \bibinfo {author} {\bibfnamefont {A.}~\bibnamefont {{Ridnaia}}}, \bibinfo {author} {\bibfnamefont {D.}~\bibnamefont {{Svinkin}}}, \bibinfo {author} {\bibfnamefont {A.}~\bibnamefont {{Tsvetkova}}}, \bibinfo {author} {\bibfnamefont {M.}~\bibnamefont {{Ulanov}}}, \bibinfo {author} {\bibfnamefont {T.}~\bibnamefont {{Cline}}}, \ and\ \bibinfo {author} {\bibnamefont {{Konus-Wind Team}}},\ }\href@noop {} {\bibfield  {journal} {\bibinfo  {journal} {GRB Coordinates Network}\ }\textbf {\bibinfo {volume} {34403}},\ \bibinfo {pages} {1} (\bibinfo {year} {2023}{\natexlab{a}})}\BibitemShut {NoStop}%
\bibitem [{\citenamefont {{Frederiks}}\ \emph {et~al.}(2023{\natexlab{b}})\citenamefont {{Frederiks}}, \citenamefont {{Lysenko}}, \citenamefont {{Ridnaia}}, \citenamefont {{Svinkin}}, \citenamefont {{Tsvetkova}}, \citenamefont {{Ulanov}}, \citenamefont {{Cline}},\ and\ \citenamefont {{Konus-Wind Team}}}]{2023GCN.35359....1F}%
  \BibitemOpen
  \bibfield  {author} {\bibinfo {author} {\bibfnamefont {D.}~\bibnamefont {{Frederiks}}}, \bibinfo {author} {\bibfnamefont {A.}~\bibnamefont {{Lysenko}}}, \bibinfo {author} {\bibfnamefont {A.}~\bibnamefont {{Ridnaia}}}, \bibinfo {author} {\bibfnamefont {D.}~\bibnamefont {{Svinkin}}}, \bibinfo {author} {\bibfnamefont {A.}~\bibnamefont {{Tsvetkova}}}, \bibinfo {author} {\bibfnamefont {M.}~\bibnamefont {{Ulanov}}}, \bibinfo {author} {\bibfnamefont {T.}~\bibnamefont {{Cline}}}, \ and\ \bibinfo {author} {\bibnamefont {{Konus-Wind Team}}},\ }\href@noop {} {\bibfield  {journal} {\bibinfo  {journal} {GRB Coordinates Network}\ }\textbf {\bibinfo {volume} {35359}},\ \bibinfo {pages} {1} (\bibinfo {year} {2023}{\natexlab{b}})}\BibitemShut {NoStop}%
\bibitem [{\citenamefont {{Thoene}}\ \emph {et~al.}(2023)\citenamefont {{Thoene}}, \citenamefont {{de Ugarte Postigo}}, \citenamefont {{Agui Fernandez}}, \citenamefont {{Izzo}}, \citenamefont {{Fynbo}}, \citenamefont {{Tanvir}}, \citenamefont {{Blazek}}, \citenamefont {{Geier}}, \citenamefont {{Cabrera-Lavers}}, \citenamefont {{Perez Toledo}},\ and\ \citenamefont {{Rivero}}}]{2023GCN.35373....1T}%
  \BibitemOpen
  \bibfield  {author} {\bibinfo {author} {\bibfnamefont {C.~C.}\ \bibnamefont {{Thoene}}}, \bibinfo {author} {\bibfnamefont {A.}~\bibnamefont {{de Ugarte Postigo}}}, \bibinfo {author} {\bibfnamefont {J.~F.}\ \bibnamefont {{Agui Fernandez}}}, \bibinfo {author} {\bibfnamefont {L.}~\bibnamefont {{Izzo}}}, \bibinfo {author} {\bibfnamefont {J.~P.~U.}\ \bibnamefont {{Fynbo}}}, \bibinfo {author} {\bibfnamefont {N.~R.}\ \bibnamefont {{Tanvir}}}, \bibinfo {author} {\bibfnamefont {M.}~\bibnamefont {{Blazek}}}, \bibinfo {author} {\bibfnamefont {S.}~\bibnamefont {{Geier}}}, \bibinfo {author} {\bibfnamefont {A.}~\bibnamefont {{Cabrera-Lavers}}}, \bibinfo {author} {\bibfnamefont {F.}~\bibnamefont {{Perez Toledo}}}, \ and\ \bibinfo {author} {\bibfnamefont {M.}~\bibnamefont {{Rivero}}},\ }\href@noop {} {\bibfield  {journal} {\bibinfo  {journal} {GRB Coordinates Network}\ }\textbf {\bibinfo {volume} {35373}},\ \bibinfo {pages} {1} (\bibinfo {year} {2023})}\BibitemShut {NoStop}%
\bibitem [{\citenamefont {{Frederiks}}\ \emph {et~al.}(2023{\natexlab{c}})\citenamefont {{Frederiks}}, \citenamefont {{Lysenko}}, \citenamefont {{Ridnaia}}, \citenamefont {{Svinkin}}, \citenamefont {{Tsvetkova}}, \citenamefont {{Ulanov}}, \citenamefont {{Cline}},\ and\ \citenamefont {{Konus-Wind Team}}}]{2023GCN.35377....1F}%
  \BibitemOpen
  \bibfield  {author} {\bibinfo {author} {\bibfnamefont {D.}~\bibnamefont {{Frederiks}}}, \bibinfo {author} {\bibfnamefont {A.}~\bibnamefont {{Lysenko}}}, \bibinfo {author} {\bibfnamefont {A.}~\bibnamefont {{Ridnaia}}}, \bibinfo {author} {\bibfnamefont {D.}~\bibnamefont {{Svinkin}}}, \bibinfo {author} {\bibfnamefont {A.}~\bibnamefont {{Tsvetkova}}}, \bibinfo {author} {\bibfnamefont {M.}~\bibnamefont {{Ulanov}}}, \bibinfo {author} {\bibfnamefont {T.}~\bibnamefont {{Cline}}}, \ and\ \bibinfo {author} {\bibnamefont {{Konus-Wind Team}}},\ }\href@noop {} {\bibfield  {journal} {\bibinfo  {journal} {GRB Coordinates Network}\ }\textbf {\bibinfo {volume} {35377}},\ \bibinfo {pages} {1} (\bibinfo {year} {2023}{\natexlab{c}})}\BibitemShut {NoStop}%
\bibitem [{\citenamefont {{Frederiks}}\ \emph {et~al.}(2024{\natexlab{a}})\citenamefont {{Frederiks}}, \citenamefont {{Lysenko}}, \citenamefont {{Ridnaia}}, \citenamefont {{Svinkin}}, \citenamefont {{Tsvetkova}}, \citenamefont {{Ulanov}}, \citenamefont {{Cline}},\ and\ \citenamefont {{Konus-Wind Team}}}]{2024GCN.37302....1F}%
  \BibitemOpen
  \bibfield  {author} {\bibinfo {author} {\bibfnamefont {D.}~\bibnamefont {{Frederiks}}}, \bibinfo {author} {\bibfnamefont {A.}~\bibnamefont {{Lysenko}}}, \bibinfo {author} {\bibfnamefont {A.}~\bibnamefont {{Ridnaia}}}, \bibinfo {author} {\bibfnamefont {D.}~\bibnamefont {{Svinkin}}}, \bibinfo {author} {\bibfnamefont {A.}~\bibnamefont {{Tsvetkova}}}, \bibinfo {author} {\bibfnamefont {M.}~\bibnamefont {{Ulanov}}}, \bibinfo {author} {\bibfnamefont {T.}~\bibnamefont {{Cline}}}, \ and\ \bibinfo {author} {\bibnamefont {{Konus-Wind Team}}},\ }\href@noop {} {\bibfield  {journal} {\bibinfo  {journal} {GRB Coordinates Network}\ }\textbf {\bibinfo {volume} {37302}},\ \bibinfo {pages} {1} (\bibinfo {year} {2024}{\natexlab{a}})}\BibitemShut {NoStop}%
\bibitem [{\citenamefont {{von Kienlin}}(2010{\natexlab{b}})}]{2010GCN.10381....1V}%
  \BibitemOpen
  \bibfield  {author} {\bibinfo {author} {\bibfnamefont {A.}~\bibnamefont {{von Kienlin}}},\ }\href@noop {} {\bibfield  {journal} {\bibinfo  {journal} {GRB Coordinates Network}\ }\textbf {\bibinfo {volume} {10381}},\ \bibinfo {pages} {1} (\bibinfo {year} {2010}{\natexlab{b}})}\BibitemShut {NoStop}%
\bibitem [{\citenamefont {{Berger}}\ and\ \citenamefont {{Chornock}}(2010)}]{2010GCN.10410....1B}%
  \BibitemOpen
  \bibfield  {author} {\bibinfo {author} {\bibfnamefont {E.}~\bibnamefont {{Berger}}}\ and\ \bibinfo {author} {\bibfnamefont {R.}~\bibnamefont {{Chornock}}},\ }\href@noop {} {\bibfield  {journal} {\bibinfo  {journal} {GRB Coordinates Network}\ }\textbf {\bibinfo {volume} {10410}},\ \bibinfo {pages} {1} (\bibinfo {year} {2010})}\BibitemShut {NoStop}%
\bibitem [{\citenamefont {{Castro-Tirado}}\ \emph {et~al.}(2015)\citenamefont {{Castro-Tirado}}, \citenamefont {{Sanchez-Ramirez}}, \citenamefont {{Lombardi}},\ and\ \citenamefont {{Rivero}}}]{2015GCN.17758....1C}%
  \BibitemOpen
  \bibfield  {author} {\bibinfo {author} {\bibfnamefont {A.~J.}\ \bibnamefont {{Castro-Tirado}}}, \bibinfo {author} {\bibfnamefont {R.}~\bibnamefont {{Sanchez-Ramirez}}}, \bibinfo {author} {\bibfnamefont {G.}~\bibnamefont {{Lombardi}}}, \ and\ \bibinfo {author} {\bibfnamefont {M.~A.}\ \bibnamefont {{Rivero}}},\ }\href@noop {} {\bibfield  {journal} {\bibinfo  {journal} {GRB Coordinates Network}\ }\textbf {\bibinfo {volume} {17758}},\ \bibinfo {pages} {1} (\bibinfo {year} {2015})}\BibitemShut {NoStop}%
\bibitem [{\citenamefont {{Golenetskii}}\ \emph {et~al.}(2015{\natexlab{c}})\citenamefont {{Golenetskii}}, \citenamefont {{Aptekar}}, \citenamefont {{Frederiks}}, \citenamefont {{Pal'Shin}}, \citenamefont {{Oleynik}}, \citenamefont {{Ulanov}}, \citenamefont {{Svinkin}}, \citenamefont {{Tsvetkova}}, \citenamefont {{Lysenko}},\ and\ \citenamefont {{Cline}}}]{2015GCN.17752....1G}%
  \BibitemOpen
  \bibfield  {author} {\bibinfo {author} {\bibfnamefont {S.}~\bibnamefont {{Golenetskii}}}, \bibinfo {author} {\bibfnamefont {R.}~\bibnamefont {{Aptekar}}}, \bibinfo {author} {\bibfnamefont {D.}~\bibnamefont {{Frederiks}}}, \bibinfo {author} {\bibfnamefont {V.}~\bibnamefont {{Pal'Shin}}}, \bibinfo {author} {\bibfnamefont {P.}~\bibnamefont {{Oleynik}}}, \bibinfo {author} {\bibfnamefont {M.}~\bibnamefont {{Ulanov}}}, \bibinfo {author} {\bibfnamefont {D.}~\bibnamefont {{Svinkin}}}, \bibinfo {author} {\bibfnamefont {A.}~\bibnamefont {{Tsvetkova}}}, \bibinfo {author} {\bibfnamefont {A.}~\bibnamefont {{Lysenko}}}, \ and\ \bibinfo {author} {\bibfnamefont {T.}~\bibnamefont {{Cline}}},\ }\href@noop {} {\bibfield  {journal} {\bibinfo  {journal} {GRB Coordinates Network}\ }\textbf {\bibinfo {volume} {17752}},\ \bibinfo {pages} {1} (\bibinfo {year} {2015}{\natexlab{c}})}\BibitemShut {NoStop}%
\bibitem [{\citenamefont {{Ruffini}}\ \emph {et~al.}(2015)\citenamefont {{Ruffini}}, \citenamefont {{Bianco}}, \citenamefont {{Enderli}}, \citenamefont {{Kovacevic}}, \citenamefont {{Li}}, \citenamefont {{Muccino}}, \citenamefont {{Moradi}}, \citenamefont {{Pisani}}, \citenamefont {{Rueda}},\ and\ \citenamefont {{Wang}}}]{2015GCN.18296....1R}%
  \BibitemOpen
  \bibfield  {author} {\bibinfo {author} {\bibfnamefont {R.}~\bibnamefont {{Ruffini}}}, \bibinfo {author} {\bibfnamefont {C.~L.}\ \bibnamefont {{Bianco}}}, \bibinfo {author} {\bibfnamefont {M.}~\bibnamefont {{Enderli}}}, \bibinfo {author} {\bibfnamefont {M.}~\bibnamefont {{Kovacevic}}}, \bibinfo {author} {\bibfnamefont {L.}~\bibnamefont {{Li}}}, \bibinfo {author} {\bibfnamefont {M.}~\bibnamefont {{Muccino}}}, \bibinfo {author} {\bibfnamefont {R.}~\bibnamefont {{Moradi}}}, \bibinfo {author} {\bibfnamefont {G.~B.}\ \bibnamefont {{Pisani}}}, \bibinfo {author} {\bibfnamefont {J.~A.}\ \bibnamefont {{Rueda}}}, \ and\ \bibinfo {author} {\bibfnamefont {Y.}~\bibnamefont {{Wang}}},\ }\href@noop {} {\bibfield  {journal} {\bibinfo  {journal} {GRB Coordinates Network}\ }\textbf {\bibinfo {volume} {18296}},\ \bibinfo {pages} {1} (\bibinfo {year} {2015})}\BibitemShut {NoStop}%
\bibitem [{\citenamefont {{Svinkin}}\ \emph {et~al.}(2020{\natexlab{b}})\citenamefont {{Svinkin}}, \citenamefont {{Golenetskii}}, \citenamefont {{Aptekar}}, \citenamefont {{Frederiks}}, \citenamefont {{Ulanov}}, \citenamefont {{Tsvetkova}}, \citenamefont {{Lysenko}}, \citenamefont {{Ridnaia}}, \citenamefont {{Cline}},\ and\ \citenamefont {{Konus-Wind Team}}}]{2020GCN.29196....1S}%
  \BibitemOpen
  \bibfield  {author} {\bibinfo {author} {\bibfnamefont {D.}~\bibnamefont {{Svinkin}}}, \bibinfo {author} {\bibfnamefont {S.}~\bibnamefont {{Golenetskii}}}, \bibinfo {author} {\bibfnamefont {R.}~\bibnamefont {{Aptekar}}}, \bibinfo {author} {\bibfnamefont {D.}~\bibnamefont {{Frederiks}}}, \bibinfo {author} {\bibfnamefont {M.}~\bibnamefont {{Ulanov}}}, \bibinfo {author} {\bibfnamefont {A.}~\bibnamefont {{Tsvetkova}}}, \bibinfo {author} {\bibfnamefont {A.}~\bibnamefont {{Lysenko}}}, \bibinfo {author} {\bibfnamefont {A.}~\bibnamefont {{Ridnaia}}}, \bibinfo {author} {\bibfnamefont {T.}~\bibnamefont {{Cline}}}, \ and\ \bibinfo {author} {\bibnamefont {{Konus-Wind Team}}},\ }\href@noop {} {\bibfield  {journal} {\bibinfo  {journal} {GRB Coordinates Network}\ }\textbf {\bibinfo {volume} {29196}},\ \bibinfo {pages} {1} (\bibinfo {year} {2020}{\natexlab{b}})}\BibitemShut {NoStop}%
\bibitem [{\citenamefont {{Belkin}}\ \emph {et~al.}(2021)\citenamefont {{Belkin}}, \citenamefont {{Pozanenko}}, \citenamefont {{Minaev}}, \citenamefont {{Krugov}},\ and\ \citenamefont {{GRB IKI FuN Collaboration}}}]{2021GCN.29281....1B}%
  \BibitemOpen
  \bibfield  {author} {\bibinfo {author} {\bibfnamefont {S.}~\bibnamefont {{Belkin}}}, \bibinfo {author} {\bibfnamefont {A.}~\bibnamefont {{Pozanenko}}}, \bibinfo {author} {\bibfnamefont {P.}~\bibnamefont {{Minaev}}}, \bibinfo {author} {\bibfnamefont {M.}~\bibnamefont {{Krugov}}}, \ and\ \bibinfo {author} {\bibnamefont {{GRB IKI FuN Collaboration}}},\ }\href@noop {} {\bibfield  {journal} {\bibinfo  {journal} {GRB Coordinates Network}\ }\textbf {\bibinfo {volume} {29281}},\ \bibinfo {pages} {1} (\bibinfo {year} {2021})}\BibitemShut {NoStop}%
\bibitem [{\citenamefont {{Ridnaia}}\ \emph {et~al.}(2021)\citenamefont {{Ridnaia}}, \citenamefont {{Frederiks}}, \citenamefont {{Golenetskii}}, \citenamefont {{Lysenko}}, \citenamefont {{Svinkin}}, \citenamefont {{Tsvetkova}}, \citenamefont {{Ulanov}}, \citenamefont {{Cline}},\ and\ \citenamefont {{Konus-Wind Team}}}]{2021GCN.30388....1R}%
  \BibitemOpen
  \bibfield  {author} {\bibinfo {author} {\bibfnamefont {A.}~\bibnamefont {{Ridnaia}}}, \bibinfo {author} {\bibfnamefont {D.}~\bibnamefont {{Frederiks}}}, \bibinfo {author} {\bibfnamefont {S.}~\bibnamefont {{Golenetskii}}}, \bibinfo {author} {\bibfnamefont {A.}~\bibnamefont {{Lysenko}}}, \bibinfo {author} {\bibfnamefont {D.}~\bibnamefont {{Svinkin}}}, \bibinfo {author} {\bibfnamefont {A.}~\bibnamefont {{Tsvetkova}}}, \bibinfo {author} {\bibfnamefont {M.}~\bibnamefont {{Ulanov}}}, \bibinfo {author} {\bibfnamefont {T.}~\bibnamefont {{Cline}}}, \ and\ \bibinfo {author} {\bibnamefont {{Konus-Wind Team}}},\ }\href@noop {} {\bibfield  {journal} {\bibinfo  {journal} {GRB Coordinates Network}\ }\textbf {\bibinfo {volume} {30388}},\ \bibinfo {pages} {1} (\bibinfo {year} {2021})}\BibitemShut {NoStop}%
\bibitem [{\citenamefont {{Minaev}}\ \emph {et~al.}(2021)\citenamefont {{Minaev}}, \citenamefont {{Pozanenko}}, \citenamefont {{Chelovekov}}, \citenamefont {{Grebenev}},\ and\ \citenamefont {{GRB IKI FuN}}}]{2021GCN.30452....1M}%
  \BibitemOpen
  \bibfield  {author} {\bibinfo {author} {\bibfnamefont {P.}~\bibnamefont {{Minaev}}}, \bibinfo {author} {\bibfnamefont {A.}~\bibnamefont {{Pozanenko}}}, \bibinfo {author} {\bibfnamefont {I.}~\bibnamefont {{Chelovekov}}}, \bibinfo {author} {\bibfnamefont {S.}~\bibnamefont {{Grebenev}}}, \ and\ \bibinfo {author} {\bibnamefont {{GRB IKI FuN}}},\ }\href@noop {} {\bibfield  {journal} {\bibinfo  {journal} {GRB Coordinates Network}\ }\textbf {\bibinfo {volume} {30452}},\ \bibinfo {pages} {1} (\bibinfo {year} {2021})}\BibitemShut {NoStop}%
\bibitem [{\citenamefont {{Volnova}}\ \emph {et~al.}(2024)\citenamefont {{Volnova}}, \citenamefont {{Tatarnikov}}, \citenamefont {{Tatarnikov}}, \citenamefont {{Pankov}}, \citenamefont {{Pozanenko}}, \citenamefont {{Belkin}},\ and\ \citenamefont {{GRB-IKI-FuN}}}]{2024GCN.36777....1V}%
  \BibitemOpen
  \bibfield  {author} {\bibinfo {author} {\bibfnamefont {A.}~\bibnamefont {{Volnova}}}, \bibinfo {author} {\bibfnamefont {A.~A.}\ \bibnamefont {{Tatarnikov}}}, \bibinfo {author} {\bibfnamefont {A.~M.}\ \bibnamefont {{Tatarnikov}}}, \bibinfo {author} {\bibfnamefont {N.}~\bibnamefont {{Pankov}}}, \bibinfo {author} {\bibfnamefont {A.}~\bibnamefont {{Pozanenko}}}, \bibinfo {author} {\bibfnamefont {S.}~\bibnamefont {{Belkin}}}, \ and\ \bibinfo {author} {\bibnamefont {{GRB-IKI-FuN}}},\ }\href@noop {} {\bibfield  {journal} {\bibinfo  {journal} {GRB Coordinates Network}\ }\textbf {\bibinfo {volume} {36777}},\ \bibinfo {pages} {1} (\bibinfo {year} {2024})}\BibitemShut {NoStop}%
\bibitem [{\citenamefont {{Roberts}}\ \emph {et~al.}(2024)\citenamefont {{Roberts}}, \citenamefont {{Meegan}},\ and\ \citenamefont {{Fermi Gamma-ray Burst Monitor Team}}}]{2024GCN.36685....1R}%
  \BibitemOpen
  \bibfield  {author} {\bibinfo {author} {\bibfnamefont {O.~J.}\ \bibnamefont {{Roberts}}}, \bibinfo {author} {\bibfnamefont {C.}~\bibnamefont {{Meegan}}}, \ and\ \bibinfo {author} {\bibnamefont {{Fermi Gamma-ray Burst Monitor Team}}},\ }\href@noop {} {\bibfield  {journal} {\bibinfo  {journal} {GRB Coordinates Network}\ }\textbf {\bibinfo {volume} {36685}},\ \bibinfo {pages} {1} (\bibinfo {year} {2024})}\BibitemShut {NoStop}%
\bibitem [{\citenamefont {{Frederiks}}\ \emph {et~al.}(2024{\natexlab{b}})\citenamefont {{Frederiks}}, \citenamefont {{Lysenko}}, \citenamefont {{Ridnaia}}, \citenamefont {{Svinkin}}, \citenamefont {{Tsvetkova}}, \citenamefont {{Ulanov}}, \citenamefont {{Cline}},\ and\ \citenamefont {{Konus-Wind Team}}}]{2024GCN.38205....1F}%
  \BibitemOpen
  \bibfield  {author} {\bibinfo {author} {\bibfnamefont {D.}~\bibnamefont {{Frederiks}}}, \bibinfo {author} {\bibfnamefont {A.}~\bibnamefont {{Lysenko}}}, \bibinfo {author} {\bibfnamefont {A.}~\bibnamefont {{Ridnaia}}}, \bibinfo {author} {\bibfnamefont {D.}~\bibnamefont {{Svinkin}}}, \bibinfo {author} {\bibfnamefont {A.}~\bibnamefont {{Tsvetkova}}}, \bibinfo {author} {\bibfnamefont {M.}~\bibnamefont {{Ulanov}}}, \bibinfo {author} {\bibfnamefont {T.}~\bibnamefont {{Cline}}}, \ and\ \bibinfo {author} {\bibnamefont {{Konus-Wind Team}}},\ }\href@noop {} {\bibfield  {journal} {\bibinfo  {journal} {GRB Coordinates Network}\ }\textbf {\bibinfo {volume} {38205}},\ \bibinfo {pages} {1} (\bibinfo {year} {2024}{\natexlab{b}})}\BibitemShut {NoStop}%
\bibitem [{\citenamefont {{Mohan}}\ \emph {et~al.}(2024)\citenamefont {{Mohan}}, \citenamefont {{Waratkar}}, \citenamefont {{Saikia}}, \citenamefont {{Swain}}, \citenamefont {{Kumar}}, \citenamefont {{Bhalerao}}, \citenamefont {{Anupama}}, \citenamefont {{Barway}}, \citenamefont {{Angail}},\ and\ \citenamefont {{GIT Team}}}]{2024GCN.38187....1M}%
  \BibitemOpen
  \bibfield  {author} {\bibinfo {author} {\bibfnamefont {T.}~\bibnamefont {{Mohan}}}, \bibinfo {author} {\bibfnamefont {G.}~\bibnamefont {{Waratkar}}}, \bibinfo {author} {\bibfnamefont {A.~P.}\ \bibnamefont {{Saikia}}}, \bibinfo {author} {\bibfnamefont {V.}~\bibnamefont {{Swain}}}, \bibinfo {author} {\bibfnamefont {R.}~\bibnamefont {{Kumar}}}, \bibinfo {author} {\bibfnamefont {V.}~\bibnamefont {{Bhalerao}}}, \bibinfo {author} {\bibfnamefont {G.~C.}\ \bibnamefont {{Anupama}}}, \bibinfo {author} {\bibfnamefont {S.}~\bibnamefont {{Barway}}}, \bibinfo {author} {\bibfnamefont {K.}~\bibnamefont {{Angail}}}, \ and\ \bibinfo {author} {\bibnamefont {{GIT Team}}},\ }\href@noop {} {\bibfield  {journal} {\bibinfo  {journal} {GRB Coordinates Network}\ }\textbf {\bibinfo {volume} {38187}},\ \bibinfo {pages} {1} (\bibinfo {year} {2024})}\BibitemShut {NoStop}%
\bibitem [{\citenamefont {{Haensel}}\ \emph {et~al.}(1986)\citenamefont {{Haensel}}, \citenamefont {{Zdunik}},\ and\ \citenamefont {{Schaefer}}}]{1986A&A...160..121H}%
  \BibitemOpen
  \bibfield  {author} {\bibinfo {author} {\bibfnamefont {P.}~\bibnamefont {{Haensel}}}, \bibinfo {author} {\bibfnamefont {J.~L.}\ \bibnamefont {{Zdunik}}}, \ and\ \bibinfo {author} {\bibfnamefont {R.}~\bibnamefont {{Schaefer}}},\ }\href@noop {} {\bibfield  {journal} {\bibinfo  {journal} {\ASAS}\ }\textbf {\bibinfo {volume} {160}},\ \bibinfo {pages} {121} (\bibinfo {year} {1986})}\BibitemShut {NoStop}%
\bibitem [{\citenamefont {{Freedman}}\ and\ \citenamefont {{McLerran}}(1978)}]{1978PhRvD..17.1109F}%
  \BibitemOpen
  \bibfield  {author} {\bibinfo {author} {\bibfnamefont {B.}~\bibnamefont {{Freedman}}}\ and\ \bibinfo {author} {\bibfnamefont {L.}~\bibnamefont {{McLerran}}},\ }\href {\doibase 10.1103/PhysRevD.17.1109} {\bibfield  {journal} {\bibinfo  {journal} {\prd}\ }\textbf {\bibinfo {volume} {17}},\ \bibinfo {pages} {1109} (\bibinfo {year} {1978})}\BibitemShut {NoStop}%
\bibitem [{\citenamefont {{Oppenheimer}}\ and\ \citenamefont {{Volkoff}}(1939)}]{1939PhRv...55..374O}%
  \BibitemOpen
  \bibfield  {author} {\bibinfo {author} {\bibfnamefont {J.~R.}\ \bibnamefont {{Oppenheimer}}}\ and\ \bibinfo {author} {\bibfnamefont {G.~M.}\ \bibnamefont {{Volkoff}}},\ }\href {\doibase 10.1103/PhysRev.55.374} {\bibfield  {journal} {\bibinfo  {journal} {Physical Review}\ }\textbf {\bibinfo {volume} {55}},\ \bibinfo {pages} {374} (\bibinfo {year} {1939})}\BibitemShut {NoStop}%
\bibitem [{\citenamefont {{Alford}}\ and\ \citenamefont {{Rajagopal}}(2002)}]{2002JHEP...06..031A}%
  \BibitemOpen
  \bibfield  {author} {\bibinfo {author} {\bibfnamefont {M.}~\bibnamefont {{Alford}}}\ and\ \bibinfo {author} {\bibfnamefont {K.}~\bibnamefont {{Rajagopal}}},\ }\href {\doibase 10.1088/1126-6708/2002/06/031} {\bibfield  {journal} {\bibinfo  {journal} {Journal of High Energy Physics}\ }\textbf {\bibinfo {volume} {2002}},\ \bibinfo {eid} {031} (\bibinfo {year} {2002})},\ \Eprint {http://arxiv.org/abs/hep-ph/0204001} {arXiv:hep-ph/0204001 [hep-ph]} \BibitemShut {NoStop}%
\bibitem [{\citenamefont {{Rajagopal}}\ and\ \citenamefont {{Wilczek}}(2001)}]{2001PhRvL..86.3492R}%
  \BibitemOpen
  \bibfield  {author} {\bibinfo {author} {\bibfnamefont {K.}~\bibnamefont {{Rajagopal}}}\ and\ \bibinfo {author} {\bibfnamefont {F.}~\bibnamefont {{Wilczek}}},\ }\href {\doibase 10.1103/PhysRevLett.86.3492} {\bibfield  {journal} {\bibinfo  {journal} {\prl}\ }\textbf {\bibinfo {volume} {86}},\ \bibinfo {pages} {3492} (\bibinfo {year} {2001})},\ \Eprint {http://arxiv.org/abs/hep-ph/0012039} {arXiv:hep-ph/0012039 [hep-ph]} \BibitemShut {NoStop}%
\bibitem [{\citenamefont {{Alford}}\ \emph {et~al.}(2001)\citenamefont {{Alford}}, \citenamefont {{Rajagopal}}, \citenamefont {{Reddy}},\ and\ \citenamefont {{Wilczek}}}]{2001PhRvD..64g4017A}%
  \BibitemOpen
  \bibfield  {author} {\bibinfo {author} {\bibfnamefont {M.}~\bibnamefont {{Alford}}}, \bibinfo {author} {\bibfnamefont {K.}~\bibnamefont {{Rajagopal}}}, \bibinfo {author} {\bibfnamefont {S.}~\bibnamefont {{Reddy}}}, \ and\ \bibinfo {author} {\bibfnamefont {F.}~\bibnamefont {{Wilczek}}},\ }\href {\doibase 10.1103/PhysRevD.64.074017} {\bibfield  {journal} {\bibinfo  {journal} {\prd}\ }\textbf {\bibinfo {volume} {64}},\ \bibinfo {eid} {074017} (\bibinfo {year} {2001})},\ \Eprint {http://arxiv.org/abs/hep-ph/0105009} {arXiv:hep-ph/0105009 [hep-ph]} \BibitemShut {NoStop}%
\bibitem [{\citenamefont {{Alford}}\ \emph {et~al.}(2004)\citenamefont {{Alford}}, \citenamefont {{Kouvaris}},\ and\ \citenamefont {{Rajagopal}}}]{2004PhRvL..92v2001A}%
  \BibitemOpen
  \bibfield  {author} {\bibinfo {author} {\bibfnamefont {M.}~\bibnamefont {{Alford}}}, \bibinfo {author} {\bibfnamefont {C.}~\bibnamefont {{Kouvaris}}}, \ and\ \bibinfo {author} {\bibfnamefont {K.}~\bibnamefont {{Rajagopal}}},\ }\href {\doibase 10.1103/PhysRevLett.92.222001} {\bibfield  {journal} {\bibinfo  {journal} {\prl}\ }\textbf {\bibinfo {volume} {92}},\ \bibinfo {eid} {222001} (\bibinfo {year} {2004})},\ \Eprint {http://arxiv.org/abs/hep-ph/0311286} {arXiv:hep-ph/0311286 [hep-ph]} \BibitemShut {NoStop}%
\bibitem [{\citenamefont {{Alford}}\ \emph {et~al.}(2005)\citenamefont {{Alford}}, \citenamefont {{Kouvaris}},\ and\ \citenamefont {{Rajagopal}}}]{2005PhRvD..71e4009A}%
  \BibitemOpen
  \bibfield  {author} {\bibinfo {author} {\bibfnamefont {M.}~\bibnamefont {{Alford}}}, \bibinfo {author} {\bibfnamefont {C.}~\bibnamefont {{Kouvaris}}}, \ and\ \bibinfo {author} {\bibfnamefont {K.}~\bibnamefont {{Rajagopal}}},\ }\href {\doibase 10.1103/PhysRevD.71.054009} {\bibfield  {journal} {\bibinfo  {journal} {\prd}\ }\textbf {\bibinfo {volume} {71}},\ \bibinfo {eid} {054009} (\bibinfo {year} {2005})},\ \Eprint {http://arxiv.org/abs/hep-ph/0406137} {arXiv:hep-ph/0406137 [hep-ph]} \BibitemShut {NoStop}%
\bibitem [{\citenamefont {{Schmitt}}\ \emph {et~al.}(2003)\citenamefont {{Schmitt}}, \citenamefont {{Wang}},\ and\ \citenamefont {{Rischke}}}]{2003PhRvL..91x2301S}%
  \BibitemOpen
  \bibfield  {author} {\bibinfo {author} {\bibfnamefont {A.}~\bibnamefont {{Schmitt}}}, \bibinfo {author} {\bibfnamefont {Q.}~\bibnamefont {{Wang}}}, \ and\ \bibinfo {author} {\bibfnamefont {D.~H.}\ \bibnamefont {{Rischke}}},\ }\href {\doibase 10.1103/PhysRevLett.91.242301} {\bibfield  {journal} {\bibinfo  {journal} {\prl}\ }\textbf {\bibinfo {volume} {91}},\ \bibinfo {eid} {242301} (\bibinfo {year} {2003})},\ \Eprint {http://arxiv.org/abs/nucl-th/0301090} {arXiv:nucl-th/0301090 [nucl-th]} \BibitemShut {NoStop}%
\bibitem [{\citenamefont {{Schmitt}}(2005)}]{2005PhRvD..71e4016S}%
  \BibitemOpen
  \bibfield  {author} {\bibinfo {author} {\bibfnamefont {A.}~\bibnamefont {{Schmitt}}},\ }\href {\doibase 10.1103/PhysRevD.71.054016} {\bibfield  {journal} {\bibinfo  {journal} {\prd}\ }\textbf {\bibinfo {volume} {71}},\ \bibinfo {eid} {054016} (\bibinfo {year} {2005})},\ \Eprint {http://arxiv.org/abs/nucl-th/0412033} {arXiv:nucl-th/0412033 [nucl-th]} \BibitemShut {NoStop}%
\bibitem [{\citenamefont {{Aguilera}}\ \emph {et~al.}(2005)\citenamefont {{Aguilera}}, \citenamefont {{Blaschke}}, \citenamefont {{Buballa}},\ and\ \citenamefont {{Yudichev}}}]{2005PhRvD..72c4008A}%
  \BibitemOpen
  \bibfield  {author} {\bibinfo {author} {\bibfnamefont {D.~N.}\ \bibnamefont {{Aguilera}}}, \bibinfo {author} {\bibfnamefont {D.}~\bibnamefont {{Blaschke}}}, \bibinfo {author} {\bibfnamefont {M.}~\bibnamefont {{Buballa}}}, \ and\ \bibinfo {author} {\bibfnamefont {V.~L.}\ \bibnamefont {{Yudichev}}},\ }\href {\doibase 10.1103/PhysRevD.72.034008} {\bibfield  {journal} {\bibinfo  {journal} {\prd}\ }\textbf {\bibinfo {volume} {72}},\ \bibinfo {eid} {034008} (\bibinfo {year} {2005})},\ \Eprint {http://arxiv.org/abs/hep-ph/0503288} {arXiv:hep-ph/0503288 [hep-ph]} \BibitemShut {NoStop}%
\bibitem [{\citenamefont {{Weissenborn}}\ \emph {et~al.}(2011)\citenamefont {{Weissenborn}}, \citenamefont {{Sagert}}, \citenamefont {{Pagliara}}, \citenamefont {{Hempel}},\ and\ \citenamefont {{Schaffner-Bielich}}}]{2011ApJ...740L..14W}%
  \BibitemOpen
  \bibfield  {author} {\bibinfo {author} {\bibfnamefont {S.}~\bibnamefont {{Weissenborn}}}, \bibinfo {author} {\bibfnamefont {I.}~\bibnamefont {{Sagert}}}, \bibinfo {author} {\bibfnamefont {G.}~\bibnamefont {{Pagliara}}}, \bibinfo {author} {\bibfnamefont {M.}~\bibnamefont {{Hempel}}}, \ and\ \bibinfo {author} {\bibfnamefont {J.}~\bibnamefont {{Schaffner-Bielich}}},\ }\href {\doibase 10.1088/2041-8205/740/1/L14} {\bibfield  {journal} {\bibinfo  {journal} {\apjl}\ }\textbf {\bibinfo {volume} {740}},\ \bibinfo {eid} {L14} (\bibinfo {year} {2011})},\ \Eprint {http://arxiv.org/abs/1102.2869} {arXiv:1102.2869 [astro-ph.HE]} \BibitemShut {NoStop}%
\bibitem [{\citenamefont {{Yuan}}\ and\ \citenamefont {{Li}}(2024)}]{2024ApJ...966....3Y}%
  \BibitemOpen
  \bibfield  {author} {\bibinfo {author} {\bibfnamefont {W.-L.}\ \bibnamefont {{Yuan}}}\ and\ \bibinfo {author} {\bibfnamefont {A.}~\bibnamefont {{Li}}},\ }\href {\doibase 10.3847/1538-4357/ad354f} {\bibfield  {journal} {\bibinfo  {journal} {\apj}\ }\textbf {\bibinfo {volume} {966}},\ \bibinfo {eid} {3} (\bibinfo {year} {2024})},\ \Eprint {http://arxiv.org/abs/2312.17102} {arXiv:2312.17102 [nucl-th]} \BibitemShut {NoStop}%
\bibitem [{\citenamefont {{Rehberg}}\ \emph {et~al.}(1996)\citenamefont {{Rehberg}}, \citenamefont {{Klevansky}},\ and\ \citenamefont {{H{\"u}fner}}}]{1996PhRvC..53..410R}%
  \BibitemOpen
  \bibfield  {author} {\bibinfo {author} {\bibfnamefont {P.}~\bibnamefont {{Rehberg}}}, \bibinfo {author} {\bibfnamefont {S.~P.}\ \bibnamefont {{Klevansky}}}, \ and\ \bibinfo {author} {\bibfnamefont {J.}~\bibnamefont {{H{\"u}fner}}},\ }\href {\doibase 10.1103/PhysRevC.53.410} {\bibfield  {journal} {\bibinfo  {journal} {\prc}\ }\textbf {\bibinfo {volume} {53}},\ \bibinfo {pages} {410} (\bibinfo {year} {1996})},\ \Eprint {http://arxiv.org/abs/hep-ph/9506436} {arXiv:hep-ph/9506436 [hep-ph]} \BibitemShut {NoStop}%
\bibitem [{\citenamefont {{Liang}}\ \emph {et~al.}(2010)\citenamefont {{Liang}}, \citenamefont {{Yi}}, \citenamefont {{Zhang}}, \citenamefont {{L{\"u}}}, \citenamefont {{Zhang}},\ and\ \citenamefont {{Zhang}}}]{2010ApJ...725.2209L}%
  \BibitemOpen
  \bibfield  {author} {\bibinfo {author} {\bibfnamefont {E.-W.}\ \bibnamefont {{Liang}}}, \bibinfo {author} {\bibfnamefont {S.-X.}\ \bibnamefont {{Yi}}}, \bibinfo {author} {\bibfnamefont {J.}~\bibnamefont {{Zhang}}}, \bibinfo {author} {\bibfnamefont {H.-J.}\ \bibnamefont {{L{\"u}}}}, \bibinfo {author} {\bibfnamefont {B.-B.}\ \bibnamefont {{Zhang}}}, \ and\ \bibinfo {author} {\bibfnamefont {B.}~\bibnamefont {{Zhang}}},\ }\href {\doibase 10.1088/0004-637X/725/2/2209} {\bibfield  {journal} {\bibinfo  {journal} {\apj}\ }\textbf {\bibinfo {volume} {725}},\ \bibinfo {pages} {2209} (\bibinfo {year} {2010})},\ \Eprint {http://arxiv.org/abs/0912.4800} {arXiv:0912.4800 [astro-ph.HE]} \BibitemShut {NoStop}%
\bibitem [{\citenamefont {{L{\"u}}}\ \emph {et~al.}(2012)\citenamefont {{L{\"u}}}, \citenamefont {{Zou}}, \citenamefont {{Lei}}, \citenamefont {{Zhang}}, \citenamefont {{Wu}}, \citenamefont {{Wang}}, \citenamefont {{Liang}},\ and\ \citenamefont {{L{\"u}}}}]{2012ApJ...751...49L}%
  \BibitemOpen
  \bibfield  {author} {\bibinfo {author} {\bibfnamefont {J.}~\bibnamefont {{L{\"u}}}}, \bibinfo {author} {\bibfnamefont {Y.-C.}\ \bibnamefont {{Zou}}}, \bibinfo {author} {\bibfnamefont {W.-H.}\ \bibnamefont {{Lei}}}, \bibinfo {author} {\bibfnamefont {B.}~\bibnamefont {{Zhang}}}, \bibinfo {author} {\bibfnamefont {Q.}~\bibnamefont {{Wu}}}, \bibinfo {author} {\bibfnamefont {D.-X.}\ \bibnamefont {{Wang}}}, \bibinfo {author} {\bibfnamefont {E.-W.}\ \bibnamefont {{Liang}}}, \ and\ \bibinfo {author} {\bibfnamefont {H.-J.}\ \bibnamefont {{L{\"u}}}},\ }\href {\doibase 10.1088/0004-637X/751/1/49} {\bibfield  {journal} {\bibinfo  {journal} {\apj}\ }\textbf {\bibinfo {volume} {751}},\ \bibinfo {eid} {49} (\bibinfo {year} {2012})},\ \Eprint {http://arxiv.org/abs/1109.3757} {arXiv:1109.3757 [astro-ph.HE]} \BibitemShut {NoStop}%
\end{thebibliography}%
\end{document}